\documentclass[a4paper,11pt]{article} 
\usepackage[margin=0.7in]{geometry} 
\usepackage{amssymb}
\usepackage{amsmath,epsfig, epsf, epic, graphicx} 
\usepackage{fancybox, lscape, subfigure} 
\usepackage{bm}
\usepackage{natbib}
\usepackage{float}
\usepackage{color}
\usepackage{xargs, xstring}
\usepackage[nokeyprefix]{refstyle}
\usepackage{varioref}
\usepackage{xr-hyper}
\usepackage[colorlinks,citecolor=blue,urlcolor=blue]{hyperref}
\usepackage[capitalize,noabbrev]{cleveref}
\usepackage[inline]{enumitem}

\newtheorem{theorem}{Theorem}

\usepackage[margin=2cm]{caption}
\usepackage{setspace}

\title{Causal Inference in high dimensions: A marriage between Bayesian modeling and good frequentist properties}

\author{Joseph Antonelli, Georgia Papadogeorgou, and Francesca Dominici}

\begin{document}
\date{}
\maketitle{}

\newcommand\independent{\protect\mathpalette{\protect\independenT}{\perp}}
\def\independenT#1#2{\mathrel{\rlap{$#1#2$}\mkern2mu{#1#2}}}

\newcommand{\bD}{\boldsymbol{D}}
\newcommand{\bPsi}{\boldsymbol{\Psi}}
\newcommand{\tob}{{(b)}}
\newcommand{\tom}{{(m)}}
\newcommandx{\est}[2][1={},2={}]{{\Delta}(\bD^{#1}, \bPsi^{#2})}

\allowdisplaybreaks

\abstract{
We introduce a framework for estimating causal effects of binary and continuous treatments in high dimensions. We show how posterior distributions of treatment and outcome models can be used together with doubly robust estimators. We propose an approach to uncertainty quantification for the doubly robust estimator which utilizes posterior distributions of model parameters and (1) results in good frequentist properties in small samples, (2) is based on a single MCMC, and (3) improves over frequentist measures of uncertainty which rely on asymptotic properties. We show that our proposed variance estimation strategy is consistent when both models are correctly specified and that it is conservative in finite samples or when one or both models are misspecified. We consider a flexible framework for modeling the treatment and outcome processes within the Bayesian paradigm that reduces model dependence, accommodates nonlinearity, and achieves dimension reduction of the covariate space. We illustrate the ability of the proposed approach to flexibly estimate causal effects in high dimensions and appropriately quantify uncertainty, and show that it performs well relative to existing approaches. Finally, we estimate the effect of continuous environmental exposures on cholesterol and triglyceride levels.
}

\section{Introduction}


There has been a rapid growth in the interest of estimating the causal effect of a treatment ($T$) on an outcome ($Y$) when the dimension of the covariate space ($\boldsymbol{X}$) is large relative to the sample size. In high-dimensions, some form of dimension reduction or variable selection is required, and traditional approaches to reducing the dimension of the parameter space can lead to biased estimators with nonregular asymptotic behavior. Recent work has focused on tailoring these approaches to the specific problem of estimating treatment effects in high dimensions. \cite{tan2020regularized} extended inverse propensity weighted estimators to this setting, and  estimated the propensity scores with a combination of regularization and calibration to improve inference on treatment effects. However, most approaches utilize both the treatment and outcome to reduce the dimension of the parameter space to reduce confounding bias. \cite{belloni2013inference} suggested using a post-selection estimator based on the union of variables selected from a first-stage treatment and outcome model, and showed that the resulting inference on treatment effects is uniformly valid. \cite{athey2016approximate} combine outcome regression models with weights that balance any remaining differences in covariates between treated and control units. Other approaches have focused on one of either the treatment or outcome model, but allowed the amount of shrinkage or regularization to depend on the parameters of the other model in a way that improves finite sample performance \citep{antonelli2017high, shortreed2017outcome, hahn2018regularization}. Relatedly, \cite{ertefaie2018variable} derived a penalization-based estimator that incorporates information from both the treatment and outcome to identify confounders, and estimates treatment effects conditional on this chosen set. 

While most of the previous approaches utilized both a treatment and outcome model, none are doubly robust in the sense that only one of the two models needs to be correctly specified for consistent estimation of treatment effects. Doubly robust estimators have been widely used in the causal inference literature \citep{bang2005doubly}, as they allow for increased robustness to model misspecification and allow for data-adaptive estimation of the treatment and outcome model. See \cite{daniel2014double} for a nice overview of doubly robust estimators and their properties. These estimators have been extended to high-dimensional settings \citep{farrell2015robust,chernozhukov2016double}, but inference for these estimators is not doubly robust: While they are consistent when only one of the two models is correctly specified, corresponding confidence intervals rely on both models being correctly specified to obtain valid inference. Other approaches that provide doubly robust point estimators but not confidence intervals include doubly robust matching estimators \citep{antonelli2016double} or targeted maximum likelihood estimators \citep{van2006targeted} combined with high-dimensional models. To address this gap in the literature, recent work has looked to construct doubly robust estimators that simultaneously admit doubly robust inference. \cite{avagyan2017honest} construct estimates of treatment and outcome model parameters using penalized estimating equations to produce so-called bias-reduced doubly robust estimators, which in turn lead to valid inference even when one model is misspecified, though this approach does not extend to cases where the parameter space is larger than the sample size. \cite{dukes2020doubly} construct tests for the null hypothesis of no treatment effect that are valid even when one model is misspecified. They assume strong conditions on the sparsity of the treatment and outcome models, but show that these conditions can be weakened when both models are correctly specified. \cite{ning2018robust} achieve doubly robust inference through covariate balancing propensity scores that target balance of variables that are associated with the outcome, while \cite{tan2018model} achieve doubly robust inference through calibrated propensity scores and weighted penalized outcome models. Both of these approaches rely on strong sparsity assumptions, as well as linearity assumptions for the outcome model. Despite providing doubly robust confidence intervals, inference for all of these approaches is based on asymptotic arguments that may not perform well in finite samples.

In this paper, we propose a doubly robust estimator that is based on Bayesian models for the treatment and outcome model, and an approach to inference on treatment effects with improved finite sample performance. Our inferential approach uses the posterior distribution of both models, and it does not rely on asymptotic approximations or on correct model specification. We show theoretically that our proposed variance estimator is consistent as long as both models are correctly specified, and that it is generally conservative in finite samples or when one or both models are misspecified. This is related to existing work that aims to improve inference for doubly robust estimators under model misspecification, though our work differs in that it explicitly focuses on finite sample variance estimation. Importantly, our results do not rely on linearity of the treatment or outcome model, and very complex models such as tree-based models or Gaussian process specifications can be used. We show that the proposed approach is directly applicable to high-dimensional settings with continuous treatments, a scenario that has been overlooked in the literature. We show empirically that the proposed procedure leads to improved performance, in particular with respect to inference in finite samples, when existing approaches that rely on asymptotics do not perform well.

\section{Notation, estimands, and identifying assumptions}
\label{sec:notation}

Let $T$ and $Y$ be the treatment and outcome of interest, respectively, while $\boldsymbol{X}$ is a $p-$dimensional vector of potential confounders. We observe an i.i.d sample $\boldsymbol{D}_i = (\boldsymbol{X}_i, T_i ,Y_i)$ for $i=1 \dots n$, and denote $\bD=(\bD_1, \bD_2, \dots, \bD_n)$. We work under the high-dimensional situation where the number of covariates exceeds the sample size, and is potentially growing with the sample size. We focus our attention on binary treatments and the average treatment effect (ATE) defined as $\Delta = E(Y(1) - Y(0))$, where $Y(t)$ is the potential outcome that would have been observed under treatment $T=t$. We assume that the stable unit treatment value assumption (SUTVA) \citep{little2000causal} holds, and that for each unit the same treatment cannot lead to different outcomes, implying $Y_i = Y_i(T_i)$.
Assuming SUTVA, the average treatment effect can be identified from observed data based on the following assumptions: \\
\indent \textit{Unconfoundedness:} $T \independent Y(t) \vert \boldsymbol{X}$ for $t=0,1$, \\
\indent \textit{Positivity:} There exist $\delta \in (0,1)$ such that $0 < \delta < P(T=1 \vert \boldsymbol{X}) < 1-\delta < 1$, \\
\noindent where $P(T=1 \vert \boldsymbol{X})$ denotes the propensity score \citep{rosenbaum1983central}. Positivity states that all subjects have a positive probability of receiving either treatment level. Unconfoundedness states that there are no unmeasured confounders and that the set of measured variables $\boldsymbol{X}$ contains all common causes of the treatment and outcome.

Throughout we focus on doubly robust estimators. Specifically, if $\boldsymbol{\Psi}$ represents the parameters of the propensity score and outcome models, and $p_{ti} = P(T_i = t \vert \boldsymbol{X_i})$, and $m_{ti} = E(Y_i \vert T_i = t, \boldsymbol{X}_i)$ represent the values of the treatment and outcome models based on the parameters $\boldsymbol{\Psi}$, 
a doubly robust estimator of the ATE for binary treatments is
\begin{align}
\est = \frac{1}{n} \sum_{i=1}^n \left[ \frac{T_i Y_i}{p_{1i}} - \frac{(T_i - p_{1i}) m_{1i}}{p_{1i}} -  \frac{(1 - T_i) Y_i}{p_{0i}} - \frac{(T_i - p_{1i}) m_{0i}}{p_{0i}} \right].
\label{eqn:DRest}
\end{align}

Even though our primary focus is on the ATE for binary treatments, we discuss continuous treatments, estimands and doubly robust estimators in this setting in Sections \ref{sec:simGlobal} and \ref{sec:app}. Identifying assumptions for continuous treatments are analogous to the ones stated above for binary treatments, though we refer readers to previous literature for details \citep{gill2001causal,hirano2004propensity,kennedy2017non}.

\vspace{-10pt}
\section{Doubly robust estimation with posterior distribution of nuisance parameters}
\label{sec:uncertainty}

Parameter values $\bPsi$ are typically not known and must be estimated. We consider the case where the propensity score and outcome models are estimated within a Bayesian framework, in which uncertainty in parameter estimation is directly acquired from the posterior distribution. We discuss one modeling approach for high dimensional data in \cref{sec:model}, but the framework presented here allows for any Bayesian modeling technique.

When the propensity score and outcome model parameters are estimated within the Bayesian framework, questions arise on how to use the posterior distribution of these models to (1) acquire estimates of treatment effects, and (2) perform inference with good frequentist properties. In this section, we discuss an approach that achieves both.

\subsection{The doubly robust estimator using posterior distributions}

First, we introduce our estimator which combines the doubly robust estimator in (\ref{eqn:DRest}) with Bayesian estimation of nuisance parameters. Let $[\bPsi | \bD]$ denote the posterior distribution, and $\big\{\bPsi^\tob \big\}_{b = 1}^B$ be $B$ draws from this posterior distribution. We define our estimator $\widehat \Delta$ as:
\begin{equation}
\widehat{\Delta} =  E_{\bPsi \vert \bD}[\est] \approx \frac1B \sum_{b=1}^B \Delta(\boldsymbol{D}, \boldsymbol{\Psi}^{(b)}),
\label{eq:our_estimator}
\end{equation}
where $\est[][\tob]$ is the quantity in (\ref{eqn:DRest}) evaluated using the observed data $\bD$ and parameters $\bPsi^\tob$. Hence, our estimator is the average value of (\ref{eqn:DRest}) with respect to the posterior distribution of model parameters. In \cref{subsec:consistent}, we discuss the estimator's asymptotic properties.

An alternative approach to combining the posterior distribution of model parameters and the doubly robust estimator in (\ref{eqn:DRest}) would substitute the nuisance parameters $p_{ti}$ and $m_{ti}$ with plug-in estimates such as their posterior means. Doing so would be in line with frequentist settings, in which doubly robust estimators are evaluated using plug-in estimates of the parameters $\bPsi$. We have found empirically that this alternative estimator leads to similar results. However, we focus our attention on the estimator in (\ref{eq:our_estimator}) since it is amenable to the variance estimation strategy presented in the following sections.

\subsection{Approach to inference}

The estimator in (\ref{eq:our_estimator}) is a straightforward combination of the doubly robust estimator in (\ref{eqn:DRest}) and the posterior distribution of $\boldsymbol{\Psi}$. Typically in Bayesian inference, the posterior distribution of \textit{model parameters} or \textit{functionals of these parameters} is sufficient for uncertainty quantification. However, the doubly robust estimator in (\ref{eqn:DRest}) is a function of {\it both} the parameters $\bPsi$ and the data $\bD$, prohibiting us from using the variance or quantiles of $\big\{ \est[][\tob] \big\}_{b = 1}^B$ to perform inference directly, and rendering inference more complicated.



In this section, we focus on deriving an inferential approach for the estimator (\ref{eq:our_estimator}) with good finite sample frequentist properties.
Frequentist operating characteristics (such as interval coverage) are focused around the estimator's sampling distribution, the distribution of point estimates over different data sets. Therefore, our {\it target variance} corresponds to the variance of our estimator's sampling distribution, which can be written as 
\begin{equation}
\text{Var}_{\bD} \big(\widehat{\Delta}\big) = \text{Var}_{\bD} \{ E_{\bPsi \vert \bD} [\est] \}.
\label{eq:target_variance}
\end{equation}

In an ideal world, this variance could be estimated by repeatedly sampling from the distribution of $\boldsymbol{D}$, calculating the posterior distribution of $\bPsi$ and the resulting $E_{\boldsymbol{\Psi} \vert \boldsymbol{D}} [\est]$ for each data set, and taking the variance of these values. We cannot do this for two reasons: 1) We do not know the distribution of the data, and 2) even if we did, it would be computationally prohibitive to estimate the posterior mean for each new data set.

Here, we detail our approach to uncertainty quantification which combines the posterior distribution of parameters based on our observed data with an efficient resampling procedure, and therefore it completely by-passes calculating the posterior distribution over multiple data sets. Theoretical properties of the proposed inferential approach are discussed in Sections \ref{subsec:consistent} and \ref{subsec:misspecified}.
First, we create $M$ new data sets, $\bD^{(1)}, \dots, \bD^{(M)}$, by sampling with replacement from the empirical distribution of the data. 
 For all possible combinations of resampled data sets and posterior samples, we calculate $ \est[\tom][\tob]$ for $m=1,\dots, M$, and $b=1,\dots, B$.
 We re-iterate here that $\boldsymbol{\Psi}^{(b)}$ denotes a sample of $\boldsymbol{\Psi}$ conditional on the observed data and it is not re-estimated for every re-sampled data set.
 The values $\est[\tom][\tob]$ can be arranged in a matrix of dimensions $M\times B$ where rows correspond to data sets, and columns correspond to posterior samples of $\boldsymbol{\Psi}$ conditional on the observed data, as shown in \cref{fig:matrix_deltas}.
\begin{figure}
\caption{Values of $\est$ for different combinations of resampled data sets and posterior samples}
\label{fig:matrix_deltas}
{
\begin{align*}
\begin{pmatrix}
\Delta(\boldsymbol{D}^{(1)}, \boldsymbol{\Psi}^{(1)}) & \Delta(\boldsymbol{D}^{(1)}, \boldsymbol{\Psi}^{(2)}) & \dots & \Delta(\boldsymbol{D}^{(1)}, \boldsymbol{\Psi}^{(B)}) \\
\Delta(\boldsymbol{D}^{(2)}, \boldsymbol{\Psi}^{(1)}) & \ddots &  & \Delta(\boldsymbol{D}^{(2)}, \boldsymbol{\Psi}^{(B)}) \\
\vdots &  & \ddots &  \vdots \\
\Delta(\boldsymbol{D}^{(M)}, \boldsymbol{\Psi}^{(1)}) & \Delta(\boldsymbol{D}^{(M)}, \boldsymbol{\Psi}^{(2)}) & \dots & \Delta(\boldsymbol{D}^{(M)}, \boldsymbol{\Psi}^{(B)})
\end{pmatrix}
\longrightarrow &
\begin{pmatrix}
E_{\bPsi \vert \bD} [\est[(1)]] \\[5pt]
E_{\bPsi \vert \bD} [\est[(2)]] \\ \vdots \\
E_{\bPsi \vert \bD} [\est[(M)]]
\end{pmatrix}\\
& \hspace{0.13\textwidth} \big\downarrow \\
& \text{Var}_{\bD^\tom}\{ E_{\bPsi \vert \bD} [\est[\tom]] \}
\end{align*}}
\end{figure}
We acquire $E_{\bPsi \vert \bD} [\est[\tom]]$ for $m=1,\dots,M$ by calculating the mean within each row. The variance of these $M$ values is an estimate of $\text{Var}_{\bD^\tom} \{ E_{\bPsi | \bD} [\est[\tom]] \}$. This variance resembles the target variance in (\ref{eq:target_variance}) but is not equal to it for two reasons. First, the new data $\bD^\tom$ are drawn from the empirical distribution of the data instead of the true joint distribution. This is acceptable in many settings and it is the main idea behind the bootstrap.
The second and most important reason is that that distribution used in the outer moment $(\bD^\tom)$ does not agree with the one in the inner moment ($\bPsi \vert \bD$). That is because the estimator $E_{\bPsi | \bD}[\est[\tom]]$  relies on the posterior distribution from the original data, $\bPsi | \bD$, instead of the posterior distribution based on the resampled data, $\bPsi | \bD^{(m)}$.
Therefore, $\text{Var}_{\bD^\tom} \{ E_{\bPsi | \bD} [\est[\tom]] \}$ 
ignores uncertainty stemming from the fact that the posterior distribution of model parameters might be different across data sets.
Hence, inference based on this quantity achieves close to the nominal level only when the uncertainty stemming from the variability in the parameters is small relative to the uncertainty stemming from the variability of the data, and it is likely to be anti-conservative in other settings (see \cref{sec:sims_more} and Appendix I). For the reasons stated above, we refer to $\text{Var}_{\bD^\tom} \{ E_{\bPsi | \bD} [\est[\tom]] \}$ as the {\it na\"ive variance}.

We propose a correction to the na\"ive variance that explicitly targets the uncertainty stemming from parameter variability, and allows us to perform inference based on a {\it single} posterior distribution (instead of $M$).
Our proposed variance estimator is
\begin{align}
\underbrace{\text{Var}_{\bD^\tom} \{ E_{\bPsi | \bD} [\est[\tom]] \}}_{\text{na\"ive variance}} +
\underbrace{\text{Var}_{\bPsi \vert \bD} [\est]}_{\text{correction}}. \label{eqn:VarEst}
\end{align}
Here, we discuss why the second term in (\ref{eqn:VarEst}) corrects for this missing uncertainty.
\cite{freedman1999wald} showed that the posterior distribution of a function of $\bPsi$ resembles the sampling distribution of its posterior mean. 
Letting $\boldsymbol{D}^{obs}$ be our observed data, this result implies that $\text{Var}_{\bPsi \vert \boldsymbol{D}^{obs}} [\est[{obs}]] \approx \text{Var}_{\bD} \{ E_{\bPsi \vert \bD} [\est[{obs}]] \}$, which is exactly the variability that the na\"ive variance ignores, the variance which stems {\it only} from the uncertainty in the estimation of the nuisance parameters.

We show that the variance estimator in (\ref{eqn:VarEst}) is consistent if both the propensity score and outcome regression models are correctly specified (\cref{subsec:consistent}), while it tends to be conservative if one or both models are misspecified (\cref{subsec:misspecified}). In Section \ref{sec:simGlobal} we empirically show that it accurately approximates the Monte Carlo variance under various scenarios.

\subsection{Consistency of the point estimator and variance estimator}
\label{subsec:consistent}

In this section we detail important results about the point estimator and our variance estimator, which justify the use of our proposed procedure. First, we show that our point estimator, $\widehat \Delta = E_{\bPsi | \bD} [\est]$, is doubly robust, and highlight that its convergence rate is a function of the posterior contraction rates of the treatment and outcome models. Second, we show that our variance estimator in (\ref{eqn:VarEst}) is consistent for the true variance if both models are correctly specified and contract at sufficiently fast rates.

We highlight a few important assumptions for these results to hold, but all other assumptions and any mathematical derivations are included in Appendices A and B. 
Let $\boldsymbol{p}_t = (p_{t1}, \dots, p_{tn})$, $\boldsymbol{m}_t = (m_{t1}, \dots, m_{tn})$, and let $\boldsymbol{p}_t^*$ and $\boldsymbol{m}_t^*$ denote their unknown, true values. Assume that $\boldsymbol{D}_i$ arises from a distribution $P_0$, and let $\mathbb{P}_n$ denote the posterior distribution from a sample of size $n$. We make the following assumption:

\textit{Contraction of treatment and outcome models}:
There exist two sequences of numbers $\epsilon_{nt} \rightarrow 0$ and $\epsilon_{ny} \rightarrow 0$, and constants $M_t > 0$ and $M_y > 0$ such that
\begin{enumerate}[label=(\alph*),leftmargin=*,topsep=0pt]
	\item $\sup\limits_{P_0} E_{P_0} \mathbb{P}_n \bigg( \frac{1}{\sqrt{n}}||\boldsymbol{p}_{t} - \boldsymbol{p}_{t}^*|| > M_t \epsilon_{nt} \mid \boldsymbol{D} \bigg) \rightarrow 0$,
    \item $\sup\limits_{P_0} E_{P_0} \mathbb{P}_n \bigg( \frac{1}{\sqrt{n}}||\boldsymbol{m}_{t} - \boldsymbol{m}_{t}^*|| > M_y \epsilon_{ny} \mid \boldsymbol{D} \bigg) \rightarrow 0$,
\end{enumerate}
where $||v|| = \sqrt{v_1^2 + \dots + v_n^2}$. This assumption states that the posterior distributions of the treatment and outcome models contract at rates $\epsilon_{nt}$ and $\epsilon_{ny}$, respectively. In low-dimensional parametric settings, if models are correctly specified, then we expect these rates to be $n^{-{1/2}}$, while slower rates are expected in high-dimensional or nonparametric settings. 

Next, we state our main result regarding point estimation. 

\begin{theorem}
Assume positivity, unconfoundedness, SUTVA, and additional regulatory assumptions found in Appendix A. If the contraction assumptions hold, then
\begin{align}
    \sup\limits_{P_0} E_{P_0} \mathbb{P}_n(|\est - \Delta^*| > M\epsilon_n \mid \boldsymbol{D}) \rightarrow 0, \label{eqn:rate}
\end{align}
with $\epsilon_n = \text{max}(n^{-1/2}, \epsilon_{nt}\epsilon_{ny})$. If only one of contraction assumptions (a) or (b) hold with contraction rate $\eta_n$, then (\ref{eqn:rate}) is satisfied with $\epsilon_n = \text{max}(n^{-1/2}, \eta_n)$.
\label{theorem:contraction}
\end{theorem}

Our setting is considerably different from frequentist settings, since our estimator is based on the full posterior distribution of the treatment and outcome models, necessitating the use of posterior contraction results. Importantly, posterior contraction in \cref{theorem:contraction} directly implies that our point estimator, $E_{\boldsymbol{\Psi} \vert \boldsymbol{D}} [\est]$, converges at the same rates. This result also implies that our estimator is doubly robust as we only need one of the contraction assumptions (a) or (b) to hold, and not necessarily both,  to ensure consistency of the proposed estimator, but that the convergence rate is {\it faster} if they {\it both} hold.

Next, we highlight our result on the variance estimation strategy. Denote the variance estimator in (\ref{eqn:VarEst}) by $\widehat{V}$, and the estimator's true variance by $V$. 

\begin{theorem}
Assume the same conditions as in \cref{theorem:contraction} as well as the additional regulatory conditions found in Appendix A. If contraction assumptions (a) and (b) both hold at faster rates than $n^{-{1/4}}$, then $\widehat{V} - V = o(n^{-1})$.
\label{theorem:variance}
\end{theorem}

\cref{theorem:variance} shows that our variance estimator is asymptotically consistent as long as both models are correctly specified and their posterior distributions contract sufficiently fast. For an empirical illustration of the asymptotic performance of our variance estimator, see Appendix E. 
Our proof shows that if both models are correctly specified, the uncertainty of our doubly-robust estimator stemming from the variability in the parameters can be asymptotically ignored. This implies that, even though our estimator uses the full posterior distribution, it is as efficient as doubly-robust estimators which use a consistent estimator of the nuisance parameters.
\cref{theorem:variance} implies that while our estimator is doubly robust, consistency of our variance estimator relies on both models being correctly specified. This is a well-known problem for doubly robust estimators, and there has been recent work at ensuring double robustness of both point estimation and inference \citep{avagyan2017honest, benkeser2017doubly, ning2018robust, tan2018model}.

While \cref{theorem:variance} holds asymptotically when both models are correctly specified, our goal is a variance estimation strategy with good {\it finite sample} performance across all scenarios, even when models are {\it misspecified}. In the following section, we discuss that our variance estimator is expected to be conservative when one model is misspecified, and in \cref{sec:simGlobal} we empirically illustrate that accounting for parameter uncertainty improves finite sample performance even when both models are correctly specified.
Therefore, our variance estimator contributes to the literature on performing valid inference using doubly robust estimators in 
\begin{enumerate*}[label=(\alph*)]
\item small samples, and 
\item when one model is misspecified.
\end{enumerate*}

\subsection{Performance when one model is misspecified}
\label{subsec:misspecified}

While uncertainty stemming from parameter estimation can be asymptotically ignored if both models are correctly specified, this does not hold when one model is misspecified, and approaches in the literature designed to alleviate this issue are based on asymptotic arguments themselves.
In contrast, our approach 
does not assume that any components of the doubly robust estimator are asymptotically negligible.

Here, we discuss that our variance estimator 
is expected to be positively biased in small samples or when one model is misspecified.
In Appendix D we show that the difference between the target and na\"ive variances is approximately
\begin{align}
    E_{\boldsymbol{D}} \bigg[E^2_{\boldsymbol{\Psi} \vert \boldsymbol{D}} \Big(\Delta(\boldsymbol{D}, \boldsymbol{\Psi})\Big) \bigg] - E_{\boldsymbol{D}}  \bigg[E^2_{\boldsymbol{\Psi} \vert \boldsymbol{D}^{obs}} \Big(\Delta(\boldsymbol{D}, \boldsymbol{\Psi})\Big) \bigg], \label{eqn:ApproxError}
\end{align}
and we study its relative magnitude to the expectation of the correction term in (\ref{eqn:VarEst}) over $\boldsymbol{D}$:
\begin{align}
    E_{\boldsymbol{D}} \bigg[ \underbrace{\text{Var}_{\bPsi \vert \bD} [\est]}_{\text{correction}} \bigg] = E_{\boldsymbol{D}}  \bigg[E_{\boldsymbol{\Psi} \vert \boldsymbol{D}} \Big(\Delta(\boldsymbol{D}, \boldsymbol{\Psi})^2\Big) \bigg] - E_{\boldsymbol{D}}  \bigg[E^2_{\boldsymbol{\Psi} \vert \boldsymbol{D}} \Big(\Delta(\boldsymbol{D}, \boldsymbol{\Psi})\Big) \bigg]. \label{eqn:Correction}
\end{align}
From Equations (\ref{eqn:ApproxError}) and (\ref{eqn:Correction}), we see that the second terms are similar, with the difference being that the inner moment in (\ref{eqn:ApproxError}) is with respect to $\bPsi | \bD^{obs}$, the observed data posterior, whereas the inner moment in (\ref{eqn:Correction}) averages over different posterior distributions $\bPsi | \bD$. However, since these quantities correspond to expectations, these terms are expected to be comparable. 
Turning our focus to the first terms and using Jensen's inequality, the first term in (\ref{eqn:ApproxError}) is smaller than or equal to the first term in (\ref{eqn:Correction}). Hence, our variance estimate is expected to be larger than the true variance in finite samples or when one or both models are misspecified. This does not guarantee conservative inference, however, as there can still be non-negligible bias that affects confidence interval coverage when one of the two models is misspecified. In Section \ref{sec:simGlobal}, we see empirically that our variance does not lead to overly conservative inference, and confidence interval coverage is at or near the nominal level. 

\vspace{-10pt}
\section{Modeling framework in high dimensions}
\label{sec:model}

While our estimation and inferential approach works in general, it is most useful in high-dimensions where accounting for uncertainty in parameter estimation can be quite difficult. We posit Bayesian high-dimensional treatment and outcome models as
\begin{equation}
\begin{aligned}
h_y^{-1}(E(Y_i \vert T_i, \boldsymbol{X}_i)) &= \beta_0 + f_t(T_i) + \sum_{j=1}^p f_j(X_{ji})\\
h_t^{-1}(E(T_i \vert \boldsymbol{X}_i)) &= \alpha_0 + \sum_{j=1}^p g_j(X_{ji}),
\end{aligned}
\label{eq:models}
\end{equation}

\subsection{Gaussian process prior specification}
\label{sec:nonpara}

The functional form of the relationships between the covariates and the treatment or outcome are unspecified in (\ref{eq:models}).
Here, we present the prior specification for the outcome model only, but analogous representations are used for the treatment model. We adopt Gaussian process priors for the unknown regression functions, $f_j()$ and $g_j()$ for $j=1,\dots,p$.
Since we only need to evaluate $f_j()$ at the $n$ observed locations, we denote the vector of values for the $j^{th}$ covariate as $\boldsymbol{X}_j = (X_{j1}, \dots, X_{jn})$ and represent our prior as:
\begin{equation}
\begin{aligned}
f_j(\boldsymbol{X_j}) &\sim (1 - \gamma_j) \delta_{\boldsymbol{0}} + \gamma_j \mathcal{N}(\boldsymbol{0}_n, \sigma^2 \tau_j^2 \boldsymbol{\Sigma}_j) & \\
\gamma_j & \sim \text{Bernoulli}(\theta) \hspace{2cm}
\theta \sim \mathcal{B}(a_{\theta},b_{\theta}) \\
\tau_j^2 & \sim \text{Gamma}(1/2, 1/2)  \hspace{1cm}
\sigma^2 \sim \text{InvGamma}(a_{\sigma^2},b_{\sigma^2}).
\end{aligned}
\label{eq:GP}
\end{equation}
Here, $\sigma^2$ is the residual variance of the model when the outcome is normally distributed, otherwise it is fixed to 1. We utilize a binary latent variable, $\gamma_j$, which indicates whether variable $j$ is important for predicting the outcome ($\gamma_j = 1$), or not ($\gamma_j = 0$).  We adopt a gamma$(1/2, 1/2)$ prior on the variance $\tau_j^2$ \citep{mitra2010two}. Finally, the $(i, i')$ entry of the covariance matrix $\boldsymbol{\Sigma}_j$ is $K(X_{ji}, X_{ji'})$, where $K(\cdot, \cdot)$ is the kernel function of the Gaussian process. We proceed with $K(z, z') =\exp\{-\frac{|z - z'|}{\phi}\}$, where $\phi$ is a bandwidth parameter that must be chosen. 

The formulation above allows for flexible modeling of the response functions $f_j()$, but it can be very computationally burdensome as the sample size increases. \cite{reich2009variable} showed that the computation burden can be alleviated by using a singular value decomposition of the kernel covariance matrices. This allows us to utilize Gaussian processes in reasonably sized data sets, but the computation can still be slow for large sample sizes. Details of this can be found in the referenced paper and in Appendix C.

\subsection{Basis expansion specification}
\label{sec:semi}

The computational burden of using Gaussian processes can be greatly alleviated by using basis functions. Even though using basis functions are less flexible for estimating $f_j(\boldsymbol{X_j})$, it greatly reduces the computational complexity which allows us to model much larger data sets. To do this, we must introduce some additional notation. Let $\boldsymbol{\widetilde{X}}_j$ represent an $n$ by $q$ matrix of basis functions, such as cubic splines. We write $f_j(\boldsymbol{X_j}) = \boldsymbol{\widetilde{X}}_j \boldsymbol{\beta}_j$ and assume: 
\begin{align*}
	&(\boldsymbol{\beta}_j \vert \gamma_j) \sim  (1 - \gamma_j) \delta_{\boldsymbol{0}_q} + \gamma_j \mathcal{N}(\boldsymbol{0}_q, \sigma^2 \sigma_{\boldsymbol{\beta}}^2 I_q)
\end{align*}
with prior specification on $\gamma_j$ and $\sigma^2$ as in (\ref{eq:GP}), and $\sigma_{\boldsymbol{\beta}}^2$ either selected via empirical Bayes, or by adopting a hyper prior.
This specification places a multivariate spike and slab prior on the {\it group} of coefficients, $\boldsymbol{\beta}_j$, that either specifies a multivariate normal prior distribution ($\gamma_j = 1$), or
forces them all to be zero and eliminates the covariate $j$ completely ($\gamma_j = 0$).

\section{Simulation studies}
\label{sec:simGlobal}

We conduct extensive simulation studies to evaluate our proposed estimation and variance procedure in the presence of binary and continuous treatments, linear and non-linear settings, with and without correct model specification, and varying dimensionality ($p/n$ ratio). A subset of these results are shown here. Additional results are shown in the Appendix, and are summarized below.

\subsection{Binary treatments}
\label{subsec:sims_binary}

We set $n=100$ and $p=500$, and generate data as:
\begin{align*}
\boldsymbol{X}_i \sim N(\boldsymbol{0}_p, \boldsymbol{\Sigma}), \hspace{1cm}
T_i \vert \boldsymbol{X}_i \sim \text{Bernoulli}(p_i)
\hspace{1cm}
Y_i \vert T_i, \boldsymbol{X}_i \sim \mathcal{N}(\mu_i, 1).
\end{align*}
We set $\Sigma_{ij} = 1$ if $i=j$ and $\Sigma_{ij} = 0.3$ if $i \neq j$. We simulate data under two scenarios for the true propensity and outcome regressions:
\begin{align*}
\textbf{Linear Simulation:} \ \ \ \ \ \
	\mu_i &= T_i + 0.75X_{1i} + X_{2i} + 0.6X_{3i} - 0.8X_{4i} - 0.7X_{5i} \\
    p_i &= \Phi(0.15 X_{1i} + 0.2 X_{2i} - 0.4 X_{5i}) \\
   \textbf{Nonlinear Simulation:} \ \ \ \ \ \ \
	\mu_i &= T_i + 0.8X_{1i} + 0.4X_{2i}^3 + 0.25 e^{|X_{2i}|} + 0.8X_{5i}^2 - 1.5\text{sin}(X_{5i}) \\
    p_i &= \Phi(0.15 X_{1i} - 0.4 X_{2i} - 0.5 X_{5i})  
\end{align*}

We fit our approach in the following manner: We consider models (\ref{eq:models})
\begin{enumerate*}[label= \alph*)]
\item specified as linear,
\item using 3 degree of freedom splines for each covariate (Section \ref{sec:semi}), or
\item using Gaussian process priors for each covariate (Section \ref{sec:nonpara}).
\end{enumerate*}
We choose the treatment and outcome model fit among a), b) and c) that minimizes the WAIC (\citep{watanabe2010asymptotic}), a Bayesian analog to model selection criteria such as AIC or BIC. Based on these model fits, we calculate the value of our estimator in (\ref{eq:our_estimator}) and its variance in (\ref{eqn:VarEst}). We refer to this estimator as \texttt{Bayes-DR}. 

In addition to our estimator, we estimate the average treatment effect using: 
\begin{enumerate*}[label = \alph*)]
\item \texttt{Double PS}: double post selection regression introduced in \cite{belloni2013inference};
\item \texttt{lasso-DR}: the doubly robust estimator introduced in \cite{farrell2015robust};
\item \texttt{De-biasing}: the residual de-biasing approach of \cite{athey2016approximate};
\item \texttt{TMLE}: Targeted maximum likelihood with lasso models; and
\item \texttt{DML}: the double machine learning approach of \cite{chernozhukov2016double} with lasso models.
\end{enumerate*}
For all competing approaches, we always use a correctly specified, linear treatment model. In the linear simulation scenario, they are implemented using a correctly specified, linear outcome model as well. That is in contrast to \texttt{Bayes-DR} that considers a collection of linear and non-linear models in the estimation of the treatment and outcome models in both linear and non-linear settings. In the nonlinear scenarios, we only compare with \texttt{TMLE} and \texttt{DML}, as the other approaches rely directly on linearity. In the non-linear settings we implement \texttt{TMLE} and \texttt{DML} by fitting a group lasso model as a screening step, and then nonlinear outcome models on the chosen covariates as a second step.
For all approaches, asymptotic standard errors were estimated, and confidence intervals were acquired based on an asymptotic normal approximation. More details of our implementation of these approaches can be found in Appendix J.

Figure \ref{fig:sim1test} shows the absolute percent bias, variance, coverage of 95\% intervals, and ratio of estimated and Monte Carlo variance in the linear and non-linear settings, for all estimators. The proposed estimator is in grey, while the alternative approaches are in black. In the linear setting, \texttt{Bayes-DR} is the only estimator that achieves interval coverage near the nominal level, while having the smallest bias {\it and} variance across all estimators. In the nonlinear simulation, \texttt{Bayes-DR} obtains the lowest MSE, and again achieves coverage close to the nominal level. 

\begin{figure}[h]
\centering
 \includegraphics[width=0.9\linewidth]{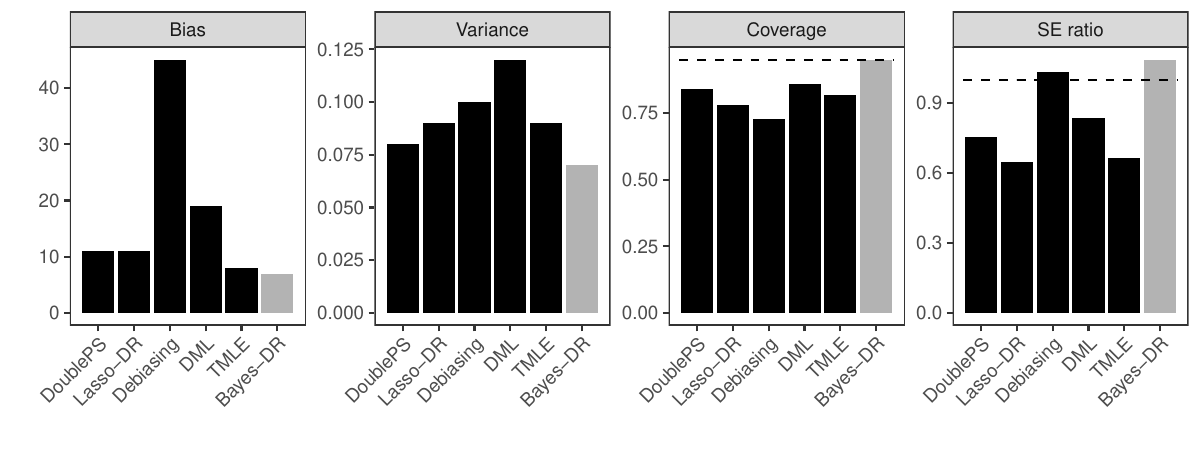} \\
  \includegraphics[width=0.9\linewidth]{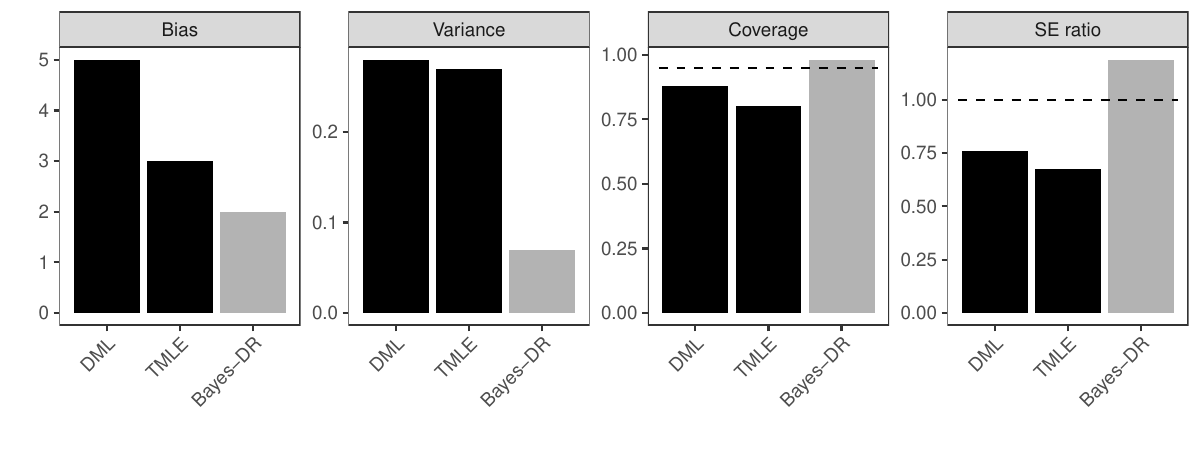}
\caption{Results from simulations with binary treatments. The top panel shows results for the linear scenario, while the bottom panel shows results for the nonlinear scenario. The first column shows absolute bias, the second column shows the variance, the third column shows 95\% interval coverages, while the fourth column is the ratio of estimated to Monte Carlo standard errors. }
\label{fig:sim1test}
\end{figure}

\subsection{Continuous treatments}
\label{subsec:sims_continuous}

We turn our attention to continuous treatments with target estimand the exposure response curve $E[Y(t)]$ for all $t$ in the support of $T$. We generate data with $n=200$, $p = 200$, and
\begin{align*}
    \boldsymbol{X}_i \sim N(\boldsymbol{0}_p, \boldsymbol{\Sigma}),
\hspace{1.5cm}
    T_i \vert \boldsymbol{X}_i \sim \mathcal{N}(\mu_{i}^t, 1), 
\hspace{1.5cm}
    Y_i \vert T_i, \boldsymbol{X}_i \sim \mathcal{N}(\mu_{i}^y, 1),
\end{align*}
\begin{align*}
    \mu_{i}^t &= 0.6 X_{1i}^2 + 0.6X_{1i} + \text{exp}(0.65|X_{2i}|) -
      0.8 X_{3i}^2, \\
	\mu_{i}^y &= 5 + 0.05T_i^3 - 0.1T_i^2 + 0.6X_{1i} + 0.4 \text{exp}(X_{1i}) + \text{log}(0.65 |X_{2i}|) +
      0.5(1 + X_{3i})^2,
\end{align*}
and $\bm \Sigma$ as in \cref{subsec:sims_binary}. Our \texttt{Bayes-DR} estimator is extended to continuous treatment settings by considering the linear, spline-based, and Gaussian process models of \cref{sec:model} for the treatment and outcome, and choosing the model fit that minimizes WAIC, as in \cref{subsec:sims_binary}. For model parameters $\bPsi$, we create the pseudo-outcome of \cite{kennedy2017non}:
\begin{align}
	\xi(\boldsymbol{D}_i, \boldsymbol{\Psi}) = \frac{Y_i - E(Y_i \vert T_i, \boldsymbol{X}_i)}{\pi(T_i\vert \boldsymbol{X}_i)} \int_{\mathcal{\boldsymbol{X}}} \pi(T_i\vert \boldsymbol{X}_i) dP_n(\boldsymbol{X}) + \int_{\mathcal{\boldsymbol{X}}} E(Y_i \vert T_i, \boldsymbol{X}_i) dP_n(\boldsymbol{X}),
\end{align}
where $P_n$ is the empirical distribution of the data and $\pi(t | x)$ is the density of $T | \boldsymbol{X}$. The pseudo-outcome is then regressed against the treatment. We use cubic polynomials to model the exposure-response curve which encaptures the true curve, though any flexible parametric approach can be accommodated. The estimated curves are then averaged over the posterior distribution as in (\ref{eq:our_estimator}) to acquire the \texttt{Bayes-DR} estimator of $E[Y(t)]$ for all $t$. Finally, we perform inference using the resampling approach described in Section \ref{sec:uncertainty}.

The competing approaches of \cref{subsec:sims_binary} are not directly applicable for continuous treatments. For this reason, we compare the \texttt{Bayes-DR} estimator to three regression-based approaches for estimating $E[Y(t)]$, which utilize the linear, spline, and Gaussian process specifications of the outcome model only, and marginalize over the covariate distribution. We refer to these regression-based estimators as \texttt{Reg-1}, \texttt{Reg-3}, and \texttt{Reg-GP}, respectively. 

\begin{figure}[h]
\centering
 \includegraphics[width=0.75\linewidth]{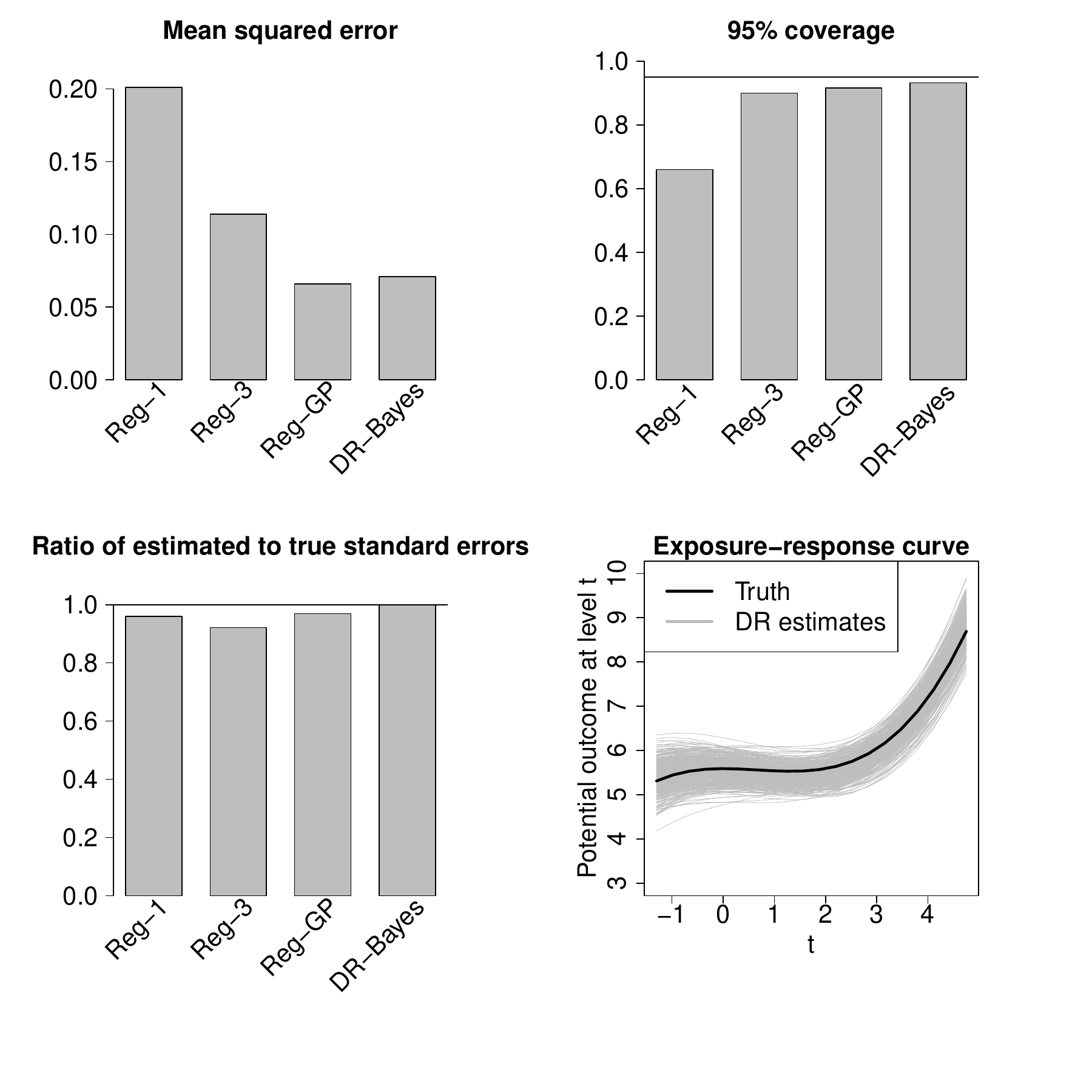}
\caption{Simulation results for continuous treatments. The top left panel presents the mean squared error, the top right panel shows the 95\% credible interval coverage, the bottom left panel shows the ratio of estimated to Monte Carlo standard errors, and the bottom right panel shows the estimates of the exposure-response curve across the 1000 simulations for the doubly robust estimator. }
\label{fig:sim2}
\end{figure}

To assess performance for estimating the whole curve, we evaluate performance metrics (bias, interval coverage) at 20 distinct exposure values, and average them. Figure \ref{fig:sim2} shows the results, averaged across 1000 simulations. The \texttt{Reg-1}  estimator does poorly in terms of MSE and interval coverage, which is expected since the outcome model is falsely assumed to be linear. The \texttt{Reg-3}, \texttt{Reg-GP}, and \texttt{Bayes-DR} approaches allow for nonlinear relationships between the covariates and treatment/outcome, and they perform well with respect to all metrics. The \texttt{Bayes-DR} estimator has close to the lowest MSE, achieves interval coverage near the nominal level, and appears unbiased in estimating the entire curve. 

\subsection{Summary of additional simulation results}
\label{sec:sims_more}

In the Appendix we present additional simulation results using different data generating mechanisms, different $p/n$ ratios, misspecified models, models that do not assume additivity, bootstrap inference for the competing approaches, and an investigation of the magnitude of the two terms in the variance estimator (\ref{eqn:VarEst}) under various data structures. 

Results under alternative data generating mechanisms are comparable to the ones shown here, and our estimation and inference approach perform well in terms of both MSE and finite sample interval coverage throughout.
As expected, as the sample size increases and the $p/n$ ratio decreases, differences between our approach and existing approaches to inference disappear as the asymptotic standard errors of existing approaches perform well.
Even though the bootstrap is not theoretically justified for every competing approach, we investigated its performance to assess whether our approach to inference performs better because it is based on resampling. We found that bootstrap intervals for competing approaches were excessively large with ratios of estimated to true standard errors well above 1. As the sample size increased, bootstrap standard errors performed well, though only in scenarios where the asymptotic intervals also perform well. 

\begin{figure}[t!]
\centering
\includegraphics[width=0.6\textwidth]{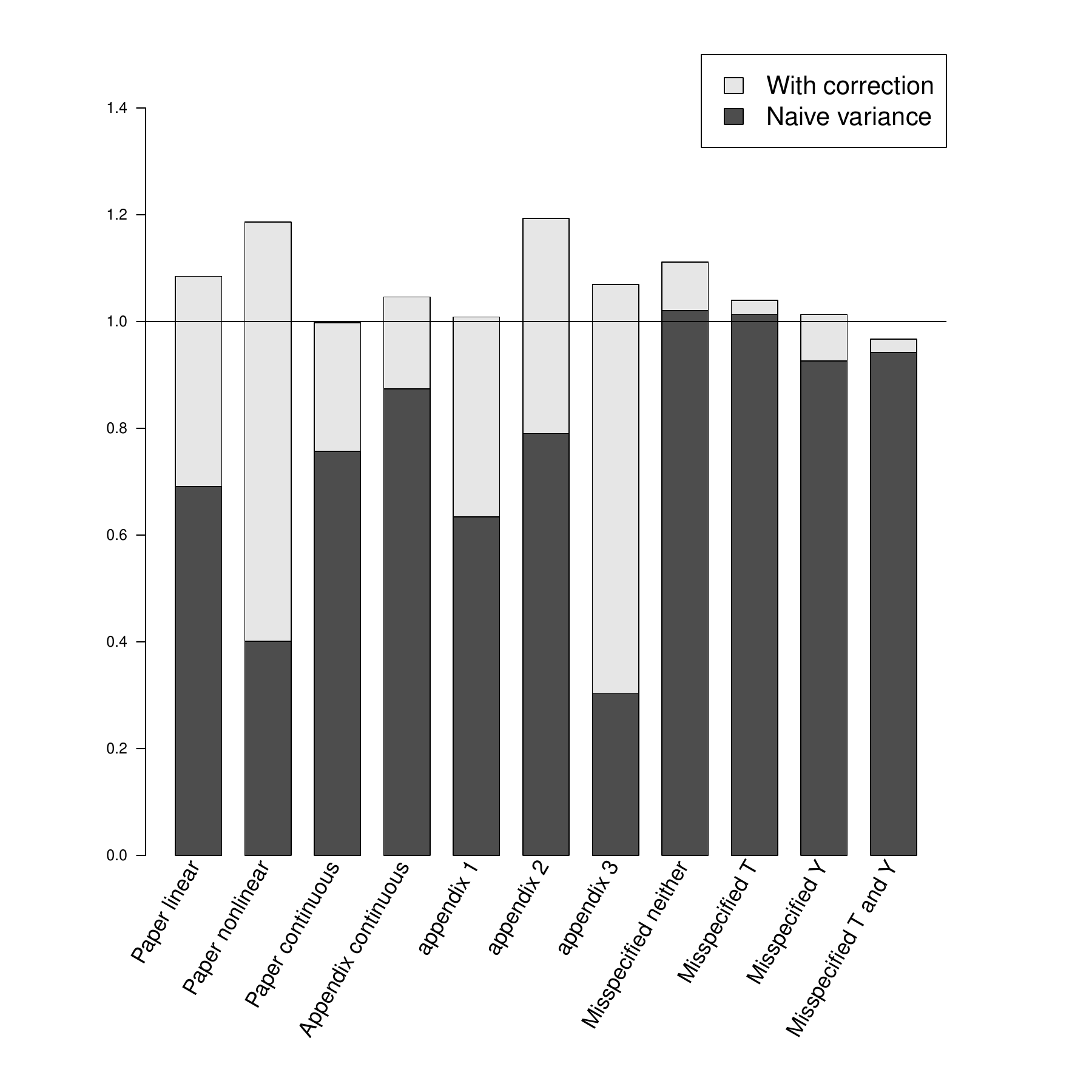}
\caption{Ratio of our variance estimator's average value to the estimator's Monte Carlo variance. Our variance estimator is separated by the contribution stemming only from the data (na\"ive variance -- dark grey) and the contribution stemming from parameter estimation (correction term -- light grey). Values near 1 indicate that our variance estimation strategy accurately reflects the estimator's true uncertainty. The horizontal axis represent various data generative mechanisms including the ones presented in \cref{sec:simGlobal}, and in the Appendix. With the order shown, simulations represent the linear binary, and non-linear binary simulations of \cref{subsec:sims_binary}, the continuous treatment simulation of \cref{subsec:sims_continuous}, an additional continuous treatment simulation from Appendix F, three additional binary treatment simulations looking at different data generating mechanisms and sparsity levels found in Appendix F, and four low-dimensional simulations with different types of model misspecification found in Appendix H.}
\label{fig:var_approximation}
\end{figure}

Lastly, \cref{fig:var_approximation} shows the contribution to our variance estimator stemming from the data only (na\"ive variance) and from parameter estimation (correction term), and how our variance estimator compares to the estimator's true variance under data generative scenarios presented here and in the Appendix. Values near 1 indicate unbiased variance estimation, and values larger (smaller) than 1 indicate that the variance estimator is larger (smaller) in expectation than the estimator's true variance. The na\"ive variance returns overly small variance estimates in many scenarios, indicating that failing to account for uncertainty in parameter estimation would lead to anti-conservative inference. In contrast, the correction term adds the proper amount of uncertainty on the na\"ive variance estimates, and returns total variance ratios much closer to 1. What's more, in scenarios where the na\"ive variance is near 1, the correction term is close to zero.

\section{Application to EWAS}
\label{sec:app}

Environmental wide association studies (EWAS) are becoming increasingly common as scientists attempt to gain a better understanding of how various chemicals and toxins affect the biological processes in the human body \citep{wild2005complementing,patel2014studying}. EWAS study the effects of a large number of exposures to which humans are invariably exposed. The National Health and Nutrition Examination Survey (NHANES) is a publicly available data source by the Centers for Disease Control and Prevention (CDC). We restrict attention to the 1999-2000, 2001-2002, 2003-2004, and 2005-2006 surveys. We use the same data as \cite{wilson2018model}, which include a large number of potential confounders from 
\begin{enumerate*}[label=\alph*)]
\item participants' questionnaires regarding their health status, and
\item clinical and laboratory tests containing information on environmental factors such as pollutants, allergens, bacterial/viral organisms, chemical toxicants, and nutrients.
\end{enumerate*}
We estimate the effects of 14 environmental agent groups (previously studied in \cite{patel2012systematic}) on three outcomes: the levels of HDL cholesterol, LDL cholesterol, and triglyceride, resulting in 42 analyses.
Exposure levels are defined as the average level across all agents within the same group. In the NHANES data, different subjects had different environmental agents measured, leading to different populations, covariate dimensions, and sample sizes for each of the 14 exposures,  and a wide range of $p/n$ ratios from 0.08 to 0.51.

\subsection{Differing levels of nonlinearity and sparsity}

\begin{figure}[h!]
\centering
 \includegraphics[width=0.75\linewidth]{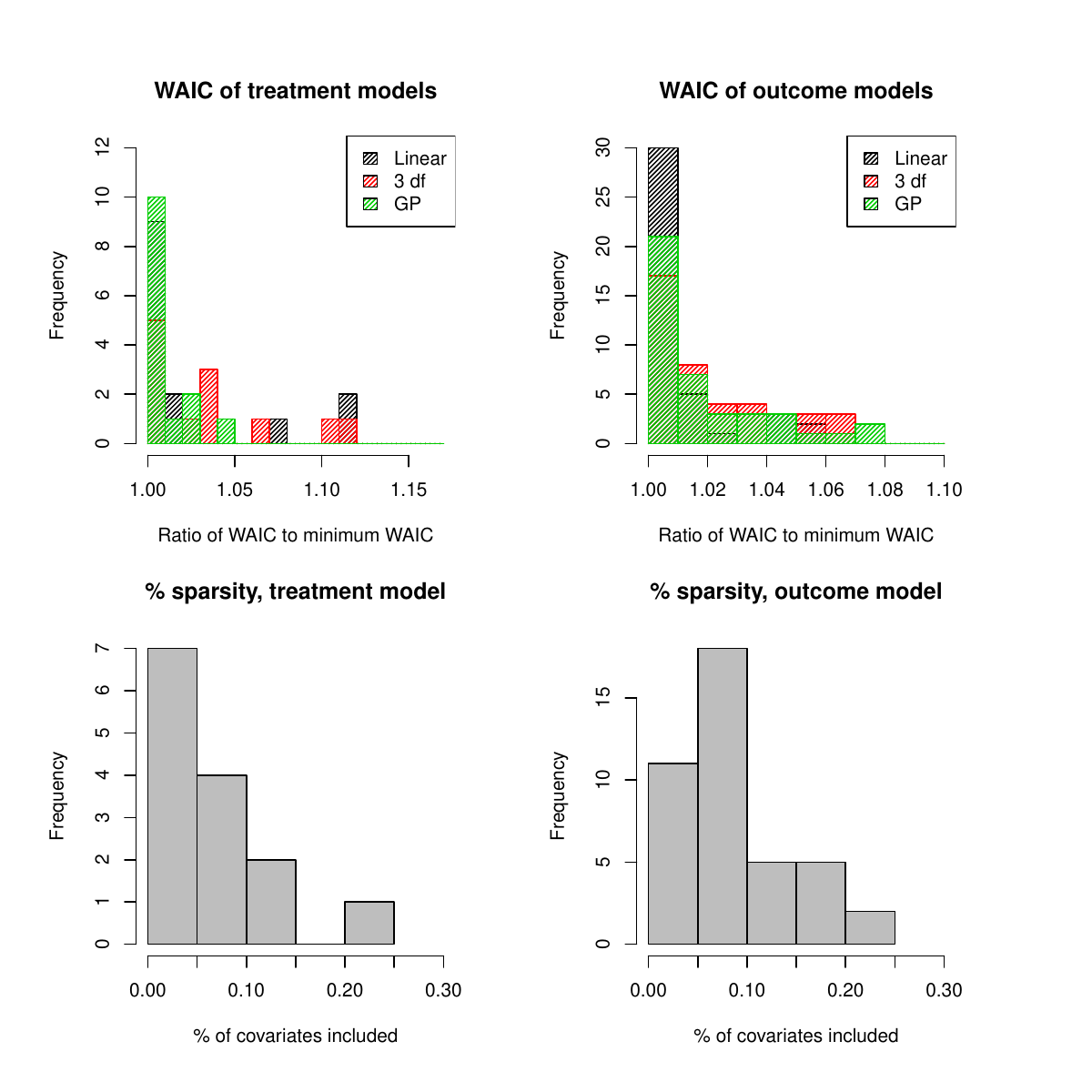}
\caption{The top panel presents the ratio of WAIC values to the minimum values for each of the three models considered. The top left panel shows the treatment model WAIC values, while the top right panel shows the WAIC for the outcome models. The bottom panel shows the percentage of covariates included in the chosen treatment and outcome model. }
\label{fig:NHANESwaic}
\end{figure}

We estimate the exposure response curves for all 42 analyses using our estimator for continuous treatments discussed in \cref{subsec:sims_continuous}. First, for each data set we fit a treatment and an outcome model under the three levels of flexibility: a linear function of the covariates, three degree of freedom splines, and Gaussian processes, and used the model with the minimum WAIC. Figure \ref{fig:NHANESwaic} shows histograms of the ratio of the WAIC values with the minimum WAIC within a given data set across the three models. A value of one indicates that a particular model had the smallest WAIC for a given data set, while larger values indicate worse fits to the data. For the treatment model, the Gaussian process prior is selected more than any other model and most of the values are less than 1.05. For the outcome model, the linear model is chosen most often across data sets, followed by the Gaussian process prior and spline model. Overall, these plots suggest that different amounts of flexibility are required across data sets, and an estimation strategy like ours which accommodates non-linear data generating processes is necessary for evaluating effects of all the environmental agents.

We also examine the extent to which our sparsity inducing priors reduced the dimension of the covariate space by studying the covariates' posterior inclusion probabilities. Figure \ref{fig:NHANESwaic} shows the percentage of covariates that have a posterior inclusion probability greater than 0.5 in the treatment and outcome models. The spike and slab priors greatly reduced the number of covariates included in each model, with less than 30\% of the covariates having posterior inclusion probabilities larger than 0.5 across data sets, and many less than 10\%. Not shown in the figure is that there are even fewer covariates included in {\it both models}, providing evidence of only weak confounding by observed variables within these data sets. This is further supported by the fact that many of the estimated exposure response curves are very similar to the curves one would obtain by not controlling for any covariates.

\subsection{Exposure response curves}

\begin{figure}[h!]
\centering
 \includegraphics[width=0.8\linewidth]{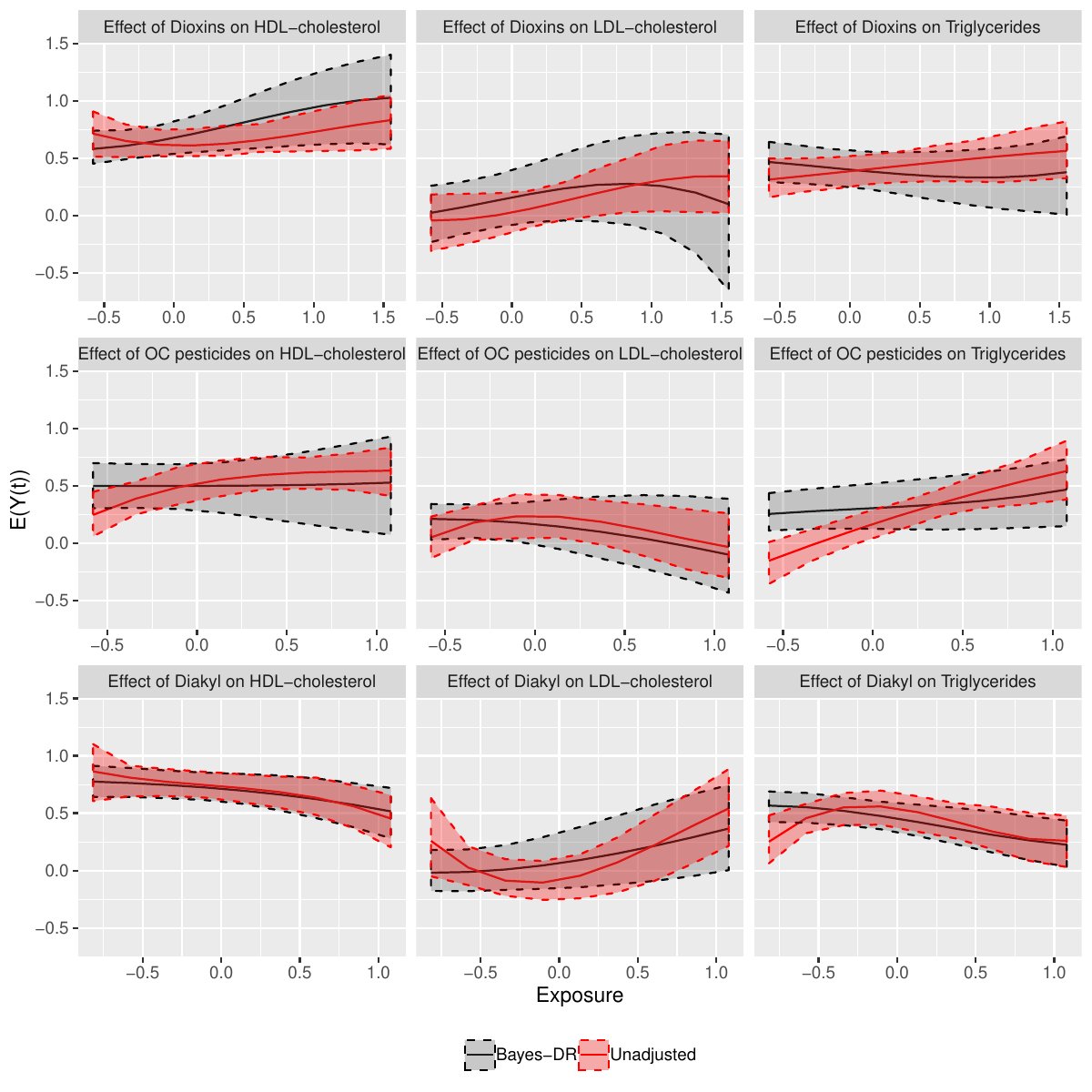}
\caption{Estimated exposure response curves from the doubly robust estimator (black line) as well as the na\"ive curve (red line), which does not adjust for any covariates.}
\label{fig:NHANEScurves}
\end{figure}

We highlight the estimated exposure response curves for three of the exposures in the analysis: Dioxins, Organochlorine pesticides, and Diakyl. The $p/n$ ratio for these three analyses was 0.41, 0.18, and 0.34, respectively. Figure \ref{fig:NHANEScurves} shows the doubly robust estimate of the exposure response curve along with the na\"ive curve estimated without adjusting for covariates. The estimated curves are fairly similar with a couple of exceptions. When comparing the results from the doubly robust estimate to the na\"ive estimate, the effect of Organochlorine pesticides on Triglycerides is much smaller, and the expected level of Triglycerides at low levels of Diakyl is much higher. In some areas of the curves, the doubly robust estimate is less uncertain, however, in general the na\"ive curves have tighter uncertainty intervals. This can happen in finite samples if there are variables that are strongly associated with the exposure and not the outcome. This could also be a feature of our strategy for variance estimation, which can be slightly conservative in finite samples. Asymptotically, we would expect both of these issues to disappear and the doubly robust estimator to be as or more efficient than the na\"ive estimator.

\section{Discussion}
\label{sec:discussion}

We have introduced an approach for causal inference that has a number of desirable features. First, it can be applied to doubly robust estimators of causal effects for binary or continuous treatments. This is particularly important as the literature on estimating the causal effect curve for continuous treatments is small, and has not been extended to high-dimensional scenarios. We showed that our approach maintains asymptotic properties such as double robustness, fast convergence rates, and consistent variance estimation, while achieving improved performance in finite samples. In particular, our inferential approach leads to nominal interval coverage when frequentist counterparts that rely on asymptotics fail to do so. Flexible Bayesian treatment and outcome models can be straightforwardly accommodated reducing the impact of model misspecification. While we focused on high-dimensional scenarios with spike and slab priors in this paper, the ideas presented apply to any modeling framework for the treatment and outcome models, and a number of estimands for which doubly-robust estimators exist. For example, even though we explored scenarios with homogeneous treatment effects, our approach is directly extendable to settings with heterogeneous treatment effects. 

Doubly robust estimation was first introduced in the Bayesian framework in \cite{saarela2016bayesian}, although there has been some debate about whether an estimator of counterfactual outcomes can utilize the propensity score within the Bayesian framework \citep{robins2015bayesian}. \cite{robins1997toward} showed that a Bayesian analysis which honors the likelihood principle can not utilize the propensity score. In this paper, we do not attempt to address these concerns, nor do we propose a ``fully'' Bayesian approach. Our purpose has been to show that Bayesian methods coupled with doubly robust estimators provide flexible alternatives with desirable finite sample properties. This is even more important in high-dimensional scenarios where model uncertainty is high and relying on asymptotic approximations does not perform well.

Lastly, we ponder on why our approach to uncertainty quantification performs better in finite samples than existing approaches rooted in asymptotic theory. Asymptotic approaches rely on certain terms vanishing as the sample size increases, and ignoring these terms can lead to anti-conservative inference in small samples. In contrast, our inferential approach relies on combining posterior distributions with a resampling procedure. Uncertainty in our estimators stemming from parameter estimation is accounted for through the posterior distribution. While this requires the user to specify a prior distribution that can affect inference, we have shown that non-informative priors perform well, and importantly, posterior distributions reflect parameter uncertainty without relying on asymptotics. Our approach relies on asymptotics solely through our resampling procedure since the bootstrap is only valid asymptotically. Since the bootstrap is used in our approach to account for uncertainty stemming from the observed data for fixed values of the parameters, and the estimator has a simple form when the parameters are fixed, it is expected that the bootstrap performs well.
As the sample size increases, any differences between our approach to inference and those rooted in asymptotics should dissipate.

\section*{Acknowledgement}
The authors would like to thank Chirag Patel for help with the NHANES data, as well as Matthew Cefalu, Rohit Patra, and Caleb Miles for incredibly helpful discussions on the manuscript. We would also like to thank the editor, associate editor, and two anonymous reviewers for outstanding feedback on the manuscript. Funding for this work was provided by National Institutes of Health (R01MH118927, ES000002, ES024332, ES007142, ES026217, ES028033, P01CA134294, R01GM111339, R35CA197449, P50MD010428, DP2MD012722), The U.S Environmental Protection Agency (83615601, 83587201-0), and The Health Effects Institute (4953-RFA14-3/16-4).

\newpage

\section*{Appendices}
\appendix
\setcounter{equation}{0}    
\renewcommand{\theequation}{A.\arabic{equation}}
\setcounter{figure}{0}    
\renewcommand{\thefigure}{A.\arabic{figure}}
\setcounter{table}{0}
\renewcommand{\thetable}{A.\arabic{table}}

\section{Proof that the variance estimator is consistent when both models are correctly specified}

For simplicity we will focus on $E(Y(t)) = \mu_t$, as it is trivial to then extend to the average treatment effect. Whenever we use a $*$ superscript, such as $\mu_t^*$, we are referring to the true value of that quantity. Throughout, we will use the following decomposition of the doubly robust estimator:
\begin{align*}
    \mu_t - \mu_t^* &= \frac{1}{n} \sum_{i=1}^n \frac{1(T_i = t)}{p_{ti}} (Y_i - m_{ti}) + m_{ti} - \mu_t^* \\
    &= \frac{1}{n} \sum_{i=1}^n (m_{ti} - m_{ti}^*) \Bigg(1 - \frac{1(T_i = t)}{p_{ti}^*} \Bigg) \\
    &+ \frac{1}{n} \sum_{i=1}^n \frac{1(T_i = t)(p_{ti} - p_{ti}^*) (m_{ti}^* - Y_i)}{p_{ti}p_{ti}^*} \\
    &+ \frac{1}{n} \sum_{i=1}^n \frac{1(T_i = t)(p_{ti} - p_{ti}^*) (m_{ti} - m_{ti}^*)}{p_{ti}p_{ti}^*} \\
    &+ \frac{1}{n} \sum_{i=1}^n \frac{1(T_i = t)}{p_{ti}^*} (Y_i - m_{ti}^*) + m_{ti}^*- \mu_t^* \\
    &= A_1 + A_2 + A_3 + B
\end{align*}
Note that it is implied everywhere when we refer to these four terms that they are a function of data $\boldsymbol{D}$ and parameters $\boldsymbol{\Psi}$. In other words, we are adopting the convention that $A_1 = A_1(\boldsymbol{D}, \boldsymbol{\Psi})$. If we let $$\Delta(\boldsymbol{D}, \boldsymbol{\Psi}) = \frac{1}{n} \sum_{i=1}^n \frac{1(T_i = t)}{p_{ti}} (Y_i - m_{ti}) + m_{ti}$$
then our variance estimator can be defined as 
$$\widehat{V} = \text{Var}_{\boldsymbol{D^{(m)}}} \{ E_{\boldsymbol{\Psi} \vert \boldsymbol{D}} [\Delta(\boldsymbol{D^{(m)}}, \boldsymbol{\Psi})] \} + \text{Var}_{\boldsymbol{\Psi} \vert \boldsymbol{D}} [\Delta(\boldsymbol{D}, \boldsymbol{\Psi})].$$
The variance of interest is $V = \text{Var}_{\boldsymbol{D}} \{ E_{\boldsymbol{\Psi} \vert \boldsymbol{D}} [\Delta(\boldsymbol{D}, \boldsymbol{\Psi})] \}$. Our goal will be to show that $\widehat{V} - V = o(n^{-1})$ when both models are correctly specified and contract at sufficiently fast rates. To do this, we will detail the asymptotic behavior of three components separately. First, we will show that $\text{Var}_{\boldsymbol{\Psi} \vert \boldsymbol{D}} [\Delta(\boldsymbol{D}, \boldsymbol{\Psi})] = o(n^{-1})$. Next, we show that $\text{Var}_{\boldsymbol{D^{(m)}}} \{ E_{\boldsymbol{\Psi} \vert \boldsymbol{D}} [\Delta(\boldsymbol{D^{(m)}}, \boldsymbol{\Psi})] \} = \text{Var}_{\boldsymbol{D}}(B) + o(n^{-1})$, and that $\text{Var}_{\boldsymbol{D}} \{ E_{\boldsymbol{\Psi} \vert \boldsymbol{D}} [\Delta(\boldsymbol{D}, \boldsymbol{\Psi})] \} = \text{Var}_{\boldsymbol{D}}(B) + o(n^{-1})$. The combination of these three results directly implies the desired result, that our variance estimator is consistent. We will also make the simplifying assumption that we are dealing with parametric models and that we can use taylor series representations for $p_{ti}$ and $m_{ti}$. We expect these results to extend to Bayesian nonparametric formulations for these parameters as long as they contract at sufficiently fast rates, though different proof techniques would be required, and we do not discuss them here. 

\subsection{Assumptions and preliminaries}
Throughout this section, we will additionally utilize the subscripts $n$ and $P_0$ to represent moments with respect to the posterior distribution and true data generating process, respectively. These are analogous to the subscripts $\boldsymbol{\Psi} \vert \boldsymbol{D}$ and $\boldsymbol{D}$, which also denote the posterior distribution and true data generating process, respectively. We will mostly restrict in this section to using $\boldsymbol{\Psi} \vert \boldsymbol{D}$ and $\boldsymbol{D}$, but in later sections when detailing posterior contraction rates, we will exclusively use $n$ and $P_0$ to conform with the literature on posterior contraction rates.  In particular $\mathbb{P}_n$ represents the posterior distribution given a sample of $n$ observations, and $E_{P_0}$ is the expected value with respect to $P_0$, the true data generating process.

\textit{Assumption 1}: Data generating process

\begin{enumerate}[label=(\alph*),leftmargin=*,topsep=0pt]
	\item $\left\{(Y_i, T_i, \boldsymbol{X}_i) \right\}_{i=1}^n$ are i.i.d samples from $P_0$
    \item The covariates $X_j$ have bounded support, in that there exists $K_x < \infty$ such that $|X_j| < K_x$ with probability 1 for all $j$.
    \item $\sup\limits_{P_0} E_{P_0}((Y - m_{ti}^*)^2) \leq K_y < \infty$.
\end{enumerate}
Assumption 1a is standard in this setting, and assumption 1b is likely to be satisfied in real applications as nearly all underlying variables are naturally bounded. Assumption 1c ensures that the residual variance of the outcome is bounded, which again should be satisfied in most applications.

\textit{Assumption 2}: Bounds on the error of posterior distributions
\begin{enumerate}[label=(\alph*),leftmargin=*,topsep=0pt]
	\item $\sup\limits_{P_0} E_{P_0} \text{Var}_n \bigg( \frac{p_{ti} - p_{ti}^*}{p_{ti}} \vert \boldsymbol{D}_i \bigg) \leq K_p < \infty$
    \item $\sup\limits_{P_0} E_{P_0} \text{Var}_n \bigg( m_{ti} - m_{ti}^* \vert \boldsymbol{D}_i \bigg) \leq K_m < \infty$
\end{enumerate}
Assumption 2a effectively states that the posterior distribution of $p_{ti}$ does not assign mass to neighborhoods of 0, and can be satisfied through prior distribution constraints. Assumption 2b states that the difference between the true conditional mean of the outcome and the corresponding posterior is bounded. This is a mild assumption in general, and is automatically satisfied if $Y$ is categorical.

\textit{Assumption 3}: Posterior contraction of treatment and outcome models 

There exist two sequence of numbers $\epsilon_{nt} \rightarrow 0$ and $\epsilon_{ny} \rightarrow 0$, and constants $M_t > 0$ and $M_y > 0$ that are independent of $\epsilon_{nt}$ and $\epsilon_{ny}$, respectively, such that
\begin{enumerate}[label=(\alph*),leftmargin=*,topsep=0pt]
	\item $\sup\limits_{P_0} E_{P_0} \mathbb{P}_n \bigg( \frac{1}{\sqrt{n}}||\boldsymbol{p}_{t} - \boldsymbol{p}_{t}^*|| > M_t \epsilon_{nt} \vert \boldsymbol{D} \bigg) \rightarrow 0$, and
    \item $\sup\limits_{P_0} E_{P_0} \mathbb{P}_n \bigg( \frac{1}{\sqrt{n}}||\boldsymbol{m}_{t} - \boldsymbol{m}_{t}^*|| > M_y \epsilon_{ny} \vert \boldsymbol{D} \bigg) \rightarrow 0$,
\end{enumerate}
where $||v|| = \sqrt{v_1^2 + \dots + v_n^2}$.
Assumption 3a and 3b state that the posterior distribution of the treatment and outcome models contract at rates $\epsilon_{nt}$ and $\epsilon_{ny}$, respectively.  Achieving rates of posterior contraction such as these typically relies on their own set of assumptions, such as conditions on the design matrix $\boldsymbol{X}$ or sparsity. We will restrict discussion of these issues to relevant papers on posterior contraction in regression models \citep{castillo2015bayesian,yang2015minimax,yoo2016supremum}. Our goal is to examine what happens to $\Delta(\boldsymbol{D}, \boldsymbol{\Psi})$ when either Assumption 3a or 3b is true.  Lastly, we need to make assumptions on specific moments of the posterior distribution, in particular that they converge sufficiently fast to zero. We will be using taylor series expansions for $p_{ti}$ and $m_{ti}$ of the following form:
\begin{align*}
    m_{ti} &= m(\boldsymbol{X}_i, \boldsymbol{\beta}) = m(\boldsymbol{X}_i, \boldsymbol{\beta}^*) + \boldsymbol{d}_1(\boldsymbol{X}_i, \boldsymbol{\beta}^*)^T (\boldsymbol{\beta} - \boldsymbol{\beta}^*) + (\boldsymbol{\beta} - \boldsymbol{\beta}^*)^T \boldsymbol{H}_1(\boldsymbol{X}_i, \boldsymbol{\widetilde{\beta}}) (\boldsymbol{\beta} - \boldsymbol{\beta}^*) \\
    p_{ti} &= g(\boldsymbol{X}_i, \boldsymbol{\alpha}) = g(\boldsymbol{X}_i, \boldsymbol{\alpha}^*) + \boldsymbol{d}_2(\boldsymbol{X}_i, \boldsymbol{\alpha}^*)^T (\boldsymbol{\alpha} - \boldsymbol{\alpha}^*) + (\boldsymbol{\alpha} - \boldsymbol{\alpha}^*)^T \boldsymbol{H}_2(\boldsymbol{X}_i, \boldsymbol{\widetilde{\alpha}}) (\boldsymbol{\alpha} - \boldsymbol{\alpha}^*),
\end{align*}
where $\boldsymbol{\widetilde{\alpha}}$ is a point between $\boldsymbol{\alpha}$ and $\boldsymbol{\alpha}^*$ and $\boldsymbol{\widetilde{\beta}}$ is a point between $\boldsymbol{\beta}$ and $\boldsymbol{\beta}^*$. We assume that the functions $m(\cdot)$ and $g(\cdot)$ have bounded first and second derivatives, and that they are sufficiently smooth so that these derivatives exist. While we only detail assumptions for $\boldsymbol{\beta}$, analogous assumptions will be made for $\boldsymbol{\alpha}$. 

\textit{Assumption 4}: Moments of the posterior distribution

\begin{enumerate}[label=(\alph*),leftmargin=*,topsep=0pt]
	\item $\text{E}_{\boldsymbol{\Psi} \vert \boldsymbol{D}} \bigg[ \frac{1}{n^2} ||\boldsymbol{m} - \boldsymbol{m}^*||_2^4 \Bigg]^{1/2}  \text{E}_{\boldsymbol{\Psi} \vert \boldsymbol{D}} \Bigg[ \frac{1}{n^2} ||\boldsymbol{p} - \boldsymbol{p}^*||_2^4 \bigg]^{1/2} = o(n^{-1})$
	\item $\text{E}_{\boldsymbol{\Psi} \vert \boldsymbol{D}} \bigg(||\boldsymbol{\beta} - \boldsymbol{\beta}^*||_2^4 \bigg) = o(n^{-1})$
	\item $\text{E}_{\boldsymbol{D}} \Bigg\{ E_{\boldsymbol{\Psi} \vert \boldsymbol{D}} \bigg(||\boldsymbol{\beta} - \boldsymbol{\beta}^*||_2^8 \bigg) \Bigg\}^{1/2} = o(n^{-1})$
\end{enumerate}

These assumptions are analogous to assumptions traditionally made in the high-dimensional and semiparametric causal inference literature \citep{belloni2013inference, farrell2015robust} that assume the convergence rates of the propensity score or outcome regression models are slightly faster than $n^{-{1/4}}$. These assumptions are nearly implied by the posterior contraction of the propensity score and outcome regression if they contract at faster than $n^{-{1/4}}$ rates. However, they are slightly stronger as they preclude the posterior distribution from assigning increasingly small probabilities to increasingly extreme values. We don't consider these to be strong assumptions as this is not likely to occur with any reasonable model for the propensity or outcome regression. Importantly, these assumptions imply other moment restrictions that we will utilize. Assumption 4b implies that $E_{\boldsymbol{\Psi} \vert \boldsymbol{D}} \bigg( ||\boldsymbol{\beta} - \boldsymbol{\beta}^*||_2^2 \bigg)^2 = o(n^{-1})$, which in turn implies that $\text{Var}_{\boldsymbol{\Psi} \vert \boldsymbol{D}} (\boldsymbol{\beta} - \boldsymbol{\beta}^*) \overset{p}{\to} \boldsymbol{0}$. Along with posterior contraction of $\boldsymbol{\beta}$, this implies that $E_{\boldsymbol{\Psi} \vert \boldsymbol{D}} (\boldsymbol{\beta} - \boldsymbol{\beta}^*) \overset{p}{\to} \boldsymbol{0}$. 

\subsection{Asymptotic behavior of $\text{Var}_{\Psi \vert D} [\Delta(D, \Psi)]$}

We will begin by detailing the asymptotic behavior of each of the four components of $\Delta(\boldsymbol{D}, \boldsymbol{\Psi})$ separately. We will analyze $\text{Var}_{\boldsymbol{\Psi} \vert \boldsymbol{D}} [A_1]$, $\text{Var}_{\boldsymbol{\Psi} \vert \boldsymbol{D}} [A_2]$, $\text{Var}_{\boldsymbol{\Psi} \vert \boldsymbol{D}} [A_3]$, and $\text{Var}_{\boldsymbol{\Psi} \vert \boldsymbol{D}} [B]$. If each of these four components separately are $o(n^{-1})$, then the entire variance must be of the same order, because all relevant covariances will be of the same or lower order. The first observation to make is that $B$ contains no unknown parameters and therefore $\text{Var}_{\boldsymbol{\Psi} \vert \boldsymbol{D}} [B] = 0$. Next we will focus on $A_1:$
\begin{align*}
    \text{Var}_{\boldsymbol{\Psi} \vert \boldsymbol{D}} [A_1] &= \text{Var}_{\boldsymbol{\Psi} \vert \boldsymbol{D}} \bigg[ \frac{1}{n} \sum_{i=1}^n \bigg(1 - \frac{1(T_i = t)}{p_{ti}^*} \bigg) (m_{ti} - m_{ti}^*) \bigg] \\
    &= \text{Var}_{\boldsymbol{\Psi} \vert \boldsymbol{D}} \bigg[ \frac{1}{n} \sum_{i=1}^n \bigg(1 - \frac{1(T_i = t)}{p_{ti}^*} \bigg) \bigg( \boldsymbol{d}_1(\boldsymbol{X}_i, \boldsymbol{\beta}^*)^T (\boldsymbol{\beta} - \boldsymbol{\beta}^*) \\
    &+ (\boldsymbol{\beta} - \boldsymbol{\beta}^*)^T \boldsymbol{H}_1(\boldsymbol{X}_i, \boldsymbol{\widetilde{\beta}}) (\boldsymbol{\beta} - \boldsymbol{\beta}^*) \bigg) \bigg] \\
    & \leq 3 \text{Var}_{\boldsymbol{\Psi} \vert \boldsymbol{D}} \bigg[ \frac{1}{n} \sum_{i=1}^n \bigg(1 - \frac{1(T_i = t)}{p_{ti}^*} \bigg) \bigg( \boldsymbol{d}_1(\boldsymbol{X}_i, \boldsymbol{\beta}^*)^T (\boldsymbol{\beta} - \boldsymbol{\beta}^*) \bigg) \bigg] \\ &+
    3 \text{Var}_{\boldsymbol{\Psi} \vert \boldsymbol{D}} \bigg[ \frac{1}{n} \sum_{i=1}^n \bigg(1 - \frac{1(T_i = t)}{p_{ti}^*} \bigg) \bigg((\boldsymbol{\beta} - \boldsymbol{\beta}^*)^T \boldsymbol{H}_1(\boldsymbol{X}_i, \boldsymbol{\widetilde{\beta}}) (\boldsymbol{\beta} - \boldsymbol{\beta}^*) \bigg) \bigg].
\end{align*}
We can now focus on these two terms separately. The first term can be expressed as
\begin{align*}
     & \text{Var}_{\boldsymbol{\Psi} \vert \boldsymbol{D}} \bigg[ \frac{1}{n} \sum_{i=1}^n \bigg(1 - \frac{1(T_i = t)}{p_{ti}^*} \bigg) \bigg( \boldsymbol{d}_1(\boldsymbol{X}_i, \boldsymbol{\beta}^*)^T (\boldsymbol{\beta} - \boldsymbol{\beta}^*) \bigg) \bigg] \\
     &= \frac{1}{n^2} \Bigg( \sum_{i=1}^n \bigg(1 - \frac{1(T_i = t)}{p_{ti}^*} \bigg) \boldsymbol{d}_1(\boldsymbol{X}_i, \boldsymbol{\beta}^*)^T \Bigg) \text{Var}_{\boldsymbol{\Psi} \vert \boldsymbol{D}} (\boldsymbol{\beta} - \boldsymbol{\beta}^*) \\
     &\times \Bigg( \sum_{i=1}^n \bigg(1 - \frac{1(T_i = t)}{p_{ti}^*} \bigg) \boldsymbol{d}_1(\boldsymbol{X}_i, \boldsymbol{\beta}^*) \Bigg) \\
     &\leq \frac{ \lambda_1^{V}}{n^2} \Bigg( \sum_{i=1}^n \bigg(1 - \frac{1(T_i = t)}{p_{ti}^*} \bigg) \boldsymbol{d}_1(\boldsymbol{X}_i, \boldsymbol{\beta}^*)^T \Bigg) \Bigg( \sum_{i=1}^n \bigg(1 - \frac{1(T_i = t)}{p_{ti}^*} \bigg) \boldsymbol{d}_1(\boldsymbol{X}_i, \boldsymbol{\beta}^*) \Bigg) \\
     &= \frac{ \lambda_1^{V}}{n} \Bigg( \frac{1}{\sqrt{n}} \sum_{i=1}^n \bigg(1 - \frac{1(T_i = t)}{p_{ti}^*} \bigg) \boldsymbol{d}_1(\boldsymbol{X}_i, \boldsymbol{\beta}^*)^T \Bigg) \Bigg(\frac{1}{\sqrt{n}} \sum_{i=1}^n \bigg(1 - \frac{1(T_i = t)}{p_{ti}^*} \bigg) \boldsymbol{d}_1(\boldsymbol{X}_i, \boldsymbol{\beta}^*) \Bigg) \\
     &= o(n^{-1}),
\end{align*}
where $\lambda_1^{V}$ is the maximum eigenvalue of $\text{Var}_{\boldsymbol{\Psi} \vert \boldsymbol{D}} (\boldsymbol{\beta} - \boldsymbol{\beta}^*)$. This expression is $o(n^{-1})$ because the multivariate CLT tells us that the two summations are $O(1)$ and we know that $\lambda_1^{V} \overset{p}{\to} 0$ by assumption 4 and the fact that $\text{Var}_{\boldsymbol{\Psi} \vert \boldsymbol{D}} (\boldsymbol{\beta} - \boldsymbol{\beta}^*) \overset{p}{\to} \boldsymbol{0}$. We can now deal with the second term as follows:
\begin{align*}
    & \text{Var}_{\boldsymbol{\Psi} \vert \boldsymbol{D}} \bigg[ \frac{1}{n} \sum_{i=1}^n \bigg(1 - \frac{1(T_i = t)}{p_{ti}^*} \bigg) \bigg((\boldsymbol{\beta} - \boldsymbol{\beta}^*)^T \boldsymbol{H}_1(\boldsymbol{X}_i, \boldsymbol{\widetilde{\beta}}) (\boldsymbol{\beta} - \boldsymbol{\beta}^*) \bigg) \bigg] \\
    &\leq \text{E}_{\boldsymbol{\Psi} \vert \boldsymbol{D}} \Bigg[ \frac{1}{n^2} \Bigg( \sum_{i=1}^n \bigg(1 - \frac{1(T_i = t)}{p_{ti}^*} \bigg) \bigg((\boldsymbol{\beta} - \boldsymbol{\beta}^*)^T \boldsymbol{H}_1(\boldsymbol{X}_i, \boldsymbol{\widetilde{\beta}}) (\boldsymbol{\beta} - \boldsymbol{\beta}^*) \bigg) \Bigg)^2 \Bigg] \\
    &\leq \text{E}_{\boldsymbol{\Psi} \vert \boldsymbol{D}} \Bigg[ \frac{1}{n^2} \Bigg( \sum_{i=1}^n \bigg(1 - \frac{1(T_i = t)}{p_{ti}^*} \bigg)^2 \sum_{i=1}^n \bigg((\boldsymbol{\beta} - \boldsymbol{\beta}^*)^T \boldsymbol{H}_1(\boldsymbol{X}_i, \boldsymbol{\widetilde{\beta}}) (\boldsymbol{\beta} - \boldsymbol{\beta}^*) \bigg)^2 \Bigg) \Bigg] \\
    &= \frac{1}{n^2} \sum_{i=1}^n \bigg(1 - \frac{1(T_i = t)}{p_{ti}^*} \bigg)^2  \text{E}_{\boldsymbol{\Psi} \vert \boldsymbol{D}} \bigg( \sum_{i=1}^n \bigg((\boldsymbol{\beta} - \boldsymbol{\beta}^*)^T \boldsymbol{H}_1(\boldsymbol{X}_i, \boldsymbol{\widetilde{\beta}}) (\boldsymbol{\beta} - \boldsymbol{\beta}^*) \bigg)^2 \Bigg) \\
    &\leq \frac{1}{n^2} \sum_{i=1}^n \bigg(1 - \frac{1(T_i = t)}{p_{ti}^*} \bigg)^2  \text{E}_{\boldsymbol{\Psi} \vert \boldsymbol{D}} \bigg( \sum_{i=1}^n \bigg( \lambda_1^{H} (\boldsymbol{\beta} - \boldsymbol{\beta}^*)^T (\boldsymbol{\beta} - \boldsymbol{\beta}^*) \bigg)^2 \Bigg) \\
    &\leq \frac{k}{n} \ \text{E}_{\boldsymbol{\Psi} \vert \boldsymbol{D}} \bigg( \sum_{i=1}^n \bigg( \lambda_1^{H} (\boldsymbol{\beta} - \boldsymbol{\beta}^*)^T (\boldsymbol{\beta} - \boldsymbol{\beta}^*) \bigg)^2 \Bigg) \\
    &\leq \frac{k'}{n} \ \text{E}_{\boldsymbol{\Psi} \vert \boldsymbol{D}} \bigg( \sum_{i=1}^n \bigg((\boldsymbol{\beta} - \boldsymbol{\beta}^*)^T (\boldsymbol{\beta} - \boldsymbol{\beta}^*) \bigg)^2 \Bigg) \\
    &\leq k' \ \text{E}_{\boldsymbol{\Psi} \vert \boldsymbol{D}} \bigg( \bigg((\boldsymbol{\beta} - \boldsymbol{\beta}^*)^T (\boldsymbol{\beta} - \boldsymbol{\beta}^*) \bigg)^2 \Bigg) \\
    &= k' \ \text{E}_{\boldsymbol{\Psi} \vert \boldsymbol{D}} \bigg(||\boldsymbol{\beta} - \boldsymbol{\beta}^*||_2^4 \bigg) \\
    &= o(n^{-1})
\end{align*}
where the last equality held by assumption 4. We also used the fact that $m(\boldsymbol{X}_i, \boldsymbol{\beta})$ has bounded second derivatives and therefore its largest eigenvalue $\lambda_1^{H}$ is also bounded. This concludes the proof that $\text{Var}_{\boldsymbol{\Psi} \vert \boldsymbol{D}} [A_1] = o(n^{-1})$. The proof for $A_2$ follows analogously, so we leave it out for brevity. Now we can turn attention to $A_3$:
\begin{align*}
    \text{Var}_{\boldsymbol{\Psi} \vert \boldsymbol{D}} [A_3] &= \text{Var}_{\boldsymbol{\Psi} \vert \boldsymbol{D}} \bigg[ \frac{1}{n} \sum_{i=1}^n \frac{1(T_i = t) (m_{ti} - m_{ti}^*) (p_{ti} - p_{ti}^*)}{p_{ti} p_{ti^*}} \bigg] \\
    &\leq \text{E}_{\boldsymbol{\Psi} \vert \boldsymbol{D}} \bigg[ \frac{1}{n^2} \Bigg( \sum_{i=1}^n \frac{1(T_i = t) (m_{ti} - m_{ti}^*) (p_{ti} - p_{ti}^*)}{p_{ti} p_{ti^*}} \Bigg)^2 \bigg] \\
    &\leq \frac{1}{n^2}  \text{E}_{\boldsymbol{\Psi} \vert \boldsymbol{D}} \bigg[ \Bigg( \sum_{i=1}^n 1(T_i = t) (m_{ti} - m_{ti}^*)^2 \Bigg)  \Bigg( \sum_{i=1}^n \bigg(\frac{(p_{ti} - p_{ti}^*)}{p_{ti} p_{ti^*}} \bigg)^2 \Bigg) \bigg] \\
    &\leq \frac{k}{n^2}  \text{E}_{\boldsymbol{\Psi} \vert \boldsymbol{D}} \bigg[ \Bigg( \sum_{i=1}^n (m_{ti} - m_{ti}^*)^2 \Bigg)  \Bigg( \sum_{i=1}^n (p_{ti} - p_{ti}^*)^2 \Bigg) \bigg] \\
    &\leq \frac{k}{n^2}  \text{E}_{\boldsymbol{\Psi} \vert \boldsymbol{D}} \bigg[ \Bigg( \sum_{i=1}^n (m_{ti} - m_{ti}^*)^2 \Bigg)^2 \Bigg]^{1/2}  \text{E}_{\boldsymbol{\Psi} \vert \boldsymbol{D}} \Bigg[ \Bigg( \sum_{i=1}^n (p_{ti} - p_{ti}^*)^2 \Bigg)^2 \bigg]^{1/2} \\
    &= k  \text{E}_{\boldsymbol{\Psi} \vert \boldsymbol{D}} \bigg[ \frac{1}{n^2} ||\boldsymbol{m} - \boldsymbol{m}^*||_2^4 \Bigg]^{1/2}  \text{E}_{\boldsymbol{\Psi} \vert \boldsymbol{D}} \Bigg[ \frac{1}{n^2} ||\boldsymbol{p} - \boldsymbol{p}^*||_2^4 \bigg]^{1/2} \\
    &= o(n^{-1}),
\end{align*}
where the last equality held because of assumption 4. This, combined with previous results on $A_1$, $A_2$, and $B$ imply that $\text{Var}_{\boldsymbol{\Psi} \vert \boldsymbol{D}} [\Delta(\boldsymbol{D}, \boldsymbol{\Psi})] = o(n^{-1})$. 

\subsection{Asymptotic behavior of $\text{Var}_{D^{(m)}} \{ E_{\Psi \vert D} [\Delta(D^{(m)}, \Psi)] \}$}

We will again begin by detailing the asymptotic behavior of each of the four components of $\Delta(\boldsymbol{D}, \boldsymbol{\Psi})$ separately. We will show that each of $\text{Var}_{\boldsymbol{D^{(m)}}} \{ E_{\boldsymbol{\Psi} \vert \boldsymbol{D}} [A_1] \}$, $\text{Var}_{\boldsymbol{D^{(m)}}} \{ E_{\boldsymbol{\Psi} \vert \boldsymbol{D}} [A_2] \}$, and $\text{Var}_{\boldsymbol{D^{(m)}}} \{ E_{\boldsymbol{\Psi} \vert \boldsymbol{D}} [A_3] \}$ are $o(n^{-1})$, and that $\text{Var}_{\boldsymbol{D^{(m)}}} \{ E_{\boldsymbol{\Psi} \vert \boldsymbol{D}} [B] \} \approx \text{Var}_{\boldsymbol{D}}[B]$. If this is true, then we have the desired result, because all relevant covariances will necessarily be $o(n^{-1})$. First let's examine the $B$ component:
\begin{align*}
    \text{Var}_{\boldsymbol{D^{(m)}}} \{ E_{\boldsymbol{\Psi} \vert \boldsymbol{D}} [B] \} &= \text{Var}_{\boldsymbol{D^{(m)}}} [B] \\
    &= \text{Var}_{\boldsymbol{D^{(m)}}} \Bigg[ \frac{1}{n} \sum_{i=1}^n \frac{1(T_i^{(m)} = t)}{{p_{ti}^*}^{(m)}} (Y_i^{(m)} - {m_{ti}^*}^{(m)}) + {m_{ti}^*}^{(m)}- \mu_t^* \Bigg] \\
    &\approx \text{Var}_{\boldsymbol{D}} \Bigg[ \frac{1}{n} \sum_{i=1}^n \frac{1(T_i = t)}{p_{ti}^*} (Y_i - m_{ti}^*) + m_{ti}^*- \mu_t^* \Bigg] \\
    &= \text{Var}_{\boldsymbol{D}}[B].
\end{align*}
The approximation stems from the fact that we are using the bootstrap to approximate the distribution of $\boldsymbol{D}$, which is justified asymptotically. Now we can focus on $A_1:$
\begin{align*}
    & \text{Var}_{\boldsymbol{D^{(m)}}} \{ E_{\boldsymbol{\Psi} \vert \boldsymbol{D}} [A_1] \} \\
    &= \text{Var}_{\boldsymbol{D^{(m)}}} \Bigg\{ E_{\boldsymbol{\Psi} \vert \boldsymbol{D}} \Bigg[ \frac{1}{n} \sum_{i=1}^n \Bigg(1 - \frac{1(T_i^{(m)} = t)}{{p_{ti}^*}^{(m)}} \Bigg) (m_{ti}^{(m)} - {m_{ti}^*}^{(m)}) \Bigg] \Bigg\} \\
    &= \text{Var}_{\boldsymbol{D^{(m)}}} \Bigg\{ E_{\boldsymbol{\Psi} \vert \boldsymbol{D}} \Bigg[ \frac{1}{n} \sum_{i=1}^n \Bigg(1 - \frac{1(T_i^{(m)} = t)}{{p_{ti}^*}^{(m)}} \Bigg) \bigg( \boldsymbol{d}_1(\boldsymbol{X}_i^{(m)}, \boldsymbol{\beta}^*)^T (\boldsymbol{\beta} - \boldsymbol{\beta}^*) \\
    & \hspace{4cm} + (\boldsymbol{\beta} - \boldsymbol{\beta}^*)^T \boldsymbol{H}_1(\boldsymbol{X}_i^{(m)}, \boldsymbol{\widetilde{\beta}}) (\boldsymbol{\beta} - \boldsymbol{\beta}^*) \bigg) \Bigg] \Bigg\} \\
    &\leq 3 \text{Var}_{\boldsymbol{D^{(m)}}} \Bigg\{ E_{\boldsymbol{\Psi} \vert \boldsymbol{D}} \Bigg[ \frac{1}{n} \sum_{i=1}^n \Bigg(1 - \frac{1(T_i^{(m)} = t)}{{p_{ti}^*}^{(m)}} \Bigg) \bigg( \boldsymbol{d}_1(\boldsymbol{X}_i^{(m)}, \boldsymbol{\beta}^*)^T (\boldsymbol{\beta} - \boldsymbol{\beta}^*) \bigg) \Bigg] \Bigg\} \\
    &+ 3 \text{Var}_{\boldsymbol{D^{(m)}}} \Bigg\{ E_{\boldsymbol{\Psi} \vert \boldsymbol{D}} \Bigg[ \frac{1}{n} \sum_{i=1}^n \Bigg(1 - \frac{1(T_i^{(m)} = t)}{{p_{ti}^*}^{(m)}} \Bigg) \bigg( (\boldsymbol{\beta} - \boldsymbol{\beta}^*)^T \boldsymbol{H}_1(\boldsymbol{X}_i^{(m)}, \boldsymbol{\widetilde{\beta}}) (\boldsymbol{\beta} - \boldsymbol{\beta}^*) \bigg) \Bigg] \Bigg\}. \\
\end{align*}
We can now work with both of these components separately. Looking at the first term, we see that
\begin{align*}
    & \text{Var}_{\boldsymbol{D^{(m)}}} \Bigg\{ E_{\boldsymbol{\Psi} \vert \boldsymbol{D}} \Bigg[ \frac{1}{n} \sum_{i=1}^n \Bigg(1 - \frac{1(T_i^{(m)} = t)}{{p_{ti}^*}^{(m)}} \Bigg) \bigg( \boldsymbol{d}_1(\boldsymbol{X}_i^{(m)}, \boldsymbol{\beta}^*)^T (\boldsymbol{\beta} - \boldsymbol{\beta}^*) \bigg) \Bigg] \Bigg\} \\
    &= E_{\boldsymbol{\Psi} \vert \boldsymbol{D}} (\boldsymbol{\beta} - \boldsymbol{\beta}^*)^T \text{Var}_{\boldsymbol{D^{(m)}}} \Bigg\{ \frac{1}{n} \sum_{i=1}^n \Bigg(1 - \frac{1(T_i^{(m)} = t)}{{p_{ti}^*}^{(m)}} \Bigg) \boldsymbol{d}_1(\boldsymbol{X}_i^{(m)}, \boldsymbol{\beta}^*)^T \bigg) \Bigg\} E_{\boldsymbol{\Psi} \vert \boldsymbol{D}} (\boldsymbol{\beta} - \boldsymbol{\beta}^*) \\
    &= \frac{1}{n} E_{\boldsymbol{\Psi} \vert \boldsymbol{D}} (\boldsymbol{\beta} - \boldsymbol{\beta}^*)^T \text{Var}_{\boldsymbol{D^{(m)}}} \Bigg\{ \frac{1}{\sqrt{n}} \sum_{i=1}^n \Bigg(1 - \frac{1(T_i^{(m)} = t)}{{p_{ti}^*}^{(m)}} \Bigg) \boldsymbol{d}_1(\boldsymbol{X}_i^{(m)}, \boldsymbol{\beta}^*)^T \bigg) \Bigg\} E_{\boldsymbol{\Psi} \vert \boldsymbol{D}} (\boldsymbol{\beta} - \boldsymbol{\beta}^*) \\
    & \approx \frac{1}{n} E_{\boldsymbol{\Psi} \vert \boldsymbol{D}} (\boldsymbol{\beta} - \boldsymbol{\beta}^*)^T \boldsymbol{\Sigma}_{A_1} E_{\boldsymbol{\Psi} \vert \boldsymbol{D}} (\boldsymbol{\beta} - \boldsymbol{\beta}^*) \\
    &\leq \frac{\lambda_1^{A_1}}{n} E_{\boldsymbol{\Psi} \vert \boldsymbol{D}} (\boldsymbol{\beta} - \boldsymbol{\beta}^*)^T  E_{\boldsymbol{\Psi} \vert \boldsymbol{D}} (\boldsymbol{\beta} - \boldsymbol{\beta}^*) \\
    &= o(n^{-1})
\end{align*}
The approximation stems from applying the multivariate CLT, and $\lambda_1^{A_1}$ is the maximum eigenvalue of $\boldsymbol{\Sigma}_{A_1}$. This is $o(n^{-1})$ because assumption 4 implies that $E_{\boldsymbol{\Psi} \vert \boldsymbol{D}} (\boldsymbol{\beta} - \boldsymbol{\beta}^*) \overset{p}{\to} 0$. Now we can move to the second term as follows:
\begin{align*}
    & \text{Var}_{\boldsymbol{D^{(m)}}} \Bigg\{ E_{\boldsymbol{\Psi} \vert \boldsymbol{D}} \Bigg[ \frac{1}{n} \sum_{i=1}^n \Bigg(1 - \frac{1(T_i^{(m)} = t)}{{p_{ti}^*}^{(m)}} \Bigg) \bigg( (\boldsymbol{\beta} - \boldsymbol{\beta}^*)^T \boldsymbol{H}_1(\boldsymbol{X}_i^{(m)}, \boldsymbol{\widetilde{\beta}}) (\boldsymbol{\beta} - \boldsymbol{\beta}^*) \bigg) \Bigg] \Bigg\} \\
     &= \text{Var}_{\boldsymbol{D^{(m)}}} \Bigg\{ \frac{1}{n} \sum_{i=1}^n \Bigg(1 - \frac{1(T_i^{(m)} = t)}{{p_{ti}^*}^{(m)}} \Bigg) E_{\boldsymbol{\Psi} \vert \boldsymbol{D}} \bigg( (\boldsymbol{\beta} - \boldsymbol{\beta}^*)^T \boldsymbol{H}_1(\boldsymbol{X}_i^{(m)}, \boldsymbol{\widetilde{\beta}}) (\boldsymbol{\beta} - \boldsymbol{\beta}^*) \bigg) \Bigg\} \\
    &\leq \text{E}_{\boldsymbol{D^{(m)}}} \Bigg\{ \frac{1}{n^2} \Bigg[ \sum_{i=1}^n \Bigg(1 - \frac{1(T_i^{(m)} = t)}{{p_{ti}^*}^{(m)}} \Bigg) E_{\boldsymbol{\Psi} \vert \boldsymbol{D}} \bigg( (\boldsymbol{\beta} - \boldsymbol{\beta}^*)^T \boldsymbol{H}_1(\boldsymbol{X}_i^{(m)}, \boldsymbol{\widetilde{\beta}}) (\boldsymbol{\beta} - \boldsymbol{\beta}^*) \bigg) \Bigg]^2 \Bigg\} \\
    &\leq \text{E}_{\boldsymbol{D^{(m)}}} \Bigg\{ \frac{1}{n^2} \sum_{i=1}^n \Bigg(1 - \frac{1(T_i^{(m)} = t)}{{p_{ti}^*}^{(m)}} \Bigg)^2 \sum_{i=1}^n \Bigg( E_{\boldsymbol{\Psi} \vert \boldsymbol{D}} \bigg( (\boldsymbol{\beta} - \boldsymbol{\beta}^*)^T \boldsymbol{H}_1(\boldsymbol{X}_i^{(m)}, \boldsymbol{\widetilde{\beta}}) (\boldsymbol{\beta} - \boldsymbol{\beta}^*) \bigg) \Bigg)^2 \Bigg\} \\
    &\leq \text{E}_{\boldsymbol{D^{(m)}}} \Bigg\{ \frac{1}{n^2} \sum_{i=1}^n \Bigg(1 - \frac{1(T_i^{(m)} = t)}{{p_{ti}^*}^{(m)}} \Bigg)^2 \sum_{i=1}^n \Bigg( E_{\boldsymbol{\Psi} \vert \boldsymbol{D}} \bigg( \lambda_1^{H_1} (\boldsymbol{\beta} - \boldsymbol{\beta}^*)^T (\boldsymbol{\beta} - \boldsymbol{\beta}^*) \bigg) \Bigg)^2 \Bigg\} \\
    &\leq \text{E}_{\boldsymbol{D^{(m)}}} \Bigg\{ \frac{k}{n^2} \sum_{i=1}^n \Bigg(1 - \frac{1(T_i^{(m)} = t)}{{p_{ti}^*}^{(m)}} \Bigg)^2 \sum_{i=1}^n E_{\boldsymbol{\Psi} \vert \boldsymbol{D}} \bigg( ||\boldsymbol{\beta} - \boldsymbol{\beta}^*||_2^2 \bigg)^2 \Bigg\} \\
    &= \frac{k}{n} E_{\boldsymbol{\Psi} \vert \boldsymbol{D}} \bigg( ||\boldsymbol{\beta} - \boldsymbol{\beta}^*||_2^2 \bigg)^2 \text{E}_{\boldsymbol{D^{(m)}}} \sum_{i=1}^n \Bigg(1 - \frac{1(T_i^{(m)} = t)}{{p_{ti}^*}^{(m)}} \Bigg)^2  \\
    & \leq k' E_{\boldsymbol{\Psi} \vert \boldsymbol{D}} \bigg( ||\boldsymbol{\beta} - \boldsymbol{\beta}^*||_2^2 \bigg)^2 \\
    &= o(n^{-1})
\end{align*}
The last equality holds because of assumption 4. Again we used that the fact that the largest eigenvalue, $\lambda_1^{H_1}$ of the hessian matrix was bounded. The proof for $A_2$ follows analogously, so we do not include it for brevity. We now can turn our attention to $A_3$:
\begin{align*}
    & \text{Var}_{\boldsymbol{D^{(m)}}} \{ E_{\boldsymbol{\Psi} \vert \boldsymbol{D}} [A_3] \} \\
    &\leq \text{E}_{\boldsymbol{D^{(m)}}} \{ E_{\boldsymbol{\Psi} \vert \boldsymbol{D}}^2 [A_3] \} \\
    &\leq \text{E}_{\boldsymbol{D^{(m)}}} \{ E_{\boldsymbol{\Psi} \vert \boldsymbol{D}} [A_3^2] \} \\
    &= \text{E}_{\boldsymbol{D^{(m)}}} \Bigg\{ E_{\boldsymbol{\Psi} \vert \boldsymbol{D}} \Bigg[ \bigg( \frac{1}{n} \sum_{i=1}^n \frac{1(T_i^{(m)} = t)(p_{ti}^{(m)} - {p_{ti}^*}^{(m)}) (m_{ti}^{(m)} - {m_{ti}^*}^{(m)})}{p_{ti}^{(m)} {p_{ti}^*}^{(m)}} \bigg)^2 \Bigg] \Bigg\} \\
    &\leq \text{E}_{\boldsymbol{D^{(m)}}} \Bigg\{ \frac{1}{n^2}  \text{E}_{\boldsymbol{\Psi} \vert \boldsymbol{D}} \Bigg[ \Bigg( \sum_{i=1}^n 1(T_i^{(m)} = t) (m_{ti}^{(m)} - {m_{ti}^*}^{(m)})^2 \Bigg)  \Bigg( \sum_{i=1}^n \bigg(\frac{(p_{ti}^{(m)} - {p_{ti}^*}^{(m)})}{p_{ti}^{(m)} {p_{ti^*}^{(m)}}} \bigg)^2 \Bigg) \Bigg]  \Bigg\}\\
    &\leq \text{E}_{\boldsymbol{D^{(m)}}} \Bigg\{  \frac{k}{n^2}  \text{E}_{\boldsymbol{\Psi} \vert \boldsymbol{D}} \Bigg[ \Bigg( \sum_{i=1}^n (m_{ti}^{(m)} - {m_{ti}^*}^{(m)})^2 \Bigg)  \Bigg( \sum_{i=1}^n (p_{ti}^{(m)} - {p_{ti}^*}^{(m)})^2 \Bigg) \Bigg] \Bigg\}\\
    &\leq \text{E}_{\boldsymbol{D^{(m)}}} \Bigg\{ \frac{k}{n^2}  \text{E}_{\boldsymbol{\Psi} \vert \boldsymbol{D}} \Bigg[ \Bigg( \sum_{i=1}^n (m_{ti}^{(m)} - {m_{ti}^*}^{(m)})^2 \Bigg)^2 \Bigg]^{1/2}  \text{E}_{\boldsymbol{\Psi} \vert \boldsymbol{D}} \Bigg[ \Bigg( \sum_{i=1}^n (p_{ti}^{(m)} - {p_{ti}^*}^{(m)})^2 \Bigg)^2 \Bigg]^{1/2} \Bigg\}\\
    &\approx \text{E}_{\boldsymbol{D}} \Bigg\{ \frac{k}{n^2}  \text{E}_{\boldsymbol{\Psi} \vert \boldsymbol{D}} \Bigg[ \Bigg( \sum_{i=1}^n (m_{ti} - {m_{ti}^*})^2 \Bigg)^2 \Bigg]^{1/2}  \text{E}_{\boldsymbol{\Psi} \vert \boldsymbol{D}} \Bigg[ \Bigg( \sum_{i=1}^n (p_{ti} - {p_{ti}^*})^2 \Bigg)^2 \Bigg]^{1/2} \Bigg\}\\
    &= k \ \text{E}_{\boldsymbol{D}} \Bigg\{\text{E}_{\boldsymbol{\Psi} \vert \boldsymbol{D}} \Bigg[ \frac{1}{n^2} ||\boldsymbol{m} - \boldsymbol{m}^*||_2^4 \Bigg]^{1/2}  \text{E}_{\boldsymbol{\Psi} \vert \boldsymbol{D}} \Bigg[ \frac{1}{n^2} ||\boldsymbol{p} - \boldsymbol{p}^*||_2^4 \Bigg]^{1/2}  \Bigg\}\\
    &= o(n^{-1}),
\end{align*}
where the last equality follows from assumption 4. This, combined with the previous results, implies that $\text{Var}_{\boldsymbol{D^{(m)}}} \{ E_{\boldsymbol{\Psi} \vert \boldsymbol{D}} [\Delta(\boldsymbol{D^{(m)}}, \boldsymbol{\Psi})] \} = \text{Var}_{\boldsymbol{D}}(B) + o(n^{-1})$.

\subsection{Asymptotic behavior of $\text{Var}_{D} \{ E_{\Psi \vert D} [\Delta(D, \Psi)] \}$}

Now we will examine the behavior of the true variance of our estimator. Again we will do this for each of the four separate components separately, starting with $B$, which is straightforward because $\text{Var}_{\boldsymbol{D}} \{ E_{\boldsymbol{\Psi} \vert \boldsymbol{D}} [B] \} = \text{Var}_{\boldsymbol{D}}[B]$. We will now show that each of the remaining three terms are $o(n^{-1})$, starting first with $A_1$:
\begin{align*}
    & \text{Var}_{\boldsymbol{D}} \{ E_{\boldsymbol{\Psi} \vert \boldsymbol{D}} [A_1] \} \\
    &= \text{Var}_{\boldsymbol{D}} \Bigg\{ E_{\boldsymbol{\Psi} \vert \boldsymbol{D}} \Bigg[ \frac{1}{n} \sum_{i=1}^n \Bigg(1 - \frac{1(T_i = t)}{p_{ti}^*} \Bigg) (m_{ti} - m_{ti}^*) \Bigg] \Bigg\} \\
    &= \text{Var}_{\boldsymbol{D}} \Bigg\{ E_{\boldsymbol{\Psi} \vert \boldsymbol{D}} \Bigg[ \frac{1}{n} \sum_{i=1}^n \bigg(1 - \frac{1(T_i = t)}{p_{ti}^*} \bigg) \bigg( \boldsymbol{d}_1(\boldsymbol{X}_i, \boldsymbol{\beta}^*)^T (\boldsymbol{\beta} - \boldsymbol{\beta}^*) \\
    &+ (\boldsymbol{\beta} - \boldsymbol{\beta}^*)^T \boldsymbol{H}_1(\boldsymbol{X}_i, \boldsymbol{\widetilde{\beta}}) (\boldsymbol{\beta} - \boldsymbol{\beta}^*) \bigg) \Bigg] \Bigg\} \\
    &\leq 3 \text{Var}_{\boldsymbol{D}} \Bigg\{ E_{\boldsymbol{\Psi} \vert \boldsymbol{D}} \Bigg[ \frac{1}{n} \sum_{i=1}^n \bigg(1 - \frac{1(T_i = t)}{p_{ti}^*} \bigg) \boldsymbol{d}_1(\boldsymbol{X}_i, \boldsymbol{\beta}^*)^T (\boldsymbol{\beta} - \boldsymbol{\beta}^*) \Bigg] \Bigg\} \\
    &+ 3\text{Var}_{\boldsymbol{D}} \Bigg\{ E_{\boldsymbol{\Psi} \vert \boldsymbol{D}} \Bigg[ \frac{1}{n} \sum_{i=1}^n \bigg(1 - \frac{1(T_i = t)}{p_{ti}^*} \bigg) \bigg( (\boldsymbol{\beta} - \boldsymbol{\beta}^*)^T \boldsymbol{H}_1(\boldsymbol{X}_i, \boldsymbol{\widetilde{\beta}}) (\boldsymbol{\beta} - \boldsymbol{\beta}^*) \bigg) \Bigg] \Bigg\} \\
\end{align*}
We will deal with these two terms separately. Starting with the first term, we can see that
\begin{align*}
    & \text{Var}_{\boldsymbol{D}} \Bigg\{ E_{\boldsymbol{\Psi} \vert \boldsymbol{D}} \Bigg[ \frac{1}{n} \sum_{i=1}^n \bigg(1 - \frac{1(T_i = t)}{p_{ti}^*} \bigg) \boldsymbol{d}_1(\boldsymbol{X}_i, \boldsymbol{\beta}^*)^T (\boldsymbol{\beta} - \boldsymbol{\beta}^*) \Bigg] \Bigg\} \\
    &\leq \text{E}_{\boldsymbol{D}} \Bigg\{ \frac{1}{n^2} \Bigg( \sum_{i=1}^n \bigg(1 - \frac{1(T_i = t)}{p_{ti}^*} \bigg) \boldsymbol{d}_1(\boldsymbol{X}_i, \boldsymbol{\beta}^*)^T E_{\boldsymbol{\Psi} \vert \boldsymbol{D}}(\boldsymbol{\beta} - \boldsymbol{\beta}^*) \Bigg)^2 \Bigg\} \\
    &= \text{E}_{\boldsymbol{D}} \Bigg\{ \frac{1}{n^2} \Bigg( \sum_{i=1}^n \bigg(1 - \frac{1(T_i = t)}{p_{ti}^*} \bigg) \boldsymbol{d}_1(\boldsymbol{X}_i, \boldsymbol{\beta}^*)^T \Bigg) E_{\boldsymbol{\Psi} \vert \boldsymbol{D}}(\boldsymbol{\beta} - \boldsymbol{\beta}^*) \\
    & \times E_{\boldsymbol{\Psi} \vert \boldsymbol{D}}(\boldsymbol{\beta} - \boldsymbol{\beta}^*)^T \Bigg( \sum_{i=1}^n \bigg(1 - \frac{1(T_i = t)}{p_{ti}^*} \bigg) \boldsymbol{d}_1(\boldsymbol{X}_i, \boldsymbol{\beta}^*) \Bigg) \Bigg\} \\
    &\leq \text{E}_{\boldsymbol{D}} \Bigg\{ \frac{\lambda_1^{A_1}}{n^2} \Bigg( \sum_{i=1}^n \bigg(1 - \frac{1(T_i = t)}{p_{ti}^*} \bigg) \boldsymbol{d}_1(\boldsymbol{X}_i, \boldsymbol{\beta}^*)^T \Bigg) \Bigg( \sum_{i=1}^n \bigg(1 - \frac{1(T_i = t)}{p_{ti}^*} \bigg) \boldsymbol{d}_1(\boldsymbol{X}_i, \boldsymbol{\beta}^*) \Bigg) \Bigg\} \\
     &\leq \text{E}_{\boldsymbol{D}} \Bigg\{ \frac{|\lambda_1^{A_1}|}{n^2} \Bigg| \Bigg( \sum_{i=1}^n \bigg(1 - \frac{1(T_i = t)}{p_{ti}^*} \bigg) \boldsymbol{d}_1(\boldsymbol{X}_i, \boldsymbol{\beta}^*)^T \Bigg) \Bigg( \sum_{i=1}^n \bigg(1 - \frac{1(T_i = t)}{p_{ti}^*} \bigg) \boldsymbol{d}_1(\boldsymbol{X}_i, \boldsymbol{\beta}^*) \Bigg) \Bigg| \Bigg\} \\
     &\leq \frac{1}{n} \text{E}_{\boldsymbol{D}} \Bigg\{ {\lambda_1^{A_1}}^2 \Bigg\}^{1/2} \text{E}_{\boldsymbol{D}} \Bigg\{ \Bigg[ \Bigg( \sum_{i=1}^n \bigg(1 - \frac{1(T_i = t)}{p_{ti}^*} \bigg) \boldsymbol{d}_1(\boldsymbol{X}_i, \boldsymbol{\beta}^*)^T \Bigg) \\
     & \times \Bigg( \sum_{i=1}^n \bigg(1 - \frac{1(T_i = t)}{p_{ti}^*} \bigg) \boldsymbol{d}_1(\boldsymbol{X}_i, \boldsymbol{\beta}^*) \Bigg) \Bigg]^2 \Bigg\}^{1/2} \\
     &= o(n^{-1})
\end{align*}
Note here that $\lambda_1^{A_1}$ is the maximum eigenvalue of $E_{\boldsymbol{\Psi} \vert \boldsymbol{D}}(\boldsymbol{\beta} - \boldsymbol{\beta}^*) E_{\boldsymbol{\Psi} \vert \boldsymbol{D}}(\boldsymbol{\beta} - \boldsymbol{\beta}^*)^T$, and this converges in probability to zero due to assumption 4. The last equality also held due to the multivariate CLT. Now we can turn attention to the second of the two terms:

\begin{align*}
    & \text{Var}_{\boldsymbol{D}} \Bigg\{ E_{\boldsymbol{\Psi} \vert \boldsymbol{D}} \Bigg[ \frac{1}{n} \sum_{i=1}^n \bigg(1 - \frac{1(T_i = t)}{p_{ti}^*} \bigg) \bigg( (\boldsymbol{\beta} - \boldsymbol{\beta}^*)^T \boldsymbol{H}_1(\boldsymbol{X}_i, \boldsymbol{\widetilde{\beta}}) (\boldsymbol{\beta} - \boldsymbol{\beta}^*) \bigg) \Bigg] \Bigg\} \\
    &= \text{Var}_{\boldsymbol{D}} \Bigg\{ \frac{1}{n} \sum_{i=1}^n \bigg(1 - \frac{1(T_i = t)}{p_{ti}^*} \bigg) E_{\boldsymbol{\Psi} \vert \boldsymbol{D}} \bigg( (\boldsymbol{\beta} - \boldsymbol{\beta}^*)^T \boldsymbol{H}_1(\boldsymbol{X}_i, \boldsymbol{\widetilde{\beta}}) (\boldsymbol{\beta} - \boldsymbol{\beta}^*) \bigg) \Bigg\} \\
    &\leq \text{E}_{\boldsymbol{D}} \Bigg\{ \frac{1}{n^2} \Bigg[ \sum_{i=1}^n \bigg(1 - \frac{1(T_i = t)}{p_{ti}^*} \bigg) E_{\boldsymbol{\Psi} \vert \boldsymbol{D}} \bigg( (\boldsymbol{\beta} - \boldsymbol{\beta}^*)^T \boldsymbol{H}_1(\boldsymbol{X}_i, \boldsymbol{\widetilde{\beta}}) (\boldsymbol{\beta} - \boldsymbol{\beta}^*) \bigg) \Bigg]^2 \Bigg\} \\
    &\leq \frac{1}{n^2} \text{E}_{\boldsymbol{D}} \Bigg\{ \Bigg[ \sum_{i=1}^n \bigg(1 - \frac{1(T_i = t)}{p_{ti}^*} \bigg)^2 \Bigg] \Bigg[ \sum_{i=1}^n E_{\boldsymbol{\Psi} \vert \boldsymbol{D}}^2 \bigg( (\boldsymbol{\beta} - \boldsymbol{\beta}^*)^T \boldsymbol{H}_1(\boldsymbol{X}_i, \boldsymbol{\widetilde{\beta}}) (\boldsymbol{\beta} - \boldsymbol{\beta}^*) \bigg) \Bigg] \Bigg\} \\
    &\leq \frac{1}{n^2} \text{E}_{\boldsymbol{D}} \Bigg\{ \Bigg[ \sum_{i=1}^n \bigg(1 - \frac{1(T_i = t)}{p_{ti}^*} \bigg)^2 \Bigg]^2 \Bigg\}^{1/2} \\
    &\times \text{E}_{\boldsymbol{D}} \Bigg\{ \Bigg[ \sum_{i=1}^n E_{\boldsymbol{\Psi} \vert \boldsymbol{D}}^2 \bigg( (\boldsymbol{\beta} - \boldsymbol{\beta}^*)^T \boldsymbol{H}_1(\boldsymbol{X}_i, \boldsymbol{\widetilde{\beta}}) (\boldsymbol{\beta} - \boldsymbol{\beta}^*) \bigg) \Bigg]^2 \Bigg\}^{1/2} \\
    &\leq \frac{k}{n} \text{E}_{\boldsymbol{D}} \Bigg\{ \Bigg[ \sum_{i=1}^n E_{\boldsymbol{\Psi} \vert \boldsymbol{D}}^2 \bigg( (\boldsymbol{\beta} - \boldsymbol{\beta}^*)^T \boldsymbol{H}_1(\boldsymbol{X}_i, \boldsymbol{\widetilde{\beta}}) (\boldsymbol{\beta} - \boldsymbol{\beta}^*) \bigg) \Bigg]^2 \Bigg\}^{1/2} \\
    &\leq \frac{k}{n} \text{E}_{\boldsymbol{D}} \Bigg\{ \Bigg[ \sum_{i=1}^n E_{\boldsymbol{\Psi} \vert \boldsymbol{D}}^2 \bigg(\lambda_1^{H} ||\boldsymbol{\beta} - \boldsymbol{\beta}^*||_2^2 \bigg) \Bigg]^2 \Bigg\}^{1/2} \\
    &\leq \frac{k'}{n} \text{E}_{\boldsymbol{D}} \Bigg\{ \Bigg[ \sum_{i=1}^n E_{\boldsymbol{\Psi} \vert \boldsymbol{D}}^2 \bigg(||\boldsymbol{\beta} - \boldsymbol{\beta}^*||_2^2 \bigg) \Bigg]^2 \Bigg\}^{1/2}\\
    &= k' \text{E}_{\boldsymbol{D}} \Bigg\{ E_{\boldsymbol{\Psi} \vert \boldsymbol{D}}^4 \bigg(||\boldsymbol{\beta} - \boldsymbol{\beta}^*||_2^2 \bigg)  \Bigg\}^{1/2}\\
    &\leq k' \text{E}_{\boldsymbol{D}} \Bigg\{ E_{\boldsymbol{\Psi} \vert \boldsymbol{D}} \bigg(||\boldsymbol{\beta} - \boldsymbol{\beta}^*||_2^8 \bigg) \Bigg\}^{1/2}\\
    &= o(n^{-1})
\end{align*}
The last equality held by assumption 4. Coupled with the earlier result, this shows that $\text{Var}_{\boldsymbol{D}} \{ E_{\boldsymbol{\Psi} \vert \boldsymbol{D}} [A_1] \} = o(n^{-1})$. An analogous proof would hold for $\text{Var}_{\boldsymbol{D}} \{ E_{\boldsymbol{\Psi} \vert \boldsymbol{D}} [A_2] \}$ so we do not show it for brevity. We can now turn attention to $A_3$. We want to show that $\text{Var}_{\boldsymbol{D}} \{ E_{\boldsymbol{\Psi} \vert \boldsymbol{D}} [A_3] \} = o(n^{-1})$. Proving this is identical to the proof for $\text{Var}_{\boldsymbol{D}^{(m)}} \{ E_{\boldsymbol{\Psi} \vert \boldsymbol{D}} [A_3] \}$ with the outer expectation changed from being with respect to $\boldsymbol{D}^{(m)}$ to $\boldsymbol{D}$. All other steps remain the same, so we will not write them out here in the interest of space. Now we have shown that $\text{Var}_{\boldsymbol{D}} \{ E_{\boldsymbol{\Psi} \vert \boldsymbol{D}} [A_1] \}$, $\text{Var}_{\boldsymbol{D}} \{ E_{\boldsymbol{\Psi} \vert \boldsymbol{D}} [A_2] \}$, and $\text{Var}_{\boldsymbol{D}} \{ E_{\boldsymbol{\Psi} \vert \boldsymbol{D}} [A_3] \}$ are all $o(n^{-1})$ while $\text{Var}_{\boldsymbol{D}} \{ E_{\boldsymbol{\Psi} \vert \boldsymbol{D}} [B] \} = \text{Var}_{\boldsymbol{D}} [B]$, which implies that $\text{Var}_{\boldsymbol{D}} \{ E_{\boldsymbol{\Psi} \vert \boldsymbol{D}} [\Delta(\boldsymbol{D}, \boldsymbol{\Psi})] \} = \text{Var}_{\boldsymbol{D}} [B] + o(n^{-1})$. This result, combined with the results of the previous sections, shows that $\widehat{V} - V = o(n^{-1})$, which is the desired result. 

\section{Proof of posterior contraction rate}

Throughout this section, we will utilize the assumptions listed above in Web Appendix A.1. We will again focus on $E(Y(t)) = \mu_t$, as it is trivial to then extend to the average treatment effect. We will use a somewhat similar decomposition of the doubly robust estimator as follows:
\begin{align*}
& \sup\limits_{P_0} E_{P_0} \mathbb{P}_n(|\mu_t - \mu_t^*| > M\epsilon_n \vert \boldsymbol{D}) \\
&= 
\sup\limits_{P_0} E_{P_0} \mathbb{P}_n \Bigg( \frac{1}{n} \Bigg|\sum_{i=1}^n \frac{1(T_i = t)}{p_{ti}} (Y_i - m_{ti}) + m_{ti} - \mu_t^* \Bigg| > M\epsilon_n \vert \boldsymbol{D} \Bigg) \\
&= \sup\limits_{P_0} E_{P_0} \mathbb{P}_n \Bigg(\Bigg|A_1 + A_2 + A_3 + B \Bigg| > M\epsilon_n \vert \boldsymbol{D} \Bigg).
\end{align*}
Where each of the four individual terms can be written as
\begin{align*}
A_1 &= \frac{1}{n} \sum_{i=1}^n (m_{ti} - \widetilde{m}_{ti}) \Bigg(1 - \frac{1(T_i = t)}{\widetilde{p}_{ti}} \Bigg) \\
A_2 &= \frac{1}{n} \sum_{i=1}^n \frac{1(T_i = t)(p_{ti} - \widetilde{p}_{ti}) (\widetilde{m}_{ti} - Y_i)}{p_{ti}\widetilde{p}_{ti}} \\
A_3 &= \frac{1}{n} \sum_{i=1}^n \frac{1(T_i = t)(p_{ti} - \widetilde{p}_{ti}) (m_{ti} - \widetilde{m}_{ti})}{p_{ti}\widetilde{p}_{ti}} \\
B &= \frac{1}{n} \sum_{i=1}^n \frac{1(T_i = t)}{\widetilde{p}_{ti}} (Y_i - \widetilde{m}_{ti}) + \widetilde{m}_{ti}- \mu_t^*.
\end{align*}
A distinction from the decomposition that was used in the previous section is that $\widetilde{p}_{ti}$ and $\widetilde{m}_{ti}$ are no longer the true values of $p_{ti}$ and $m_{ti}$. We are now letting $\widetilde{p}_{ti}$ and $\widetilde{m}_{ti}$ be the limiting values of $p_{ti}$ and $m_{ti}$, i.e. the value that their posterior distribution contracts around. Therefore we are still assuming that the treatment and outcome model contract at rates $\epsilon_{nt}$ and $\epsilon_{ny}$, though for now we are not assuming that they contract around the correct values. We will show what happens in situations where one or both models are correctly specified, i.e. $\widetilde{p}_{ti} = p_{ti}^*$ or $\widetilde{m}_{ti} = m_{ti}^*$. We can now write the probability as
\begin{align*}
\sup\limits_{P_0} E_{P_0} \mathbb{P}_n(|\mu_t - \mu_t^*| > M\epsilon_n \vert \boldsymbol{D}) &= 
\sup\limits_{P_0} E_{P_0} \mathbb{P}_n(|A_1 + A_2 + A_3 + B| > M\epsilon_n \vert \boldsymbol{D}) \\
&\leq  \sup\limits_{P_0} E_{P_0} \mathbb{P}_n(|A_1| > \frac{M}{4}\epsilon_n \vert \boldsymbol{D}) +
\sup\limits_{P_0} E_{P_0} \mathbb{P}_n(|A_2| > \frac{M}{4}\epsilon_n \vert \boldsymbol{D}) \\
&+  \sup\limits_{P_0} E_{P_0} \mathbb{P}_n(|A_3| > \frac{M}{4}\epsilon_n \vert \boldsymbol{D}) +
\sup\limits_{P_0} E_{P_0} \mathbb{P}_n(|B| > \frac{M}{4}\epsilon_n \vert \boldsymbol{D}),
\end{align*}
so it suffices to show under which situations each of the four components above contracts at the $\epsilon_n$ rate. We will begin with the $B$ component, which does not depend on either the posterior distribution of the treatment or outcome model as it is simply the doubly robust estimator of $\mu_t$ evaluated at the limiting values for the propensity score and outcome regression minus the parameter of interest.
\begin{align*}
	\sup\limits_{P_0} E_{P_0} \mathbb{P}_n(|B| > \frac{M}{4}\epsilon_n \vert \boldsymbol{D}) &= \sup\limits_{P_0} E_{P_0} \mathbb{P}_n \Bigg( \Bigg| \frac{1}{n} \sum_{i=1}^n \frac{1(T_i = t)}{\widetilde{p}_{ti}} (Y_i - \widetilde{m}_{ti}) + \widetilde{m}_{ti}- \mu_t^* \Bigg| > \frac{M}{4}\epsilon_n \vert \boldsymbol{D} \Bigg) \\
    &= \sup\limits_{P_0} E_{P_0} 1 \Bigg( \Bigg| \frac{1}{n} \sum_{i=1}^n \frac{1(T_i = t)}{\widetilde{p}_{ti}} (Y_i - \widetilde{m}_{ti}) + \widetilde{m}_{ti}- \mu_t^* \Bigg| > \frac{M}{4}\epsilon_n \Bigg) \\
    &= \sup\limits_{P_0} P_{P_0} \Bigg( \Bigg| \frac{1}{n} \sum_{i=1}^n \frac{1(T_i = t)}{\widetilde{p}_{ti}} (Y_i - \widetilde{m}_{ti}) + \widetilde{m}_{ti}- \mu_t^* \Bigg| > \frac{M}{4}\epsilon_n \Bigg).
\end{align*}
The second equality holds because all of the components of $B$ are components of $\boldsymbol{D}$ and are therefore just constants when conditioning on $\boldsymbol{D}$. The quantity inside of the absolute values is easily shown to have expectation 0 with respect to $P_0$ if either $\widetilde{m}_{ti} = m_{ti}^*$ or $\widetilde{p}_{ti} = p_{ti}^*$, which highlights the doubly robust aspect of the estimator. Therefore we can apply Chebyshev's inequality to say:
\begin{align*}
\sup\limits_{P_0} E_{P_0} \mathbb{P}_n(|B| > \frac{M}{4}\epsilon_n \vert \boldsymbol{D}) &\leq \sup\limits_{P_0} \frac{16 \text{Var}_{P_0}(B)}{M^2 \epsilon_n^2} \\
    &=\sup\limits_{P_0} \frac{16 \sigma_B^2}{M^2 \epsilon_n^2 n}
\end{align*}
where $\sigma_B^2 = \text{Var}_{P_0} \Bigg( \frac{1(T_i = t)}{p_{ti}^*} (Y_i - m_{ti}^*) \Bigg) \leq K_{B} < \infty$, for some constant $K_{B}$ by assumption 1 and positivity. Clearly if $\epsilon_n > n^{-1/2}$ then this probability goes to zero and we have the desired result. Now we can move to the $A_1$ and $A_2$ terms, which have analogous proofs for showing contraction rates, so we will only show the proof for $A_1$ for brevity. First, let's consider the case where the treatment model is misspecified, i.e. $\widetilde{p}_{ti} \neq p_{ti}^*$. In this case, we can do the following algebraic manipulations:
\begin{align*}
    & \sup\limits_{P_0} E_{P_0} \mathbb{P}_n(|A_1| > \frac{M}{4}\epsilon_n \vert \boldsymbol{D}) \\
    &= \sup\limits_{P_0} E_{P_0} \mathbb{P}_n \Bigg( \bigg| \frac{1}{n} \sum_{i=1}^n (m_{ti} - \widetilde{m}_{ti}) \bigg(1 - \frac{1(T_i = t)}{\widetilde{p}_{ti}} \bigg) \bigg| > \frac{M}{4}\epsilon_n \vert \boldsymbol{D} \Bigg) \\
    &\leq \sup\limits_{P_0} E_{P_0} \mathbb{P}_n \Bigg( \sqrt{\frac{1}{n} \sum_{i=1}^n (m_{ti} - \widetilde{m}_{ti})^2} \sqrt{\frac{1}{n} \sum_{i=1}^n\bigg(1 - \frac{1(T_i = t)}{\widetilde{p}_{ti}} \bigg)^2} > \frac{M}{4}\epsilon_n \vert \boldsymbol{D} \Bigg) \\
    &\leq \sup\limits_{P_0} E_{P_0} \mathbb{P}_n \Bigg( \sqrt{\frac{1}{n} \sum_{i=1}^n (m_{ti} - \widetilde{m}_{ti})^2} > \frac{M \epsilon_n}{4 K} \vert \boldsymbol{D} \Bigg) \\
     &= \sup\limits_{P_0} E_{P_0} \mathbb{P}_n \Bigg( \frac{1}{\sqrt{n}}||\boldsymbol{m}_{t} - \widetilde{\boldsymbol{m}}_{t}|| > \frac{M \epsilon_n}{4 K} \vert \boldsymbol{D} \Bigg). \\
\end{align*}
We can see that this last term is simply the posterior contraction rate of the outcome model. Therefore the $A_1$ term contracts at whatever rate the outcome model contracts at when the propensity score model is misspecified. Note that this does not rely on the outcome model being correctly specified, it simply needs it to contract around a value at a certain rate. Now, we can look at the more friendly scenario, where the propensity score model is correctly specified. For this term we need to introduce some additional notation. We will define
\begin{align*}
    f_n(\boldsymbol{D}_i) &= m_{ti} \Bigg(1 - \frac{1(T_i = t)}{\widetilde{p}_{ti}} \Bigg) \\
    f_0(\boldsymbol{D}_i) &= \widetilde{m}_{ti} \Bigg(1 - \frac{1(T_i = t)}{\widetilde{p}_{ti}} \Bigg)
\end{align*}
We will utilize lemma 19.24 from \cite{van2000asymptotic}, which requires that we show
\begin{align*}
    \int \big(f_n(\boldsymbol{D}_i) - f_0(\boldsymbol{D}_i) \big)^2 d P(\boldsymbol{D}_i) \overset{p}{\to} 0,
\end{align*}
where $P(\boldsymbol{D}_i)$ represents the distribution of the observed data. We can write this quantity as
\begin{align*}
    \int \big(f_n(\boldsymbol{D}_i) - f_0(\boldsymbol{D}_i) \big)^2 d P(\boldsymbol{D}_i) &= \int \Bigg[ (m_{ti} - \widetilde{m}_{ti}) \bigg(1 - \frac{1(T_i = t)}{\widetilde{p}_{ti}} \bigg) \Bigg]^2 d P(\boldsymbol{D}_i) \\
    &= \int (m_{ti} - \widetilde{m}_{ti})^2 \bigg(1 - \frac{1(T_i = t)}{\widetilde{p}_{ti}} \bigg)^2 d P(\boldsymbol{D}_i) \\
    &\leq K_{A_1} \int (m_{ti} - \widetilde{m}_{ti})^2 d P(\boldsymbol{D}_i),
\end{align*}
which converges to zero by the posterior contraction of $m_{ti}$ around $\widetilde{m}_{ti}$. The inequality held for some constant $K_{A_1} < \infty$ by the positivity assumption. If the posterior contraction of $m_{ti}$ is defined with the supremum over $P_0$, then we can show a stronger result, which is that 
\begin{align*}
    \sup\limits_{P_0} \int \big(f_n(\boldsymbol{D}_i) - f_0(\boldsymbol{D}_i) \big)^2 d P(\boldsymbol{D}_i) \overset{p}{\to} 0.
\end{align*}
Lemma 19.24 of \cite{van2000asymptotic} then tells us that 
\begin{align*}
    \sup\limits_{P_0} E_{P_0} \mathbb{P}_n \Bigg( \Bigg| \frac{1}{\sqrt{n}} \sum_{i=1}^n (m_{ti} - \widetilde{m}_{ti}) \Bigg(1 - \frac{1(T_i = t)}{\widetilde{p}_{ti}} \Bigg) \Bigg| > \epsilon \vert \boldsymbol{D} \Bigg) \to 0
\end{align*}
for any $\epsilon > 0$. Note this this is normalized by $1/\sqrt{n}$ instead of $1/n$ as is defined in $A_1$. Due to this, we can also say that
\begin{align*}
    \sup\limits_{P_0} E_{P_0} \mathbb{P}_n \Bigg( \Bigg| \frac{1}{n} \sum_{i=1}^n (m_{ti} - \widetilde{m}_{ti}) \Bigg(1 - \frac{1(T_i = t)}{\widetilde{p}_{ti}} \Bigg) \Bigg| > \epsilon n^{-{1/2}} \vert \boldsymbol{D} \Bigg) \to 0,
\end{align*}
and therefore we have posterior contraction at the $\epsilon_n = n^{-{1/2}}$ rate. Note that this lemma requires the function $m_{ti}$ to be in a Donsker class of functions, which precludes some highly flexible models as well as many high-dimensional models. For a reference on the entropy of high-dimensional models, which can help to determine if certain function classes are Donsker, see \cite{plan2013one}. The entropy is a function of the sparsity of the model and therefore certain sparsity conditions will be required to satisfy this assumption. For a discussion of the sparsity restrictions that this places, as well as further references on this matter, see \cite{chernozhukov2016double}. Note that if one wants to drop this assumption, sample splitting can be used, though we will not discuss that possibility here. In total, this implies that $A_1$ contracts at $\epsilon_n = n^{-{1/2}}$ if the propensity score model is correctly specified, and contracts at $\epsilon_{ny}$, the contraction rate of the outcome model (regardless of correct specification of the outcome model) otherwise. An analogous proof holds for $A_2$ so we omit the details here. $A_2$ will contract at the $\epsilon_n = n^{-{1/2}}$ rate if the outcome model is correctly specified, and will contract at $\epsilon_{nt}$, the contraction rate of the treatment model (regardless of correct specification of the treatment model) otherwise. 
We will now proceed to $A_3$ where the specific rates of contraction for the two models will simultaneously play a role.
\begin{align*}
	&\sup\limits_{P_0}E_{P_0}  \mathbb{P}_n(|A_3| > \frac{M}{4}\epsilon_n \vert \boldsymbol{D}) \\
    &= \sup\limits_{P_0} E_{P_0} \mathbb{P}_n \Bigg(\Bigg|\frac{1}{n} \sum_{i=1}^n \frac{1(T_i = t)(p_{ti} - \widetilde{p}_{ti}) (m_{ti} - \widetilde{m}_{ti})}{p_{ti}\widetilde{p}_{ti}} \Bigg| > \frac{M}{4}\epsilon_n \vert \boldsymbol{D} \Bigg) \\
    &\leq \sup\limits_{P_0} E_{P_0} \mathbb{P}_n \Bigg( \sqrt{\frac{1}{n} \sum_{i=1}^n \Bigg( \frac{1(T_i = t)(p_{ti} - \widetilde{p}_{ti})}{p_{ti}\widetilde{p}_{ti}} \Bigg)^2} \sqrt{\frac{1}{n} \sum_{i=1}^n (m_{ti} - \widetilde{m}_{ti})^2} > \frac{M}{4}\epsilon_n \vert \boldsymbol{D} \Bigg) \\
    &\leq \sup\limits_{P_0} E_{P_0} \mathbb{P}_n \Bigg( \sqrt{\frac{1}{n} \sum_{i=1}^n K_{A_{31}}(p_{ti} - \widetilde{p}_{ti})^2}\sqrt{\frac{1}{n} \sum_{i=1}^n (m_{ti} - \widetilde{m}_{ti})^2} > \frac{M}{4}\epsilon_n \vert \boldsymbol{D} \Bigg) \\
    &= \sup\limits_{P_0} E_{P_0} \mathbb{P}_n \Bigg( \frac{1}{n} ||\boldsymbol{p}_{t} - \widetilde{\boldsymbol{p}}_{t}|| \ ||\boldsymbol{m}_{t} - \widetilde{\boldsymbol{m}}_{t}|| > \frac{M \epsilon_n}{4 \sqrt{K_{A_{31}}}} \vert \boldsymbol{D} \Bigg).
\end{align*}
The first inequality comes from the Cauchy-Schwartz inequality, and the second inequality holds true for some constant $0 < K_{A_{31}} < \infty$ from assumption 2 and positivity. We further decompose this probability as
\begin{align*}
    & \sup\limits_{P_0} E_{P_0} \mathbb{P}_n \Bigg( \frac{1}{n} ||\boldsymbol{p}_{t} - \widetilde{\boldsymbol{p}}_{t}|| \ ||\boldsymbol{m}_{t} - \widetilde{\boldsymbol{m}}_{t}|| > \frac{M \epsilon_n}{4 \sqrt{K_{A_{31}}}} \vert \boldsymbol{D} \Bigg) \\
    &= \sup\limits_{P_0} E_{P_0} \Bigg[ \mathbb{P}_n \Bigg( \frac{1}{n} ||\boldsymbol{p}_{t} - \widetilde{\boldsymbol{p}}_{t}|| \ ||\boldsymbol{m}_{t} - \widetilde{\boldsymbol{m}}_{t}|| > \frac{M \epsilon_n}{4 \sqrt{K_{A_{31}}}} \vert \boldsymbol{D}, \frac{1}{\sqrt{n}} ||\boldsymbol{m}_{t} - \widetilde{\boldsymbol{m}}_{t}|| > \epsilon_n^{\nu_1}  \Bigg) \\
    & \times \mathbb{P}_n \Bigg( \frac{1}{\sqrt{n}} ||\boldsymbol{m}_{t} - \widetilde{\boldsymbol{m}}_{t}|| > \epsilon_n^{\nu_1} \vert \boldsymbol{D} \Bigg) \\
    &+ \mathbb{P}_n \Bigg( \frac{1}{n} ||\boldsymbol{p}_{t} - \widetilde{\boldsymbol{p}}_{t}|| \ ||\boldsymbol{m}_{t} - \widetilde{\boldsymbol{m}}_{t}|| > \frac{M \epsilon_n}{4 \sqrt{K_{A_{31}}}} \vert \boldsymbol{D}, \frac{1}{\sqrt{n}} ||\boldsymbol{m}_{t} - \widetilde{\boldsymbol{m}}_{t}|| \leq \epsilon_n^{\nu_1}  \Bigg) \\
    & \times \mathbb{P}_n \Bigg( \frac{1}{\sqrt{n}} ||\boldsymbol{m}_{t} - \widetilde{\boldsymbol{m}}_{t}|| \leq \epsilon_n^{\nu_1} \vert \boldsymbol{D} \Bigg) \Bigg] \\
    &\leq \sup\limits_{P_0} E_{P_0} \Bigg[ \mathbb{P}_n \Bigg( \frac{1}{\sqrt{n}} ||\boldsymbol{m}_{t} - \widetilde{\boldsymbol{m}}_{t}|| > \epsilon_n^{\nu_1} \vert \boldsymbol{D} \Bigg) \\
    &+ \mathbb{P}_n \Bigg( \frac{1}{n} ||\boldsymbol{p}_{t} - \widetilde{\boldsymbol{p}}_{t}|| \ ||\boldsymbol{m}_{t} - \widetilde{\boldsymbol{m}}_{t}|| > \frac{M \epsilon_n}{4 \sqrt{K_{A_{31}}}} \vert \boldsymbol{D}, \frac{1}{\sqrt{n}} ||\boldsymbol{m}_{t} - \widetilde{\boldsymbol{m}}_{t}|| \leq \epsilon_n^{\nu_1}  \Bigg) \Bigg] \\
    &\leq \sup\limits_{P_0} E_{P_0} \Bigg[ \mathbb{P}_n \Bigg( \frac{1}{\sqrt{n}} ||\boldsymbol{m}_{t} - \widetilde{\boldsymbol{m}}_{t}|| > \epsilon_n^{\nu_1} \vert \boldsymbol{D} \Bigg) + \mathbb{P}_n \Bigg( \frac{1}{\sqrt{n}} ||\boldsymbol{p}_{t} - \widetilde{\boldsymbol{p}}_{t}|| \ > \frac{M \epsilon_n}{4 \sqrt{K_{A_{31}}} \epsilon_n^{\nu_1}} \vert \boldsymbol{D} \Bigg) \Bigg] \\
    &= \sup\limits_{P_0} E_{P_0} \Bigg[ \mathbb{P}_n \Bigg( \frac{1}{\sqrt{n}} ||\boldsymbol{m}_{t} - \widetilde{\boldsymbol{m}}_{t}|| > \epsilon_n^{\nu_1} \vert \boldsymbol{D} \Bigg) + \mathbb{P}_n \Bigg( \frac{1}{\sqrt{n}} ||\boldsymbol{p}_{t} - \widetilde{\boldsymbol{p}}_{t}|| \ > \frac{M \epsilon_n^{1 - \nu_1}}{4 \sqrt{K_{A_{31}}}} \vert \boldsymbol{D} \Bigg) \Bigg] \\
\end{align*}
The first term converges to zero as long as the outcome model contracts at the $\epsilon_n^{\nu_1}$ rate, while the second term converges to zero if the treatment model contracts at the $\epsilon_n^{1 - \nu_1}$ rate. This shows that $A_3$ contracts at the $\epsilon_n$ rate as long as the product of the treatment and outcome model contraction rates is $\epsilon_n$. 

We have now shown the contraction rates for each component of the doubly robust estimator, so we can now see what the overall contraction rate is. If both models are correctly specified, then the $A_1, A_2,$ and $B$ terms contract at $\epsilon_n = n^{-{1/2}}$, while $A_3$ contracts at $\epsilon_{nt} \epsilon_{ny}$. This shows that the doubly robust estimator contracts at either $n^{-{1/2}}$ or $\epsilon_{nt} \epsilon_{ny}$, whichever is slower. If one model is misspecified, then the story is somewhat different. The $B$ term still contracts at $n^{-{1/2}}$, the $A_3$ term still contracts at $\epsilon_{nt} \epsilon_{ny}$, and one of the $A_1$ or $A_2$ terms contracts at $n^{-{1/2}}$, depending on which model is misspecified. Let's suppose that the treatment model is misspecified. In this case, the $A_1$ term will contract at $\epsilon_{ny}$, the contraction rate of the outcome model. When one model is misspecified, the contraction rate of the doubly robust estimator will either be $n^{-{1/2}}$ or the contraction rate of the correctly specified model, whichever is slower. Typically $n^{-{1/2}}$ will only occur under misspecification of one of the models if the correctly specified model is finite-dimensional and parametric. 

\section{Details of posterior sampling}

Here we will present the details required for posterior sampling from both the semiparametric and nonparametric priors utilized. Throughout we will denote the full observed data as $\boldsymbol{D}_i = (Y_i, T_i, \boldsymbol{X}_i)$. First we will present the posterior sampling for the semiparametric prior that models the conditional associations between the treatment/outcome and covariates using splines with $d$ degrees of freedom. We will be always be working with $\boldsymbol{X}$ being standardized to have mean zero and variance 1, which is crucial when using spike and slab priors. Throughout, we will show how to estimate the outcome model, but sampling from the treatment model is analagous with straightforward alterations. Finally, we will be working with the latent outcome $Y_i^*$, where in the case of continuous data, $Y_i^* = Y_i$. If $Y_i$ is binary, then at every iteration of our MCMC we draw $Y_i^*$ from a truncated normal distribution with mean set to $\beta_0 + f_t(T_i) + \sum_{j=1}^p f_j(X_{ji})$ and variance set to 1. If $Y_i = 1$ then this distribution is truncated below by 0 and if $Y_i = 0$ then it is truncated above by 0. Once we have obtained $Y_i^*$, then posterior sampling can continue using the latent outcome as if we had linear regression, even if the outcome is binary. 

\subsubsection*{MCMC sampling for semiparametric prior}

Below we detail the full conditional updates for all parameters in the model.

\begin{enumerate}
	\item If $Y_i$ is binary then set $\sigma^2 = 1$, and if the outcome is continuous draw $\sigma^2$ from an inverse-gamma distribution with parameters $a^*$ and $b^*$, defined as:
         \begin{align*}
     	a^* &= a_{\sigma^2} + \frac{n}{2} + \frac{d \sum_{j=1}^p \gamma_j}{2} \\
        b^* &= b_{\sigma^2} + \frac{\sum_{i=1}^n \left( Y_i^* - \beta_0 - f_t(T_i) - \sum_{j=1}^p  f_{j}(\boldsymbol{X_{ji}}) \right)^2}{2}  + \sum_{j=1}^p \sum_{k=1}^d \frac{\beta_{jk}^2}{2\sigma_{\boldsymbol{\beta}}^2}
     \end{align*}
     \item While not discussed in the main text, we will be placing a $\mathcal{IG}(a_{\sigma_{\boldsymbol{\beta}}^2}, b_{\sigma_{\boldsymbol{\beta}}^2}$) prior on $\sigma^2_{\boldsymbol{\beta}}$ and therefore we can update from the full conditional:
     \begin{align*}
     	\sigma^2_{\boldsymbol{\beta}} \vert \bullet \sim \mathcal{IG}\left(a_{\sigma_{\boldsymbol{\beta}}^2} + \frac{d \sum_{j=1}^p \gamma_j}{2}, b_{\sigma_{\boldsymbol{\beta}}^2} + \sum_{j=1}^p \sum_{k=1}^d \frac{\beta_{jk}^2}{2\sigma^2} \right)
     \end{align*}
     \item Update $\theta$ from the full conditional:
     \begin{align*}
     	\theta \vert \bullet \sim \mathcal{B}\left(a_{\theta} + \sum_{j=1}^p \gamma_j, b_{\theta} + \sum_{j=1}^p (1 - \gamma_j) \right)
     \end{align*}
     \item To update $\gamma_j$ for $j=1 \dots p$ we need to look at the conditional posterior that has marginalized over $\boldsymbol{\beta}_j$. Specifically, if we allow $\boldsymbol{\Lambda}$ to represent all parameters in the model except for $(\gamma_j, \boldsymbol{\beta}_j)$ then we can update $\gamma_j$ from the following conditional distribution:
     \begin{align*}
     	p(\gamma_j = 1 \vert \boldsymbol{D}, \boldsymbol{\Lambda}) &= \frac{p(\boldsymbol{\beta_{j}} = \boldsymbol{0}, \gamma_j = 1 \vert \boldsymbol{D}, \boldsymbol{\Lambda})}{p(\boldsymbol{\beta_{j}} = \boldsymbol{0} \vert \gamma_j = 1,  \boldsymbol{D}, \boldsymbol{\Lambda})} \\
	&= \frac{p( \boldsymbol{D}, \boldsymbol{\Lambda} \vert \boldsymbol{\beta_{j}} = \boldsymbol{0}, \gamma_j = 1) p(\boldsymbol{\beta_{j}} = \boldsymbol{0}, \gamma_j = 1)}{p( \boldsymbol{D}, \boldsymbol{\Lambda}) p(\boldsymbol{\beta_{j}} = \boldsymbol{0} \vert \gamma_j = 1,  \boldsymbol{D}, \boldsymbol{\Lambda})} \\
&= \frac{p( \boldsymbol{D}, \boldsymbol{\Lambda} \vert \boldsymbol{\beta_{j}} = \boldsymbol{0}) p(\boldsymbol{\beta_{j}} = \boldsymbol{0}, \gamma_j = 1)}{p( \boldsymbol{D}, \boldsymbol{\Lambda}) p(\boldsymbol{\beta_{j}} = \boldsymbol{0} \vert \gamma_j = 1,  \boldsymbol{D}, \boldsymbol{\Lambda})} \\
& \propto \frac{p(\boldsymbol{\beta_{j}} = \boldsymbol{0}, \gamma_j = 1)}{p(\boldsymbol{\beta_{j}} = \boldsymbol{0} \vert \gamma_j = 1,  \boldsymbol{D}, \boldsymbol{\Lambda})} \\
&=  \frac{\theta \ \Phi(\boldsymbol{0}; \boldsymbol{0}, \boldsymbol{\Sigma_{\beta}})}{\Phi(\boldsymbol{0}; \boldsymbol{M}, \boldsymbol{V})}
     \end{align*}
     where $\Phi()$ represents the multivariate normal density function. $\boldsymbol{M}$ and $\boldsymbol{V}$ represent the conditional posterior mean and variance for $\boldsymbol{\beta}_j$ when $\gamma_j = 1$ and can be defined as
    \begin{align*}
    	\boldsymbol{M} = \left(\frac{\widetilde{\boldsymbol{X}}_j^T \widetilde{\boldsymbol{X}}_j}{\sigma^2} + \boldsymbol{\Sigma_{\beta}}^{-1} \right)^{-1}  \boldsymbol{\widetilde{X}}_j^T \widetilde{\boldsymbol{Y}} , \ \ \ \ \boldsymbol{V} = \left(\frac{\widetilde{\boldsymbol{X}}_j^T \widetilde{\boldsymbol{X}}_j}{\sigma^2} + \boldsymbol{\Sigma_{\beta}}^{-1} \right)^{-1},
    \end{align*}
    where $\widetilde{\boldsymbol{Y}} = \boldsymbol{Y}^* - \beta_0 - f_t(\boldsymbol{T}) - \sum_{k \neq p} f_k(\boldsymbol{X}_k)$ and $\boldsymbol{\Sigma_{\beta}}$ is a $d-$dimensional diagonal matrix with $\sigma^2\sigma^2_{\boldsymbol{\beta}}$ on the diagonals.
     \item For $j=1 \dots p$, if $\gamma_j =1$ update $\boldsymbol{\beta}_j$ from a multivariate normal distribution with mean $\boldsymbol{M}$ and variance $\boldsymbol{V}$ as defined above. If $\gamma_j = 0$, then set $\boldsymbol{\beta}_j = \boldsymbol{0}$. 
     \item We will jointly update $\beta_0$ and $f_t(\boldsymbol{T})$. For now we will let $f_t(\boldsymbol{T}) = \beta_t T$, though the full conditional will take the same form even if we model $f_t(\boldsymbol{T})$ with polynomials or splines. Define $\boldsymbol{Z}_t = [\boldsymbol{1}', \boldsymbol{T}]$, then the full conditional is of the form
     \begin{align*}
     	(\beta_0, \beta_t) \vert \bullet \sim MVN \left( \left(\frac{\boldsymbol{Z}_t^T \boldsymbol{Z}_t}{\sigma^2} + \boldsymbol{\Sigma}_t^{-1} \right)^{-1}  \boldsymbol{Z}_t^T \widetilde{\boldsymbol{Y}} , 
        \left(\frac{\boldsymbol{Z}_t^T \boldsymbol{Z}_t}{\sigma^2} + \boldsymbol{\Sigma}_t^{-1} \right)^{-1}\right)
     \end{align*}
     where $\widetilde{\boldsymbol{Y}} = \boldsymbol{Y}^* - \sum_{j=1}^p f_j(\boldsymbol{X}_j)$ and $\boldsymbol{\Sigma_{t}}$ is a diagonal matrix with $K$ on the diagonals, with $K$ large so that the treatment effect is not heavily shrunk towards zero.
\end{enumerate}

\subsubsection*{MCMC sampling with gaussian process priors}

Now we will detail the posterior sampling for the model defined in Section 2.2. 

\begin{enumerate}
	\item Update $(\theta,\beta_0, \beta_t)$ using the same updates as above for the semiparametric prior specification.
    \item To update $\gamma_j$ for $j=1 \dots p$ we need to look at the conditional posterior that has marginalized over $f_j(\boldsymbol{X}_j)$. Specifically, if we allow $\boldsymbol{\Lambda}$ to represent all parameters in the model except for $(\gamma_j, f_j(\boldsymbol{X}_j))$ then we can update $\gamma_j$ from the following conditional distribution:
     \begin{align*}
     	p(\gamma_j = 1 \vert \boldsymbol{D}, \boldsymbol{\Lambda}) &= \frac{p(f_j(\boldsymbol{X}_j) = \boldsymbol{0}, \gamma_j = 1 \vert \boldsymbol{D}, \boldsymbol{\Lambda})}{p(f_j(\boldsymbol{X}_j) = \boldsymbol{0} \vert \gamma_j = 1,  \boldsymbol{D}, \boldsymbol{\Lambda})} \\
	&= \frac{p( \boldsymbol{D}, \boldsymbol{\Lambda} \vert f_j(\boldsymbol{X}_j) = \boldsymbol{0}, \gamma_j = 1) p(f_j(\boldsymbol{X}_j) = \boldsymbol{0}, \gamma_j = 1)}{p( \boldsymbol{D}, \boldsymbol{\Lambda}) p(f_j(\boldsymbol{X}_j) = \boldsymbol{0} \vert \gamma_j = 1,  \boldsymbol{D}, \boldsymbol{\Lambda})} \\
&= \frac{p( \boldsymbol{D}, \boldsymbol{\Lambda} \vert f_j(\boldsymbol{X}_j) = \boldsymbol{0}) p(f_j(\boldsymbol{X}_j) = \boldsymbol{0}, \gamma_j = 1)}{p( \boldsymbol{D}, \boldsymbol{\Lambda}) p(f_j(\boldsymbol{X}_j) = \boldsymbol{0} \vert \gamma_j = 1,  \boldsymbol{D}, \boldsymbol{\Lambda})} \\
& \propto \frac{p(f_j(\boldsymbol{X}_j) = \boldsymbol{0}, \gamma_j = 1)}{p(f_j(\boldsymbol{X}_j) = \boldsymbol{0} \vert \gamma_j = 1,  \boldsymbol{D}, \boldsymbol{\Lambda})} \\
&=  \frac{\theta \ \Phi(\boldsymbol{0}; \boldsymbol{0}, \sigma^2 \tau^2 \boldsymbol{\Sigma_j})}{\Phi(\boldsymbol{0}; \boldsymbol{M}, \boldsymbol{V})}
     \end{align*}
     where $\Phi()$ represents the multivariate normal density function. $\boldsymbol{M}$ and $\boldsymbol{V}$ represent the conditional posterior mean and variance for $f_j(\boldsymbol{X}_j)$ when $\gamma_j = 1$ and can be defined as
    \begin{align*}
    	\boldsymbol{M} = \left(\boldsymbol{I}_n + \frac{1}{\tau_j^2}\boldsymbol{\Sigma_j}^{-1} \right)^{-1} \widetilde{\boldsymbol{Y}} , \ \ \ \ \boldsymbol{V} = \left(\boldsymbol{I}_n + \frac{1}{\tau_j^2}\boldsymbol{\Sigma_j}^{-1} \right)^{-1},
    \end{align*}
    where $\widetilde{\boldsymbol{Y}} = \boldsymbol{Y}^* - \beta_0 - f_t(\boldsymbol{T}) - \sum_{k \neq p} f_k(\boldsymbol{X}_k)$.
    \item For $j=1 \dots p$, if $\gamma_j =1$ update $f_j(\boldsymbol{X}_j)$ from a multivariate normal distribution with mean $\boldsymbol{M}$ and variance $\boldsymbol{V}$ as defined above. If $\gamma_j = 0$, then set $f_j(\boldsymbol{X}_j) = \boldsymbol{0}$.
    \item If $\gamma_j = 0$, update $\tau_j^2$ from it's prior distribution, which is a Gamma($1/2,1/2$). If $\gamma_j = 1$, update $\tau_j^2$ from the following distribution:
    \begin{align*}
    	\mathcal{IG} \left(\frac{n+1}{2}, \frac{1}{2} + \frac{f_j(\boldsymbol{X}_j)^T \boldsymbol{\Sigma}_j^{-1} f_j(\boldsymbol{X}_j)} {2 \sigma^2} \right)
    \end{align*}
    	\item If $Y_i$ is binary then set $\sigma^2 = 1$, and if the outcome is continuous draw $\sigma^2$ from an inverse-gamma distribution with parameters $a^*$ and $b^*$ defined as:
         \begin{align*}
     	a^* &= a_{\sigma^2} + \frac{n (1 + \sum_{j=1}^p \gamma_j)}{2} \\
        b^* &= b_{\sigma^2} + \frac{\sum_{i=1}^n \left( Y_i^* - \beta_0 - f_t(T_i) - \sum_{j=1}^p  f_{j}(\boldsymbol{X_{ji}}) \right)^2}{2}  + \sum_{j=1}^p \frac{\gamma_j f_j(\boldsymbol{X}_j)^T \boldsymbol{\Sigma}_j^{-1} f_j(\boldsymbol{X}_j)} {2 \tau_j^2}
     \end{align*}
\end{enumerate}

One thing to note is that in the conditional updates for $(\gamma_j, f_j(\boldsymbol{X}_j))$, we must calculate $\left(\boldsymbol{I}_n + \frac{1}{\tau_j^2}\boldsymbol{\Sigma_j}^{-1} \right)^{-1}$, which means inverting an $n$ by $n$ matrix at every MCMC iteration. To avoid this, we can first compute the singular value decomposition, $\boldsymbol{\Sigma}_j = \boldsymbol{A B A}^T$, where $\boldsymbol{A}$ is a matrix of eigenvectors and $\boldsymbol{B}$ is a diagonal matrix of eigenvalues. From this, it can be shown that $\left(\boldsymbol{I}_n + \frac{1}{\tau_j^2}\boldsymbol{\Sigma_j}^{-1} \right)^{-1} = \boldsymbol{A} \big( \boldsymbol{I}_n + \frac{\boldsymbol{B}^{-1}}{\tau_j^2}\big)^{-1} \boldsymbol{A}^T$, which only requires inverting a diagonal matrix and can be computed much faster.

\section{Understanding the variance estimator}

Throughout we use the subscript $\boldsymbol{D}$ to denote the distribution of the observed data, $\boldsymbol{D}^{(m)}$ to denote the empirical distribution of the data from which we draw new data sets, and $\boldsymbol{D}^{obs}$ is the observed data.

In this section, we explicitly write out the error in using $\text{Var}_{\boldsymbol{D}^{(m)}}  \big[E_{\boldsymbol{\Psi} \vert \boldsymbol{D}^{obs}} \big(\Delta(\boldsymbol{D}^{(m)}, \boldsymbol{\Psi})\big) \big]$ as the estimator for the variance $\text{Var}_{\boldsymbol{D}} \big[E_{\boldsymbol{\Psi} \vert \boldsymbol{D}} \big(\Delta(\boldsymbol{D}, \boldsymbol{\Psi})\big) \big]$. This will provide insight as to why our variance estimator tends to recover the true variance or is conservative, leading to nominal coverage rates.
We can write out their difference as follows:
\begin{align*}
	\text{Difference} &= \text{Var}_{\boldsymbol{D}} \bigg[E_{\boldsymbol{\Psi} \vert \boldsymbol{D}} \Big(\Delta(\boldsymbol{D}, \boldsymbol{\Psi})\Big) \bigg] - \text{Var}_{\boldsymbol{D}^{(m)}}  \bigg[E_{\boldsymbol{\Psi} \vert \boldsymbol{D}^{obs}} \Big(\Delta(\boldsymbol{D}^{(m)}, \boldsymbol{\Psi})\Big) \bigg] \\
	&= E_{\boldsymbol{D}} \bigg[E^2_{\boldsymbol{\Psi} \vert \boldsymbol{D}} \Big(\Delta(\boldsymbol{D}, \boldsymbol{\Psi})\Big) \bigg] - E^2_{\boldsymbol{D}} \bigg[E_{\boldsymbol{\Psi} \vert \boldsymbol{D}} \Big(\Delta(\boldsymbol{D}, \boldsymbol{\Psi})\Big) \bigg] \\
	& \hspace{20pt} - E_{\boldsymbol{D}^{(m)}}  \bigg[E^2_{\boldsymbol{\Psi} \vert \boldsymbol{D}^{obs}} \Big(\Delta(\boldsymbol{D}^{(m)}, \boldsymbol{\Psi})\Big) \bigg] +
	 E^2_{\boldsymbol{D}^{(m)}}  \bigg[E_{\boldsymbol{\Psi} \vert \boldsymbol{D}^{obs}} \Big(\Delta(\boldsymbol{D}^{(m)}, \boldsymbol{\Psi})\Big) \bigg] \\
	 &\approx E_{\boldsymbol{D}} \bigg[E^2_{\boldsymbol{\Psi} \vert \boldsymbol{D}} \Big(\Delta(\boldsymbol{D}, \boldsymbol{\Psi})\Big) \bigg] - E^2_{\boldsymbol{D}} \bigg[E_{\boldsymbol{\Psi} \vert \boldsymbol{D}} \Big(\Delta(\boldsymbol{D}, \boldsymbol{\Psi})\Big) \bigg] \\
	&\hspace{20pt} - E_{\boldsymbol{D}}  \bigg[E^2_{\boldsymbol{\Psi} \vert \boldsymbol{D}^{obs}} \Big(\Delta(\boldsymbol{D}, \boldsymbol{\Psi})\Big) \bigg] +
	 E^2_{\boldsymbol{D}}  \bigg[E_{\boldsymbol{\Psi} \vert \boldsymbol{D}^{obs}} \Big(\Delta(\boldsymbol{D}, \boldsymbol{\Psi})\Big) \bigg] \\
	 &\approx  E_{\boldsymbol{D}} \bigg[E^2_{\boldsymbol{\Psi} \vert \boldsymbol{D}} \Big(\Delta(\boldsymbol{D}, \boldsymbol{\Psi})\Big) \bigg] - E_{\boldsymbol{D}}  \bigg[E^2_{\boldsymbol{\Psi} \vert \boldsymbol{D}^{obs}} \Big(\Delta(\boldsymbol{D}, \boldsymbol{\Psi})\Big) \bigg].
\end{align*}
\noindent The first approximation stems from the fact that we replaced $E_{\boldsymbol{D}^{(m)}}$ with $E_{\boldsymbol{D}}$. This is a reasonable approximation as the distribution of $\boldsymbol{D}^{(m)}$ is the empirical distribution of a sample from the distribution of $\boldsymbol{D}$. The second approximation stems from the fact that the average of the posterior mean using the observed data posterior, $E_{\boldsymbol{D}}  E_{\boldsymbol{\Psi} \vert \boldsymbol{D}^{obs}}$ should closely resemble $E_{\boldsymbol{D}} E_{\boldsymbol{\Psi} \vert \boldsymbol{D}}$ as the posterior mean from one sample is a reasonable estimator of the posterior mean averaged over repeated samples.

We now look at the average of the correction term, the term added to the initial variance. We show that on average this term closely resembles the difference above, thereby producing a variance estimator that is of the correct magnitude, though likely conservative in finite samples or when one model is misspecified.
\begin{align*}
	E_{\boldsymbol{D}}  \bigg[\text{Var}_{\boldsymbol{\Psi} \vert \boldsymbol{D}} \Big(\Delta(\boldsymbol{D}, \boldsymbol{\Psi})\Big) \bigg]
	= E_{\boldsymbol{D}}  \bigg[E_{\boldsymbol{\Psi} \vert \boldsymbol{D}} \Big(\Delta(\boldsymbol{D}, \boldsymbol{\Psi})^2\Big) \bigg] - E_{\boldsymbol{D}}  \bigg[E^2_{\boldsymbol{\Psi} \vert \boldsymbol{D}} \Big(\Delta(\boldsymbol{D}, \boldsymbol{\Psi})\Big) \bigg] 
\end{align*}

\noindent There are a couple of important things to note in order to understand how this term provides a conservative estimator of the difference term above. Note that the second term in the right side of the expression for the correction is nearly identical to the second term of the difference. The one difference is that the posterior distribution in the difference term is the posterior of the observed data, denoted by $\boldsymbol{\Psi} \vert \boldsymbol{D}^{obs}$, whereas the correction term averages over different posterior distributions $\boldsymbol{\Psi} \vert \boldsymbol{D}$. The posterior mean from the observed data should provide a reasonable estimator of the average posterior mean and therefore these terms should be similar. Now we can look at the first terms in both the difference and correction expressions. Using Jensen's inequality, we can see that
\begin{align*}
    E_{\boldsymbol{D}}  \bigg[E_{\boldsymbol{\Psi} \vert \boldsymbol{D}} \Big(\Delta(\boldsymbol{D}, \boldsymbol{\Psi})^2\Big) \bigg] \geq
    E_{\boldsymbol{D}} \bigg[E^2_{\boldsymbol{\Psi} \vert \boldsymbol{D}} \Big(\Delta(\boldsymbol{D}, \boldsymbol{\Psi})\Big) \bigg].
\end{align*}
This shows that the first term in the correction is at least as large as the first term in the difference. This shows that the correction term will tend to be somewhat conservative. In Web Appendix I we evaluate how this correction does across all simulation scenarios looked at, and we see that it always leads to estimates of the variance that are close to the sampling distribution variance as desired, and generally are only slightly conservative. Further, our asymptotic results show that this conservativeness will disappear asymptotically if both models are correctly specified. 

\section{Asymptotic simulations}

In this section we aim to confirm our theoretical results that our variance estimator is consistent when both models are correctly specified, and to see the extent to which it is conservative in finite samples or when one model is misspecified. For this illustration we will fix $p=10$ and vary the sample size $n \in \{100,250,500,750,1000,2500,5000,10000 \}$. We will utilize the following data generating process for the treatment and outcome models
\begin{align*}
\boldsymbol{X}_i &\sim N(\boldsymbol{0}_p, \boldsymbol{\Sigma}), \hspace{1cm}
T_i \vert \boldsymbol{X}_i \sim \text{Bernoulli}(p_i)
\hspace{1cm}
Y_i \vert T_i, \boldsymbol{X}_i \sim \mathcal{N}(\mu_i, \boldsymbol{I}_n) \\
	\mu_i &= T_i + 0.8X_{1i} + 0.5X_{2i} \\
    p_i &= \Phi(0.4X_{1i} - 0.5X_{2i} + 0.4X_{5i}), \\
\end{align*}
where covariates are drawn from independent standard normal distributions. To evaluate the performance of variance estimation we will look at the ratio of the estimated to true standard errors. Specifically we will take the ratio of the average estimated standard error and the standard deviation of the estimates across all simulations. If this number is equal to 1, that would suggest that our variance is estimating the true variance correctly. Values above 1 indicate conservativeness, while values below 1 indicate anti-conservative inference. We will look at four different doubly robust estimators:
\begin{enumerate}
    \item A correctly specified treatment and outcome model
    \item A correctly specified treatment and outcome model without the second term added to the variance
    \item A misspecified outcome model by excluding the first four covariates, and a correctly specified treatment model
    \item A misspecified treatment model by excluding the first four covariates, and a correctly specified outcome model
\end{enumerate}

\begin{figure}[H]
\centering
 \includegraphics[width=0.55\linewidth]{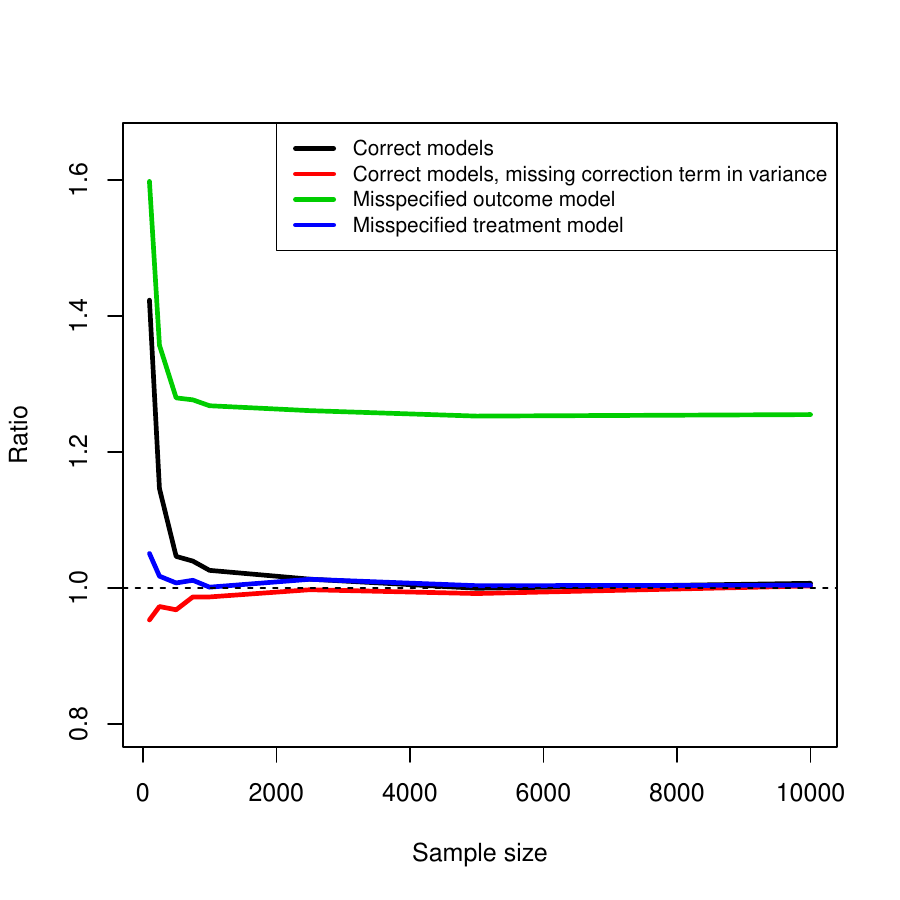}
\caption{Results of the asymptotic simulation study.}
\label{fig:simAsymptotic}
\end{figure}

The results can be seen in \ref{fig:simAsymptotic}. We see that our variance estimator converges to the truth as the sample size grows when both models are correctly specified, though is somewhat conservative for small sample sizes. When the outcome model is misspecified we see that the variance estimator is conservative and that this does not go away asymptotically. If we were to not include our variance correction term (the second term in our variance estimator) then we get anti-conservative inference in small samples, but accurate inference asymptotically. Lastly, if the treatment model is misspecified, we are conservative in small samples, but this decreases as the sample size increases. This simulation confirms our theoretical results that our inference is valid when both models are misspecified. Interestingly, the variance estimator seems to be consistent even when the treatment model is misspecified. We have seen this repeatedly in simulations, though there is no theory to prove this fact. Overall the simulation shows that our variance estimator is consistent, but that it can still provide valid, though conservative, inference when models are misspecified or the sample size is small. 

\section{Additional simulation scenarios}

Here we will run a number of additional simulation scenarios to assess the performance of the proposed approach. The first three scenarios will be in the binary treatment setting, while the fourth will look at a continuous exposure response curve. The first two scenarios will be from sparse data generating models that have different functional forms from the simulations in the paper, while the third scenario looks at a non sparse setting to see how the method performs when the assumption of sparsity does not hold. The fourth simulation differs from the continuous treatment simulation of the paper in that it has linear relationships between the covariates and treatment / outcome instead of nonlinear ones. 

\subsection{Scenario 1}

\begin{figure}[p]
\centering
 \includegraphics[width=0.8\linewidth]{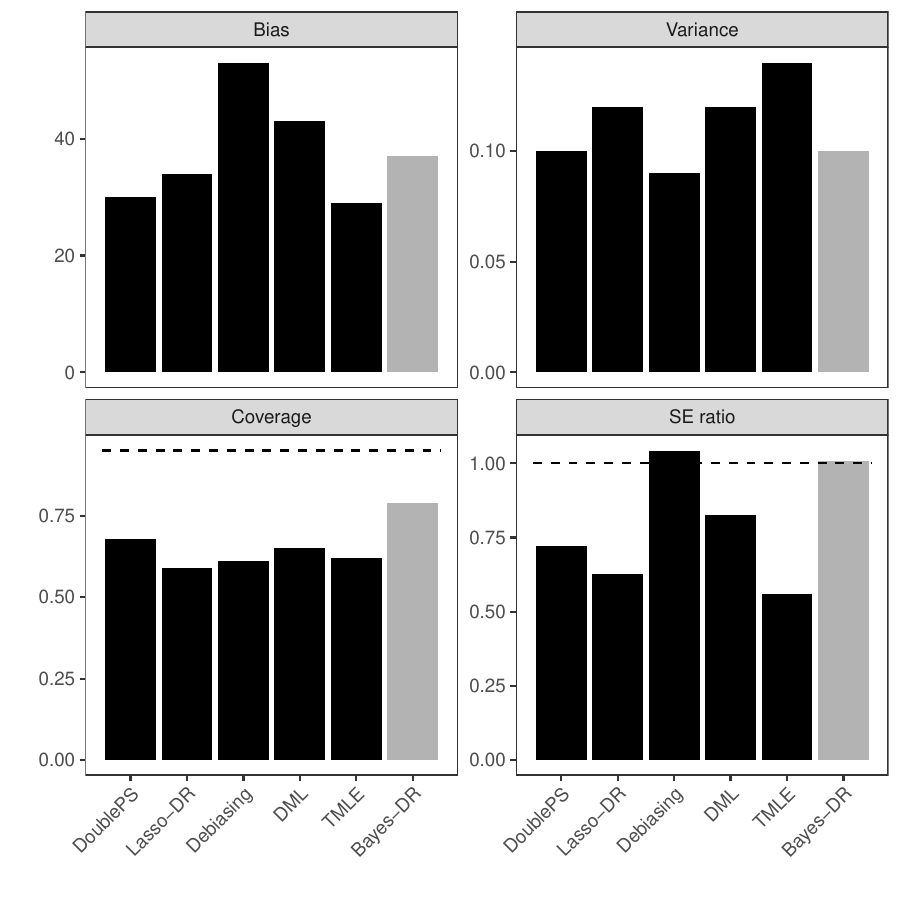}
\caption{Results from the first additional simulation scenario. The top left panel shows absolute bias, the top right panel shows the variance, the bottom left panel shows 95\% interval coverages, while the bottom right panel is the ratio of estimated to Monte Carlo standard errors.  }
\label{fig:addSim1}
\end{figure}

Here we will run an identical simulation to the linear simulation scenario from the manuscript with $n=100$ and $p=500$, except now we will generate data from the following treatment and outcome models:
\begin{align*}
	\mu_i &= T_i + 0.45X_{1i} + 0.7 X_{2i} - 0.6X_{3i} + 1.3 X_{4i} - 0.5X_{5i} \\
    p_i &= \Phi(0.35 X_{1i} + 0.2 X_{2i} - 0.3 X_{3i} - 0.4 X_{5i})
\end{align*}
The results of this simulation study can be seen in \ref{fig:addSim1}. We see that the proposed approach does very well in terms of variance, as only the debiasing estimator is lower. All approaches had non-negligible bias so none of them obtain the nominal coverage rate. Our approach, however, appears to be accounting for more of the finite sample uncertainty in the estimation of the causal effect, which leads to a much higher coverage rate. Our approach has a ratio of the true to Monte Carlo standard errors that is equal to 1, while the remaining approaches with the exception of the debiasing approach are far below 1. 

\subsection{Scenario 2}

Now, similarly to an experiment run by \cite{belloni2013inference}, we will generate data such that $\mu_i = T_i + \boldsymbol{X_i \beta}$, and $p_i = \Phi(\boldsymbol{X_i \alpha})$, where $\boldsymbol{\beta} = \boldsymbol{\alpha} = (1, 1/4, 1/9, \dots, 1/p^2)$. This situation is not strictly sparse as none of the coefficients are exactly zero, though it is approximately sparse in the sense that a small number of confounders can remove essentially all of the confounding bias. We will set $n=100$ and $p=300$. 

\begin{figure}[p]
\centering \includegraphics[width=0.8\linewidth]{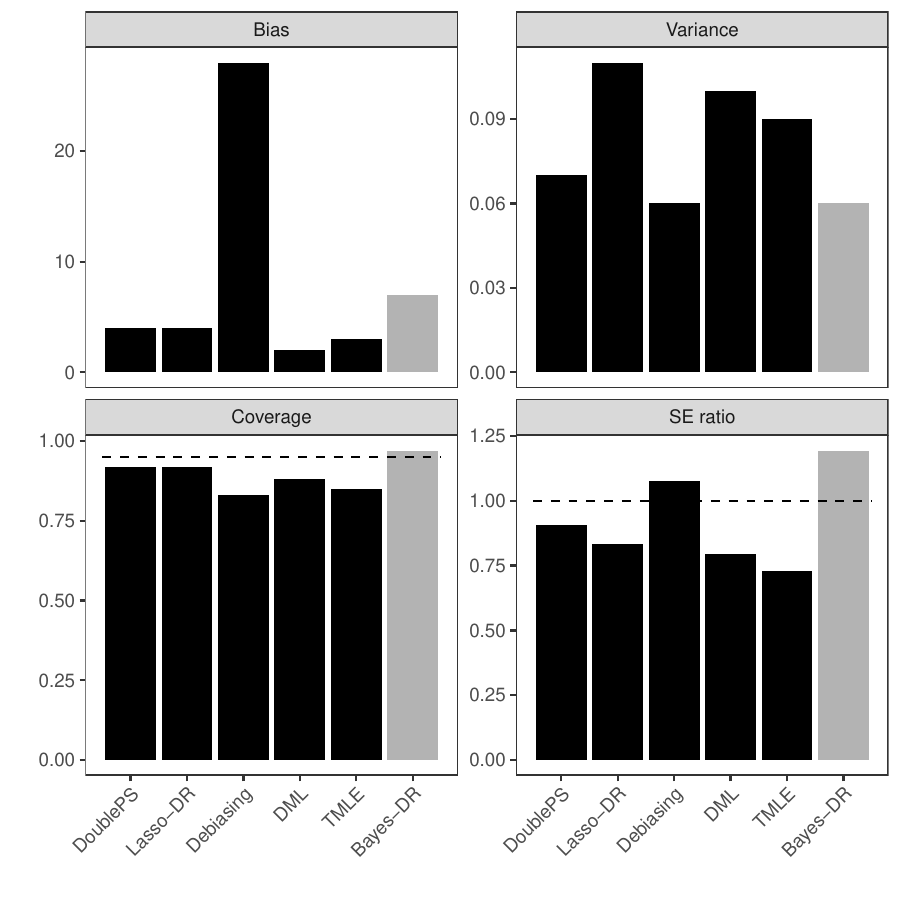}
\caption{Results from the second additional simulation scenario. The top left panel shows absolute bias, the top right panel shows the variance, the bottom left panel shows 95\% interval coverages, while the bottom right panel is the ratio of estimated to Monte Carlo standard errors.  }
\label{fig:addSim2}
\end{figure}

The results of this simulation study can be seen in \ref{fig:addSim2}. Most approaches have small bias with the exception of debiasing. The Double PS and Lasso-DR procedures both obtain reasonable levels of coverage only slightly below the nominal level, which is caused by the fact that the ratio of their estimated to true standard errors is below 1. Our approach is slightly conservative in this setting as we obtain coverages slightly above 95\% with a ratio of true to estimated standard errors that is 1.19.

\subsection{Scenario 3}

Next, we will look at scenario that comes from \cite{athey2016approximate}, where the propensity score is dense. First we define 20 clusters, $\{\mathbf{c_1},\dots, \mathbf{c_{20}} \}$ where $\mathbf{c_k} \sim \mathcal{N}(0, I_{p x p})$. Second, we draw $\mathbf{C}_i$ uniformly at random from one of the 20. Third, we draw the covariates from a multivariate normal distribution centered at $\mathbf{C}_i$ with the identity matrix as the covariance. Fourth, we set $T_i = 1$ with probability 0.25 for the first 10 clusters, and $T_i = 1$ with probability 0.75 for the remaining clusters. Finally, we generate data from the outcome model defined as $Y_i = 10 T_i + \boldsymbol{X \beta} + \epsilon_i$, where $\boldsymbol{\beta} \propto (1, \frac{1}{2}, \dots, \frac{1}{p})$ and is normalized such that $|| \boldsymbol{\beta}||_2^2 = 18$. Here we will again set $n=100$ and $p=300$. Intuitively, this is a simulation scenario in which the outcome model is approximately sparse, though the treatment model is dense as all of the covariates are associated with the treatment. 

\begin{figure}[p]
\centering
 \includegraphics[width=0.8\linewidth]{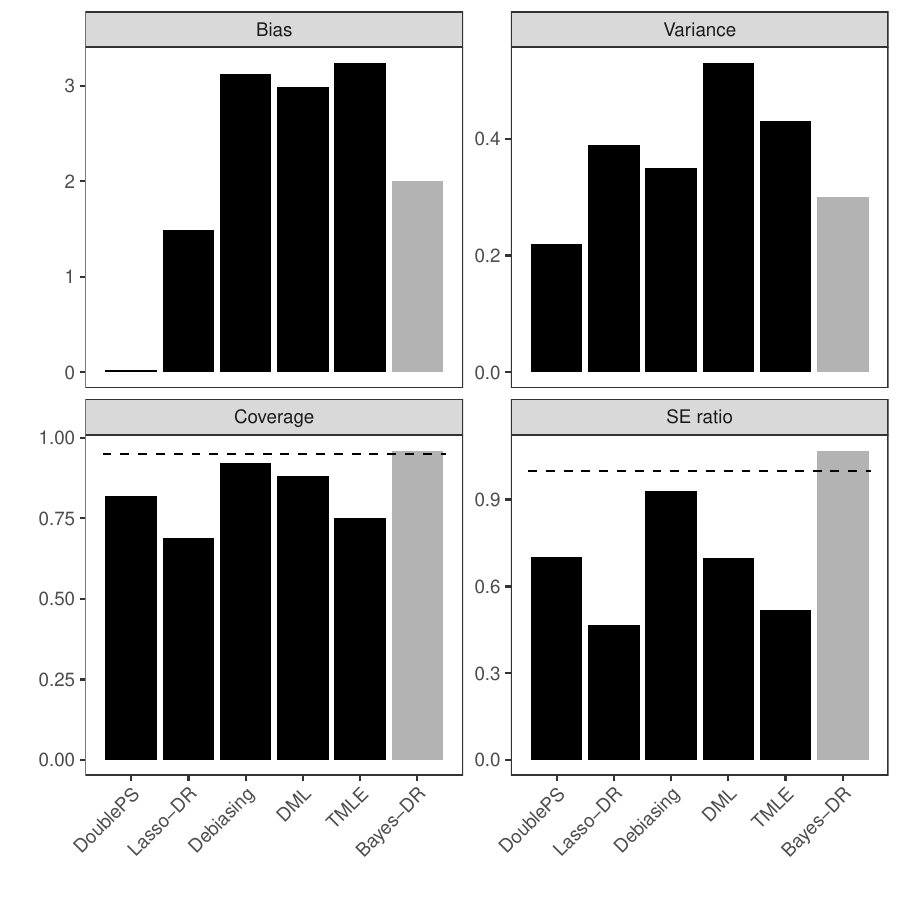}
\caption{Results from the third additional simulation scenario. The top left panel shows absolute bias, the top right panel shows the variance, the bottom left panel shows 95\% interval coverages, while the bottom right panel is the ratio of estimated to Monte Carlo standard errors.  }
\label{fig:addSim3}
\end{figure}

The results of this simulation study can be seen in \ref{fig:addSim3}. The Double PS approach works the best in this setting in terms of bias and variance. This is probably due to the fact that it uses an outcome model in the end to perform inference, and therefore is less affected by the dense propensity model. Our approach has a higher variance due to the dense propensity score model, but importantly still obtains 95\% interval coverage as we seem to be accurately estimating the uncertainty in our estimator.

\subsection{Scenario 4}

Here we will set $n=200$ and $p=200$. We will generate data from the following models:
\begin{align*}
	Y_i \vert T_i, \boldsymbol{X}_i &\sim \mathcal{N}(\mu_{i}^y, 1) \\
    T_i \vert \boldsymbol{X}_i &\sim \mathcal{N}(\mu_{i}^t, 1) \\
    \boldsymbol{X}_i &\sim N(\boldsymbol{0}_p, \boldsymbol{I}_n),
\end{align*}
where
\begin{align*}
	\mu_{i}^y &= 5 - 0.1T_i + 0.05T_i^3 + 0.5X_{1i} + 0.5X_{2i} - 0.3X_{5i} \\
    \mu_{i}^t &= 0.4X_{1i} + 0.6X_{2i} - 0.5X_{4i},
\end{align*}

\begin{figure}[p]
\centering
 \includegraphics[width=0.75\linewidth]{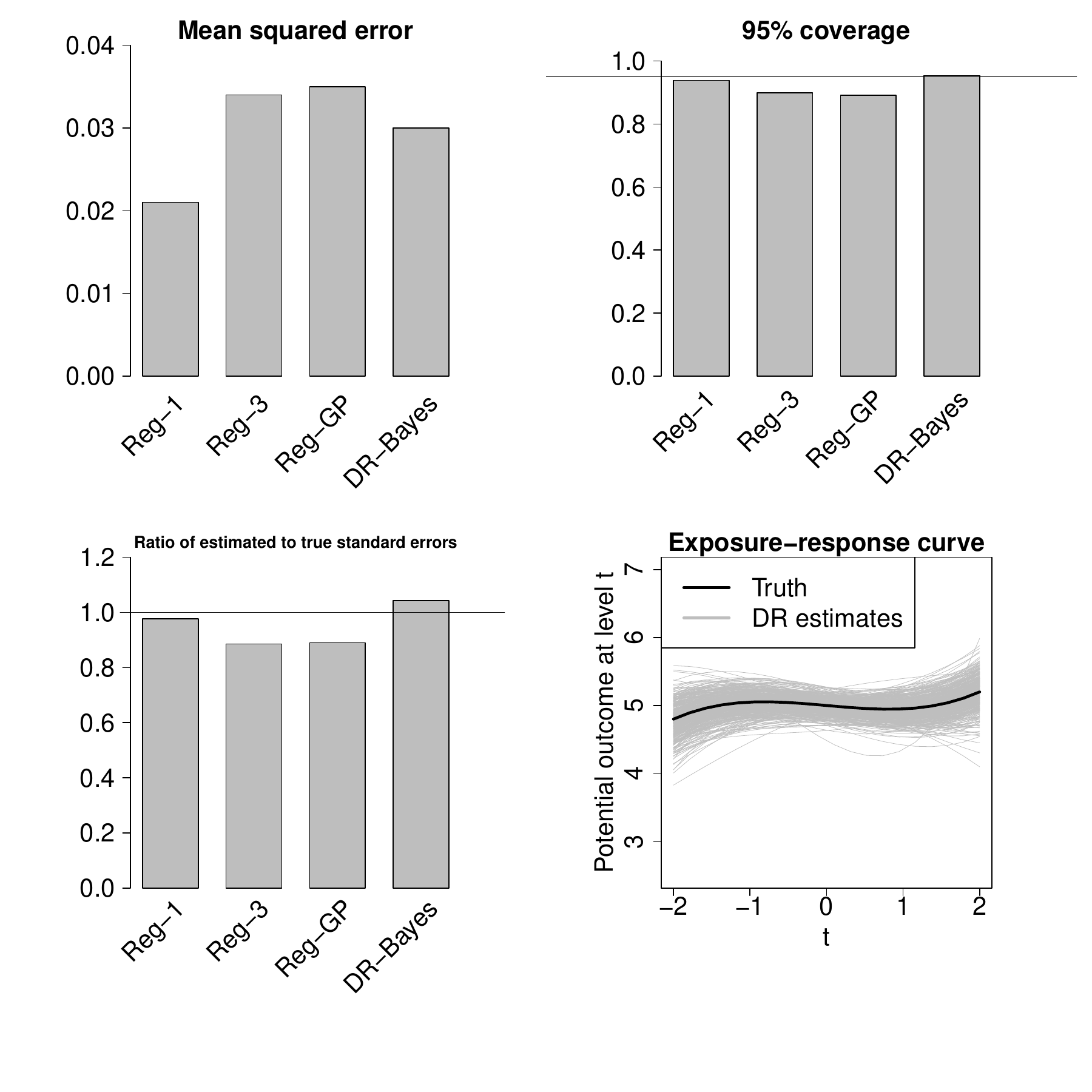}
\caption{Results from the fourth additional simulation scenario. The top left panel shows mean squared error, the top right panel shows the 95\% interval coverages, the bottom left panel is the ratio of estimated to Monte Carlo standard errors, and the bottom right panel shows the true exposure-response curve and the estimates of it across simulated data sets.}
\label{fig:addSim4}
\end{figure}

The results can be seen in \ref{fig:addSim4}. In this scenario, the confounding structure is linear and therefore all of the approaches are able to achieve small amounts of bias. Unsurprisingly, the model assuming linearity does the best of all the approaches, while the performance gets slightly worse in terms of MSE as the amount of nonlinearity increases. Importantly though, the DR-Bayes estimator is still able to achieve 95\% interval coverages across the range of the exposure, and the ratio of estimated to true standard errors is quite close to 1. This indicates that our strategy for variance estimation is performing well and leading to valid inference. 

\section{Simulation without additivity of models}

Here we show that additivity of the treatment and outcome models is not required for our approach as long as there exists a Bayesian modeling approach that incorporates interactions among the variables considered. To highlight this issue, we will use the SoftBart approach of \cite{linero2018bayesian}, which is an extension of the commonly used Bayesian Additive Regression Trees (BART, \cite{chipman2010bart}) to allow for smoother functions and high-dimensional covariate spaces. As this is a tree-based approach, it naturally allows for interactions among the high-dimensional covariate space, thereby alleviating the additivity assumptions used in the simulations of the main manuscript. We will use SoftBart to estimate both the treatment and outcome models, and we will compare with linear models that assume additivity and impose sparsity via spike-and-slab priors. Both models will be embedded within our estimation and inferential strategy to combine posterior samples with resampling. 

We will set $n=150$ and $p=200$, and generate covariates from independent, standard normal distributions. We will generate data from the following models:
\begin{align*}
	\mu_i &= T_i + 0.8X_{1i} + 0.5X_{2i} +  0.8X_{1i}X_{2i} + 0.6X_{3i}X_{4i} \\
    p_i &= \Phi(0.5X_{1i} - 0.4X_{2i} + 0.3X_{1i}X_{2i} + 0.4X_{3i}).
\end{align*}
The additive model will be misspecified for both the treatment and outcome models as there are nonzero interaction terms in each model, while the SoftBart approach allows for such interactions between the covariates. The results can be found in \ref{tab:additive}. We see that when the additive model is used for the treatment and outcome models that there is a substantial bias of 29\%. When SoftBart is used for the two models, the bias drops to only 6\% and the MSE is cut by more than half. In both cases, we end up with reasonable ratios for the estimated to true standard errors, and the SoftBart approach, due to it's lack of bias, achieves close to nominal coverage rates. 
\begin{table}[ht]
\centering
\begin{tabular}{|l|llll|}
  \hline
  & Coverage & SE ratio & MSE & absolute bias \\ 
  \hline
SoftBart & 0.924 & 0.96 & 0.08 & 6 \\ 
  Additive & 0.852 & 1.17 & 0.17 & 29 \\ 
   \hline
\end{tabular}
\caption{Results from simulation with interactions among the covariates. SE ratio represents the ratio of the average estimated standard errors to the Monte Carlo standard errors of the estimates.}
\label{tab:additive}
\end{table}

\section{Inference when models are misspecified}

\begin{figure}[H]
\centering
 \includegraphics[width=0.5\linewidth]{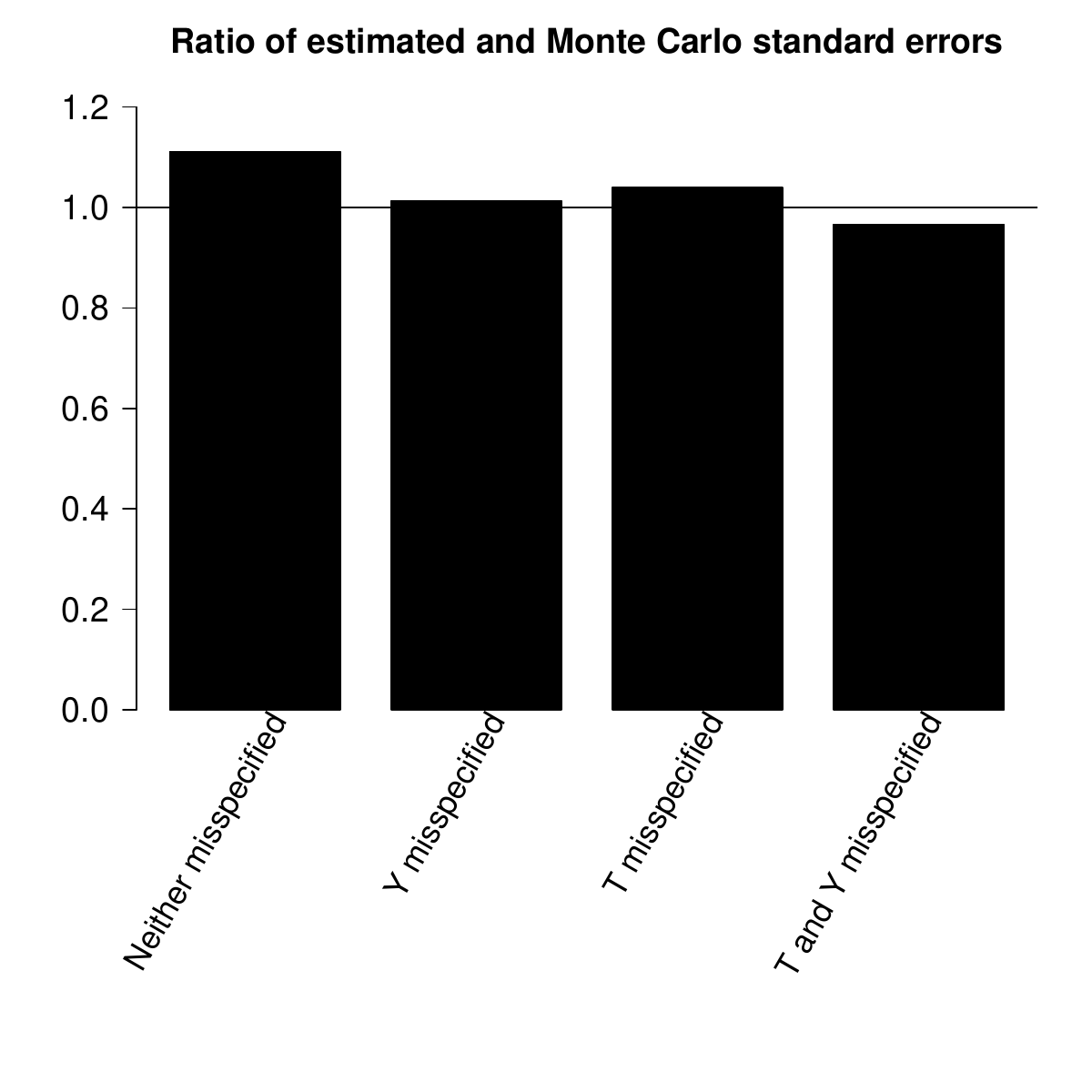}
\caption{Ratio of the estimated to Monte Carlo standard errors when different models are misspecified.}
\label{fig:misspecification}
\end{figure}

Here we will evaluate the extent to which our approach to inference will work if one or both of the models is misspecified. Clearly if both models are misspecified then the approach will likely not attain 95\% coverage, but we can still evaluate the ratio of estimated to true standard errors in this setting. We will restrict to $p=10$ in this setting as our question of interest is based more in model misspecification, which we don't want to conflate with the high-dimensional aspect of the procedure. Our data generating models are either linear or squared functions of the covariates, though we assume linear models for both the treatment and outcome in all cases. The ratio of the estimated to Monte Carlo standard errors can be seen in Figure \ref{fig:misspecification}. We can see that in any of the four scenarios considered, our approach to inference does a good job of estimating the variance. 

\section{Assessing impact of variance correction}

\begin{figure}[!b]
\centering
 \includegraphics[width=0.7\linewidth]{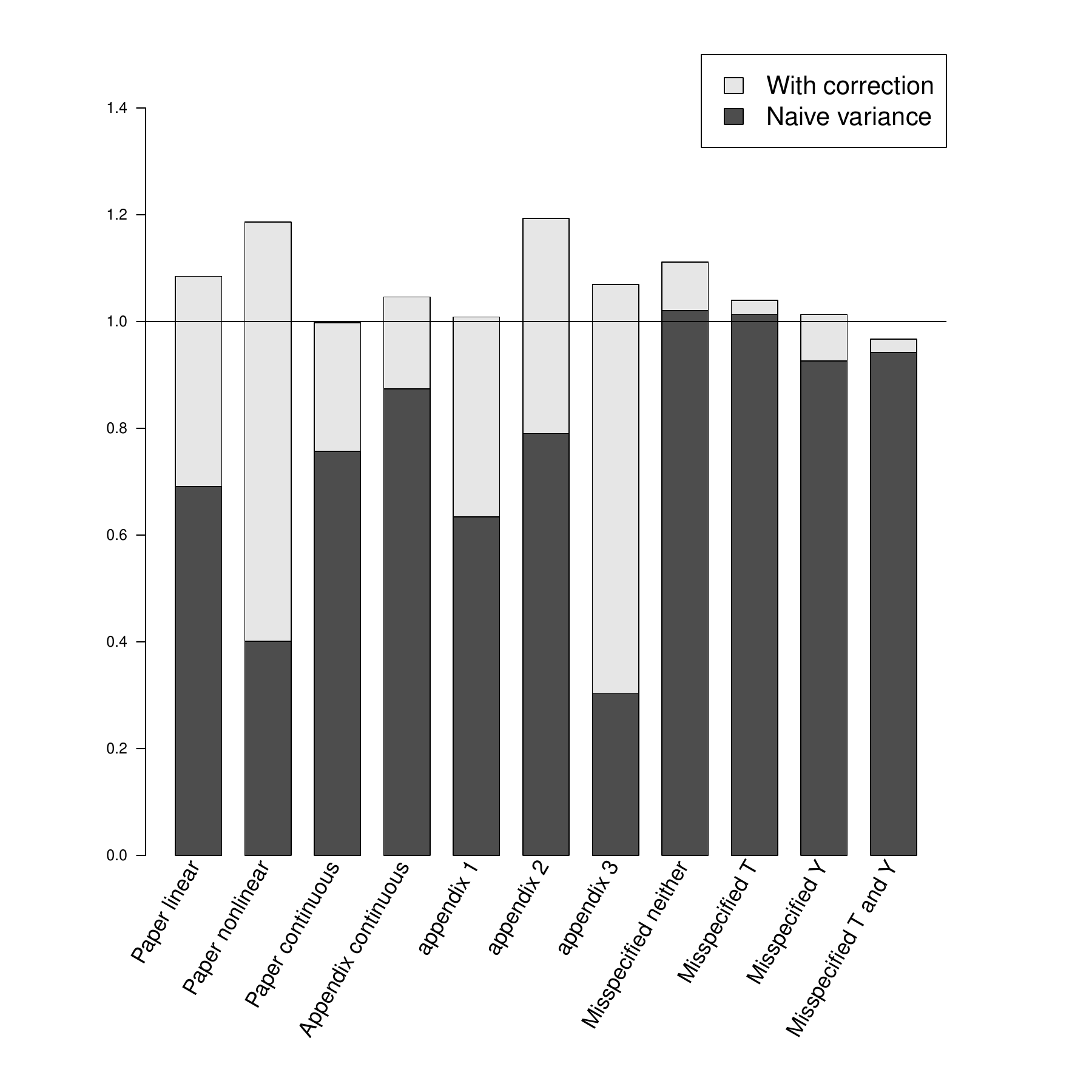}
\caption{Ratio of the estimated to Monte Carlo standard errors when only the initial variance is used and when the correction is added.}
\label{fig:approximation}
\end{figure}

In Section 3 of the manuscript we detailed our approach to variance estimation which entailed an approximation to the variance of interest, plus a correction that attempts to eliminate anti-conservative variances induced by error in the na\"ive variance. Here, we summarize the impact of this correction on estimation across all scenarios looked at in the paper.  \ref{fig:approximation} shows the ratio of the estimated to true standard errors when only the approximate variance is used, as well as the ratio when the correction is applied on top of the na\"ive variance. We see that in some scenarios, such as the misspecification scenarios that were low-dimensional,  the approximate variance is quite good and is close to achieving a ratio of 1. In other scenarios, such as the nonlinear simulation from the paper or the third appendix simulation, the approximation is nowhere near the correct variance, achieving a ratio below 0.4. Importantly though, the correction brings these variances up to a ratio close to, or above, 1 in nearly all scenarios. When the approximation is already doing quite well such as the misspecification scenarios, the correction only adds a small amount as the initial variance is already adequate. In scenarios where the approximation is poor, the variance correction dramatically increases the variance estimator to the correct levels. This shows that even though the approximate variance is asymptotically justified when both models are correctly specified, it is crucial to include the correction term which can fix inference in finite samples or when one model is misspecified. 

\section{Implementation of existing approaches}

First we will detail how we implement the competing approaches for the linear simulation study. For all estimators, linear models are assumed, in contrast to our approach that attempts to find the degree of nonlinearity required. The residual de-biasing approach is implemented using the \texttt{balanceHD} R package, which is available at \url{github.com/swager/balanceHD}. This R package estimates the treatment effect and provides confidence intervals, with which we perform inference. For both TMLE and double machine learning we explore two possible approaches for estimating nuisance parameters. One approach uses lasso models to estimate the nuisance parameters, while the other uses post-lasso models to estimate nuisance parameters. Post-lasso uses an initial variable selection step from a lasso model, and then calculates an unpenalized estimator of the nuisance parameters. For any particular simulation in the paper and appendix, we only report the best of these two approaches with respect to MSE and interval coverage. The TMLE approach is implemented using the \texttt{tmle} R package \citep{tmle}. In the super learner library for TMLE, we only included SL.glmnet, as the true model is contained within this high-dimensional linear model. If we are using post-lasso estimators then we include an initial variable screening step from a lasso regression and then include only SL.glm in the super learner library. This package provides both estimates and confidence intervals, which is how we performed inference. For the double machine learning approach, we used sample splitting with $K=5$ splits and linear models based on the \texttt{glmnet} package \citep{glmnet} for both the treatment and outcome model. Alternatively we find an initial lasso estimate to identify important variables, and then we fit an unpenalized version of these models. We performed inference using the asymptotic standard errors derived in \cite{chernozhukov2016double}. Confidence intervals were created by adding and subtracting 1.96 times this estimated standard error from the point estimate. For the double post selection approach, we fit lasso models for both the treatment and outcome using glmnet with the tuning parameter chosen via cross validation. We then take the union of the selected variables from the two lasso regressions, and re-fit an unpenalized outcome regression model using the union of the selected covariates as confounders. Standard errors are calculated using the asymptotic standard errors derived in \cite{belloni2013inference}, and confidence intervals are created by again adding and subtracting 1.96 times the standard error estimates. We used a similar approach to implementing the lasso-DR approach of \cite{farrell2015robust}. We fit lasso models using glmnet and cross validation and identified important covariates for both the treatment and outcome models. Then, given these respective set of covariates for each model, we calculate unpenalized estimators of the outcome and treatment model. These are then used to calculate the doubly robust estimator described in our manuscript. Inference is again done using the asymptotic standard errors derived in \cite{farrell2015robust}.

For the nonlinear simulation section, we restricted attention to TMLE and double machine learning as these were the most readily available to include nonlinear terms into the model. For both approaches, we ran an initial variable selection step for both the treatment and outcome model, and then identified important covariates as those that are in either the treatment or outcome model. For TMLE, we take these covariates and use the TMLE package with a super learner that includes generalized additive models. Whenever the model only required linearity, we only included GLM into the super learner to avoid using the overly flexible models when they are not necessary. For double machine learning, we take the reduced set of covariates and fit either linear models (when the truth is linear), or nonlinear additive models using spline representations of the covariates with 3 degrees of freedom. Otherwise, the implementation of the double machine learning approach is the same as for the linear case.

\section{Using the bootstrap for competing approaches}

Here, we will assess whether bootstrapping the competing approaches can provide improved inference over the asymptotic intervals used in the manuscript. Bootstrapping is justified and has been used for both TMLE \citep{schnitzer2015double} and double machine learning \citep{knaus2018double}. For the other estimators, it is not clear whether the standard nonparametric bootstrap would provide valid inference, however, here we will perform inference using the nonparametric bootstrap for all competing approaches to evaluate whether it provides better finite sample performance. This is to address the question of whether our approach is only doing better in finite samples because we are resampling, while the other approaches are not. We compared the bootstrap and asymptotic standard errors for all simulation studies with binary treatments in which all estimators were used. In particular, the ratio of the average estimated standard error to the true standard deviation of the sampling distribution of the estimators can be found in Table \ref{tab:BootSim}.

\begin{table}[t]
\caption{Ratio of average estimated standard error and true standard error. The top line shows this ratio for the proposed approach to variance estimation, while the remaining rows highlight both the bootstrap based and asymptotic based standard errors for each of the remaining approaches.}
\label{tab:BootSim}
\centering
 \resizebox{.9\textwidth}{!}{
\begin{tabular}{| r | rrrr |}
  \hline
 & PaperLinear & Appendix1 & Appendix2 & Appendix3 \\ 
  \hline
Proposed approach & 1.08 & 1.01 & 1.19 & 1.07 \\ 
\hline
  DoublePS, asymptotic SE & 0.75 & 0.73 & 0.90 & 0.70 \\ 
  DoublePS, bootstrap SE & 72.74 & 94.35 & 76.65 & 79.27 \\ 
  \hline
  Lasso-DR, asymptotic SE & 0.65 & 0.63 & 0.83 & 0.52 \\ 
  Lasso-DR, bootstrap SE & $>$ 100 & $>$ 100 & $>$ 100 & $>$ 100 \\ 
  \hline
  Debiasing, asymptotic SE & 1.03 & 1.05 & 1.08 & 0.93 \\ 
  Debiasing, bootstrap SE & 0.95 & 0.94 & 0.95 & 1.12 \\ 
  \hline
  DML, asymptotic SE & 0.90 & 0.93 & 0.86 & 0.83 \\ 
  DML, bootstrap SE & 1.11 & 1.10 & 1.05 & 1.16 \\ 
  \hline
  DML post selection, asymptotic SE & 0.84 & 0.83 & 0.80 & 0.70 \\ 
  DML post selection, bootstrap SE & $>$ 100 & $>$ 100 & $>$ 100 & $>$ 100 \\
  \hline
  TMLE, asymptotic SE & 0.49 & 0.47 & 0.62 & 0.30 \\ 
  TMLE, bootstrap SE & 2.04 & 1.90 & 1.70 & 1.84 \\ 
  \hline
  TMLE post selection, asymptotic SE & 0.66 & 0.57 & 0.74 & 0.53 \\ 
  TMLE post selection, bootstrap SE & 1.81 & 1.44 & 1.62 & 1.54 \\
   \hline
\end{tabular}
}
\end{table}

We see that the bootstrap provides extremely large standard error estimates (ratios close to, or greater than, 100) for the DoublePS, Lasso-DR, and DML post selection estimators. The bootstrap also provides standard errors that are too large for TMLE estimators, with ratios well above 1. The only estimators in which the bootstrap performs reasonably well are the debiasing and DML estimators. The debiasing estimator already has a good ratio of estimated to true standard errors using the asymptotic standard errors. The poor coverage for the debiasing estimator is due to bias, and the bootstrap variance can not fix this for improved interval coverage. The DML estimator with lasso models for nuisance parameters is the only estimator in which the bootstrap provides improved inference with ratios slightly above 1. The asymptotic standard errors have ratios below 1 for this estimator, and the bootstrap improves the ratios in each simulation study. The DML estimator based on lasso models, however, performs much worse than the DML post selection estimator in terms of bias and variance. Therefore the DML estimator has coverages of 0.82, 0.60, 0.97, and 0.72 for these four simulations, despite the improved variance estimation.

To assess whether these approaches to inference correct themselves as the sample size increases, we tried the same test, but with $n=500$ and $p=20$. The results can be seen in Table \ref{tab:BootSimLargeN}. We see that when we are in a scenario with a larger sample size and more favorable $p/n$ ratio that both methods to inference provide reasonable standard errors. The bootstrap no longer provides overly large standard errors, and the asymptotic standard errors have standard error ratios close to 1.

\begin{table}[t]
\caption{Ratio of average estimated standard error and true standard error for the simulation with $n=500$ and $p=20$.}
\label{tab:BootSimLargeN}
\centering
 \resizebox{.9\textwidth}{!}{
\begin{tabular}{| r | rrrr |}
  \hline
 & PaperLinear & Appendix1 & Appendix2 & Appendix3 \\ 
  \hline
  DoublePS, asymptotic SE & 0.94 & 0.99 & 0.99 & 0.99 \\ 
  DoublePS, boostrap SE & 0.95 & 1.00 & 1.00 & 0.98 \\ 
   \hline
  Lasso-DR, asymptotic SE & 0.97 & 1.02 & 1.07 & 1.06 \\ 
  Lasso-DR, boostrap SE & 0.97 & 1.03 & 1.01 & 1.01 \\ 
   \hline
  Debiasing, asymptotic SE & 0.99 & 1.05 & 1.05 & 1.02 \\ 
  Debiasing, boostrap SE & 0.95 & 1.00 & 0.99 & 0.99 \\ 
   \hline
  DML, asymptotic SE & 0.92 & 0.94 & 0.94 & 0.92 \\ 
  DML, boostrap SE & 0.94 & 0.98 & 0.99 & 0.96 \\
   \hline
  DML post selection, asymptotic SE & 0.92 & 0.96 & 0.97 & 0.96 \\ 
  DML post selection, boostrap SE & 0.95 & 1.00 & 1.00 & 0.99 \\ 
   \hline
  TMLE, asymptotic SE & 0.89 & 0.91 & 0.90 & 0.92 \\ 
  TMLE, boostrap SE & 0.95 & 1.00 & 0.99 & 0.99 \\ 
   \hline
  TMLE post selection, asymptotic SE & 0.94 & 0.98 & 0.98 & 0.93 \\ 
  TMLE post selection, boostrap SE & 0.96 & 1.01 & 1.00 & 0.94 \\ 
   \hline
\end{tabular}
}
\end{table}

\section{Illustration of how asymptotics suffer in high-dimensions}

In this section we compare the finite sample variance of our proposed estimator and the one proposed by \cite{farrell2015robust}. We simulate data from sparse, linear models for both the treatment and outcome. We apply our doubly robust estimator with Bayesian linear models and sparsity inducing priors as described in Section 3 and 4 of the manuscript. To build a doubly robust estimator, \cite{farrell2015robust} fit lasso (or group lasso) regressions \citep{tibshirani1996regression,yuan2006model} on both a treatment and outcome model to identify covariates that are associated with the treatment and outcome, respectively. Then, they re-fit non-penalized estimators of the treatment and outcome models using only the covariates identified by the original lasso regressions. From these two regressions they can calculate the doubly robust estimator defined in Equation 1 where $p_{ti}$ and $m_{ti}$ are estimated using the non-penalized regression models. The authors derived some important theoretical results that demonstrate that their proposed double robust estimator is consistent and asymptotically normal. Our goal of this brief illustration is to elucidate why utilizing Bayesian methods to account for parameter uncertainty, which do not rely on asymptotics, can provide a more accurate assessment of the finite sample uncertainty, especially in high-dimensional scenarios. Here we focus on the estimator from \cite{farrell2015robust} as it uses the exact same doubly robust estimator, with the main difference coming in how inference is performed. As seen in the simulation study of the main manuscript, these ideas extend to other estimators rooted in asymptotics.

\begin{figure}[H]
\centering
 \includegraphics[width = 0.8\linewidth]{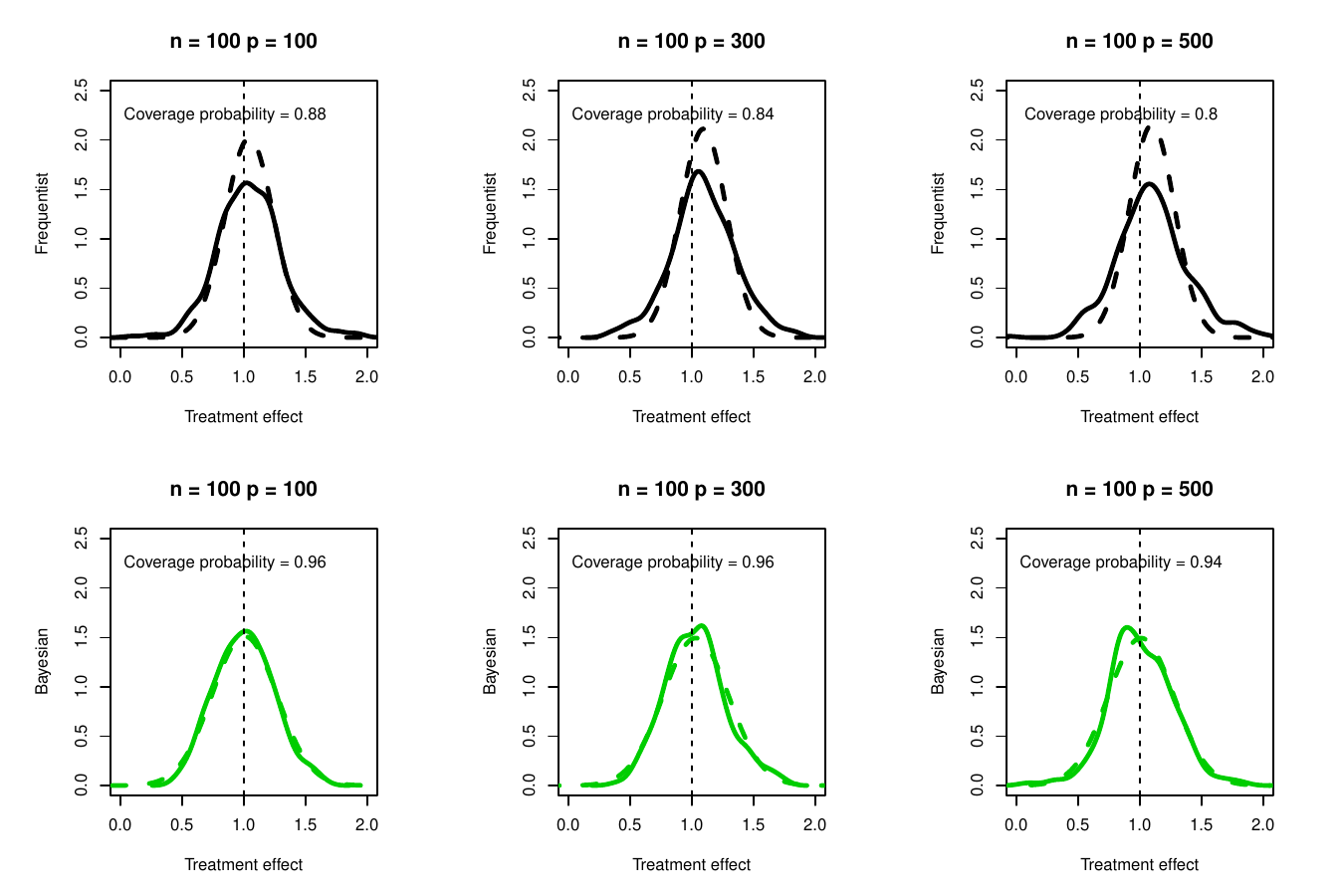}
\caption{Comparison of empirical and assumed distributions for the doubly robust estimator of \cite{farrell2015robust} and our approach from Section 3. The solid lines are the empirical sampling distributions, while the dashed lines are normal distributions with standard deviation equal to the average estimated standard deviation across the simulations}
\label{fig:BayesVsFreq}
\end{figure}

For each of the two doubly robust estimators, we plot two lines. First, we show the sampling distribution of the estimator as taken by the empirical distribution of the estimates across a large number of simulated datasets. Next, we plot a normal density centered at the mean of the estimates across all datasets with a standard deviation that is the average estimated standard error across all datasets. If the estimated standard errors are correct, then this average standard error should be the same as the standard deviation of the sampling distribution and the two curves should look similar. \ref{fig:BayesVsFreq} shows the results for $n=100$ and $p \in \{100,300,500\}$. The top row shows the results for the estimator based on asymptotic confidence intervals and the dashed line has much smaller tails than the solid line, indicating that the asymptotic distribution used for inference is not properly accounting for the uncertainty in the estimator. This phenomenon gets worse as $p$ grows larger, and we see that the coverage probabilities decrease from 88\% to 80\%. Our approach to the same estimator, however, maintains the correct coverage probabilities for any dimension of the data, and the dashed and solid lines are very similar, showing that the uncertainty in the estimator is fully accounted for. 

\bibliographystyle{authordate1}
\bibliography{SSLconfounding}

\end{document}